\documentclass[aps, prx, twocolumn, footinbib, superscriptaddress]{revtex4-2}

\usepackage{soul}
\soulregister\cite7
\soulregister\ref7
\soulregister\eqref7
\usepackage{xcolor}
\usepackage{comment}
\usepackage[T1]{fontenc}
\usepackage{amsmath,amsfonts,amssymb,amsthm,bbm,bm,braket}
\usepackage{wasysym}
\usepackage{enumerate}
\usepackage{graphicx}
\usepackage{mathtools}
\usepackage{color}
\usepackage{lipsum}  
\usepackage{phfqit}  
\usepackage{tikz}
\usepackage{derivative}
\usetikzlibrary{shadings, decorations.pathreplacing, arrows.meta, decorations.text}
\definecolor{linkblue}{rgb}{0.0,0.2,0.6}
\usepackage[colorlinks=true,linkcolor=linkblue,citecolor=linkblue,urlcolor=linkblue,bookmarksnumbered=true]{hyperref}

\definecolor{Red}{rgb}{1,0,0}
\definecolor{Blue}{rgb}{0.5,0,0.9}
\definecolor{Green}{rgb}{0,0.5,0}
\definecolor{Magenta}{rgb}{1,0,0.56}
\definecolor{Orange}{rgb}{1,0.64,0}

\definecolor{LightGreen}{rgb}{0.20,0.68,0.33}
\definecolor{DarkGreen}{rgb}{0.00,0.32,0.10}

\definecolor{blue}{rgb}{0,0,1}
\definecolor{grey}{rgb}{0.6,0.6,0.6}
\definecolor{spinRed}{RGB}{227, 26, 28}
\definecolor{spinBlue}{RGB}{31, 120, 180}
\definecolor{scLead}{RGB}{166, 206, 227}
\definecolor{rotGreen}{RGB}{0, 0, 0}
\definecolor{gap1}{RGB}{227, 149, 23}
\definecolor{gap2}{RGB}{20, 186, 219}

\newcommand\dif[1]{\mathrm{d}#1\,}

\begin{document}

\title{Spin-controlled thermophase and noise-optimized thermometry \\in multiorbital quantum dots}

\author{Markus Plautz}
\email{markus.plautz@ntnu.no}
\author{Jeroen Danon}
\author{Jacob Linder}
\email{jacob.linder@ntnu.no}
\affiliation{Center for Quantum Spintronics, Department of Physics, Norwegian \\ University of Science and Technology, NO-7491 Trondheim, Norway}
\date{\today}

\begin{abstract}
Converting temperature differences into electrical signals underlies both thermoelectric energy conversion and thermometry, which is essential for managing heat in cryogenic quantum technologies.
Superconducting circuits typically show a weak thermoelectric response,
since their quasiparticle spectrum is nearly particle-hole symmetric. In
a Josephson junction, a temperature difference can instead be converted
into a superconducting phase difference in an open circuit.
We investigate this thermophase effect in a setup where the junction consists of a two-orbital quantum dot with Rashba spin-orbit coupling (SOC) and Zeeman splitting. A second orbital near
the superconducting coherence peaks supports quasiparticle transport
over a broad gate range. Zeeman splitting can enhance the thermophase
by suppressing pair transport, while spin-orbit hybridization shifts
the orbital resonances and can reverse the sign of the response.
We assess the junction as a phase-based thermometer using a noise
budget that includes readout and current noise
together with fluctuations of the orbital energies and the magnetic
field. The optimal operating points do not maximize the thermophase but balance a large quasiparticle response against sensitivity to parameter fluctuations. Under experimentally motivated noise assumptions, we obtain projected
noise-equivalent temperatures of order
$10\,\mu\mathrm{K}/\sqrt{\mathrm{Hz}}$ and
determine the range over which a linear calibration remains accurate. Gate-tunable quantum-dot junctions thus offer a phase-based route to sensing temperature differences in superconducting circuits at sub-kelvin temperatures.
\end{abstract}

\maketitle

\noindent
\section{Introduction}\label{sec:introduction}

Managing heat at cryogenic temperatures is central to the energy budget
of quantum technologies, where the available cooling power is
small~\cite{krinnerEngineeringCryogenicSetups2019,pekolaColloquiumQuantumHeat2021}.
Measuring temperature and heat flow at the nanoscale is therefore
important for cryogenic detectors and thermal
circuits~\cite{giazottoOpportunitiesMesoscopicsThermometry2006}.
Superconducting circuits are natural candidates for this task since they
operate in this regime and offer phase-sensitive readout~\cite{gasparinettiFastElectronThermometry2015,sairaDispersiveThermometryJosephson2016,zgirskiNanosecondThermometryJosephson2018}. However, the
nearly particle-hole symmetric quasiparticle spectrum of a conventional
superconductor suppresses the linear thermoelectric response of
superconducting junctions. Electron-like and hole-like excitations then
carry opposing currents that largely cancel. A substantial response
requires a junction that treats these excitations differently.

A junction that combines spin splitting with
spin-selective tunneling provides one route to this end~\cite{machonNonlocalThermoelectricEffects2013,ozaetaPredictedVeryLarge2014,linderSpinCaloritronicsSuperconductors2016,kolendaObservationThermoelectricCurrents2016,bergeretColloquiumNonequilibriumEffects2018,heikkilaThermalElectricSpin2019}. This mechanism has recently enabled magnetically controllable thermoelectric heat engines and spin valves based on superconducting-ferromagnetic hybrids \cite{araujoSuperconductingSpintronicHeat2024, tueroLargeTunableThermoelectric2025}.
Quantum dots provide an alternative through straightforwardly tunable transmission resonances~\cite{hwangCrossThermoelectricCoupling2015,kleeorinLargeTunableThermophase2016,kampPhasedependentHeatCharge2019}.
Magnetic impurities, magnetic textures or spin-orbit coupling also offer
control over thermoelectric transport in superconducting
hybrids~\cite{kalenkovTheoryLargeThermoelectric2012,bathenSpinSeebeckEffect2017}.
Josephson junctions further link thermal transport to the superconducting
phase~\cite{fornieriPhasecoherentCaloritronicsSuperconducting2017}. This link has enabled
phase-controlled heat flow~\cite{giazottoJosephsonHeatInterferometer2012} as well
as thermoelectric energy conversion once particle-hole symmetry is
broken~\cite{marchegianiPhasetunableThermoelectricityJosephson2020,germaneseBipolarThermoelectricJosephson2022}.

In an open-circuit Josephson junction, a temperature bias can
produce a phase difference rather than a voltage. The thermally driven
quasiparticle current is compensated by a supercurrent, establishing
a stationary thermophase~\cite{ginzburgThermoelectricEffectsSuperconducting1991,guttmanThermoelectricThermophaseEffects1997}.
Large thermophases have been predicted in ferromagnetic
junctions~\cite{giazottoVeryLargeThermophase2015}
and gate-tunable single-quantum-dot junctions~\cite{kleeorinLargeTunableThermophase2016}.
Quantum dots are particularly attractive for this purpose. Their discrete
levels make the transmission strongly energy dependent, while gate
voltages shift these levels \textit{in situ}~\cite{dohTunableSupercurrentSemiconductor2005,jarilloherreroQuantumSupercurrentTransistors2006,defranceschiHybridSuperconductorQuantum2010}. 
Studies of double-quantum-dot Josephson junctions further show that additional levels and
interdot coupling modify the thermophase~\cite{kumarPhaseThermalDrivenTransport2023,kumarJosephsonThermophaseEffect2025}.

Semiconductor quantum dots offer additional controls beyond the level
position. Dots in InAs or InSb nanowires typically have strong spin-orbit coupling~\cite{fasthDirectMeasurementSpinOrbit2007,nadj-pergeSpinOrbitQubit2010,hanVariableOrbitalDependent2023}
and large, level-dependent~$g$
factors~\cite{csonkaGiantFluctuationsGate2008,nilssonGiantLevelDependent2009,muMeasurementsAnisotropicGfactors2020}. Spin-orbit coupling hybridizes the orbitals, while a magnetic field
splits their spin states, as sketched in Fig.~\ref{fig:setup}(a).
The effects of these ingredients on the Josephson current have been studied
~\cite{dellannaJosephsonCurrentQuantum2007,
zazunovAnomalousJosephsonCurrent2009,limJosephsonCurrentCarbon2011,brunettiAnomalousJosephsonCurrent2013,yokoyamaAnomalousJosephsonEffect2014}. Spin-resolved Andreev states and spin-dependent phase shifts have also
been observed in semiconductor quantum-dot
junctions~\cite{leeSpinresolvedAndreevLevels2014,szombatiJosephsonF0junctionNanowire2016,whiticarZeemandrivenParityTransitions2021,bargerbosSpectroscopySpinsplitAndreev2023}. To the best of our knowledge, however, how spin-orbit coupling and Zeeman splitting in a multiorbital dot
shape the thermophase has not been addressed so far.

A thermophase can be detected with a superconducting interferometer, as
sketched in Fig.~\ref{fig:setup}(b). The quantum-dot junction is placed
in a superconducting loop together with a conventional reference
Josephson junction with a suppressed thermoelectric response. 
A SQUID detects the circulating current in the loop,
while an external flux is tuned until this current vanishes. The
temperature-induced shift of this compensating flux then gives the
open-circuit thermophase directly (Sec.~\ref{sec:measuring}). Related interferometric schemes have been used to measure the
current-phase relation~\cite{dellaroccaMeasurementCurrentphaseRelation2007,
spantonCurrentphaseRelationsFewmode2017,nicheleRelatingAndreevBound2020}
and anomalous phase shifts~\cite{szombatiJosephsonF0junctionNanowire2016,
assoulineSpinorbitInducedPhaseshift2019,mayerGateControlledAnomalous2020,
strambiniJosephsonPhaseBattery2020} of Josephson junctions.
Thermoelectric circulating currents in superconducting loops have
likewise been detected magnetically~\cite{zavaritskiiThermoelectric1974,
vanharlingenExperimentalStudyThermoelectricity1980}.

For thermometry, however, a large phase response alone does not ensure
good temperature resolution. In the interferometric setup, the resolution is
limited by the flux noise of the readout and by current fluctuations in
the junction circuit. Fluctuations of the gate voltages or of the
magnetic field add further noise because they shift the calibration of
the thermometer. The response must therefore be weighed against these
noise sources, as in other superconducting
thermometers~\cite{giazottoFerromagneticInsulatorBasedSuperconductingJunctions2015,sairaDispersiveThermometryJosephson2016}.

Here, we investigate thermophase generation through a two-orbital quantum
dot. Spin-orbit coupling acts  on the dot
and in the tunneling amplitudes, while a Zeeman field acts on the dot.
We show how these controls affect quasiparticle transport and the Josephson response differently, changing
both the magnitude and the sign of the thermophase Seebeck coefficient.
We then assess the junction as a thermometer using the noise-equivalent
temperature, which combines
readout and current noise with fluctuations of the orbital energies and
the magnetic field. Comparing the signal and noise identifies
operating points that would be missed by maximizing the Seebeck
coefficient alone. Finally, we determine the range over which a local
linear calibration remains accurate. Using experimentally motivated noise assumptions, we find optimized operating points reaching noise-equivalent temperatures of order~$10\,\mu\mathrm{K}/\sqrt{\mathrm{Hz}}$, comparable to state-of-the-art electron thermometers based on superconducting junctions~\cite{gasparinettiFastElectronThermometry2015,sairaDispersiveThermometryJosephson2016,karimiNoninvasiveThermometerBased2018}. Taken together, our results identify a spinful multiorbital Josephson junction as a tunable platform for cryogenic thermometry using the thermophase response.
 
\section{Setup and methods}\label{sec:model}
 
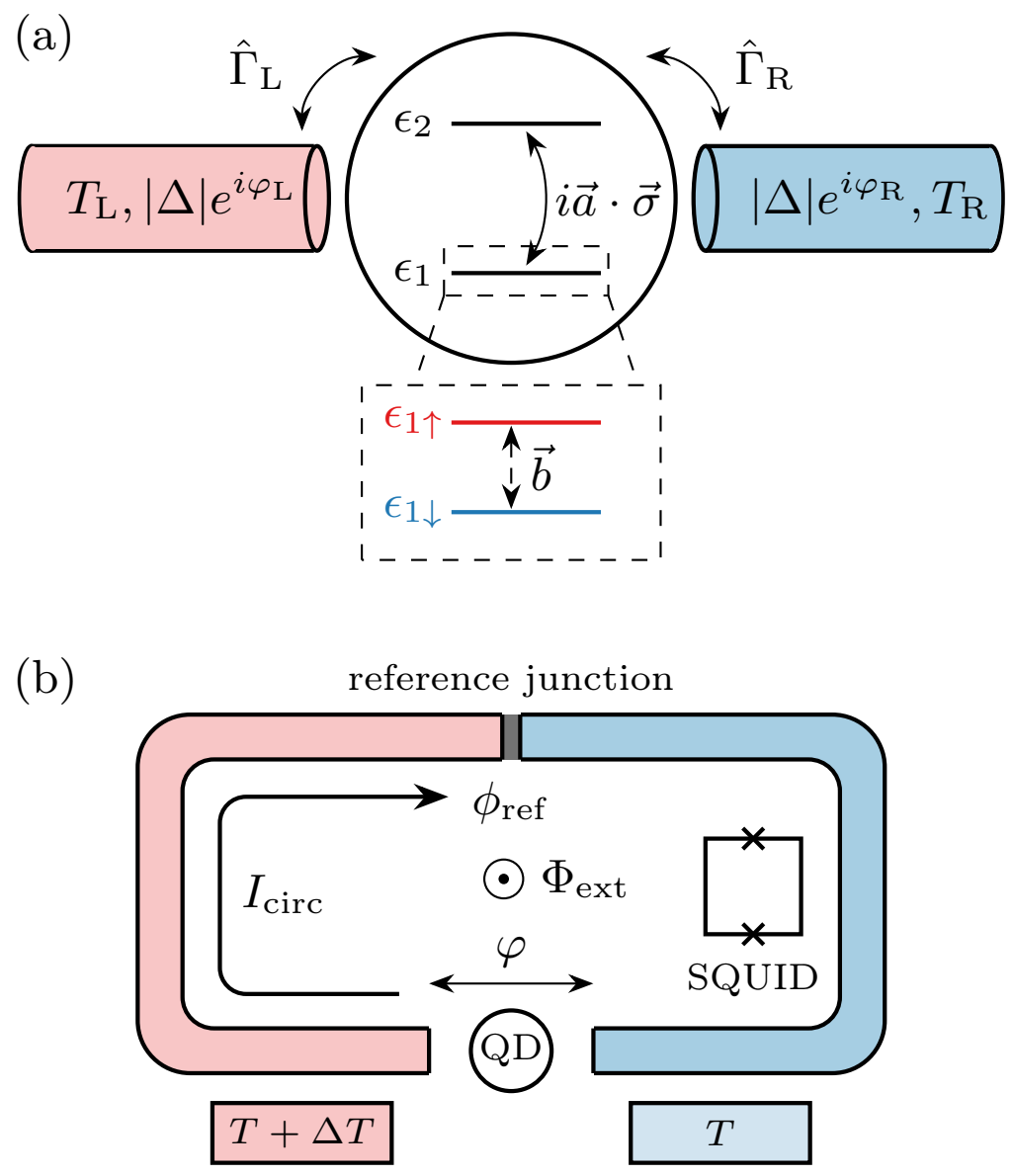
\begin{figure}[t]
    \centering 
    \begin{tikzpicture}[scale=1.0]
        \path (-3.45,0) -- (3.45,0);
        \node[anchor=north west] at (-3.45,1.35) {(a)};

        \draw[color=black, thick] (-3.21,0) ellipse (0.08 and 0.35);
        \draw[color=black, thick] (-3.22, -0.35) -- ++(1.9, 0);
        \draw[color=black, thick] (-3.22, 0.35) -- ++(1.9, 0);

        \fill[color=spinRed!25] (-3.2,-0.34) rectangle (-1.3,0.34);
        \fill[color=spinRed!25] (-3.2,0) ellipse (0.08 and 0.35);
        \fill[color=spinRed!25] (-1.3,0) ellipse (0.08 and 0.35);
        \draw[color=black, thick] (-1.3,0) ellipse (0.08 and 0.35);
        \node at (-2.2,0) {\small $T_{\rm L}, \abs{\Delta}e^{i\varphi_{\mathrm{L}}}$};

        \draw[color=black, thick] (3.21,0) ellipse (0.08 and 0.35);
        \fill[color=scLead] (1.3,0) ellipse (0.08 and 0.35);
        \fill[color=scLead] (3.2,0) ellipse (0.08 and 0.34);
        
        \fill[color=scLead] (1.3,-0.34) rectangle (3.2,0.34);
        \draw[color=black, thick] (1.3, -0.35) -- ++(1.9, 0);
        \draw[color=black, thick] (1.3, 0.35) -- ++(1.9, 0);
        \draw[color=black, thick] (1.3,0) ellipse (-0.08 and 0.35);
        \node at (2.4,0) {\small $\abs{\Delta}e^{i\varphi_{\mathrm{R}}}, T_{\rm R}$};

        \draw[thick] (0,0) circle (1.1cm);
        \draw[color=black, thick] (-0.4, 0.5) -- ++(1, 0);
        \draw[color=black, thick] (-0.4, -0.5) -- ++(1, 0);

        \node at (-0.65, 0.5) {\small $\epsilon_{2}$};
        \node at (-0.65, -0.5) {\small $\epsilon_{1}$};
        \node at (0.65, 0) {\small $i\vec{a}\cdot\vec{\sigma}$};

        \draw[color=black, dashed] (-0.45,-0.65) rectangle (0.65,-0.32);
        \draw[color=black, dashed] (-0.45,-0.65) -- ++(-0.2, -0.6);
        \draw[color=black, dashed] (0.65,-0.65) -- ++(0.2, -0.6);
        \draw[color=black, dashed] (-1,-2.42) rectangle (1, -1.25);

        \draw[<->, >=Stealth] (0.1,-0.45) arc[start angle=-30, end angle=30, radius=0.9cm];

        \draw[color=spinRed, thick] (-0.4, -1.5) -- ++(1, 0);
        \draw[color=spinBlue, thick] (-0.4, -2.1) -- ++(1, 0);
        \draw[<->, >=Stealth, densely dashed] (0, -1.52) -- ++(0, -0.56);
        \node at (0.2, -1.8) {\small $\vec{b}$};
        \node[color=spinBlue] at (-0.65, -2.1) {\small $\epsilon_{1\downarrow}$};
        \node[color=spinRed] at (-0.65, -1.5) {\small $\epsilon_{1\uparrow}$};
        \draw[<->, >=Stealth] (1.4,0.45) arc[start angle=0, end angle=90, radius=0.5cm];
        \draw[<->, >=Stealth] (-1.4,0.45) arc[start angle=180, end angle=90, radius=0.5cm];
        \node at (1.7, 0.9) {\small $\hat{\Gamma}_{\mathrm{R}}$};
        \node at (-1.7, 0.9) {\small $\hat{\Gamma}_{\mathrm{L}}$};

    \end{tikzpicture}

    \vspace{0.5cm}

    \begin{tikzpicture}[scale=1.0, >=Stealth]
        \path (-3.45,0) -- (3.45,0);
        \node[anchor=north west] at (-3.45,1.75) {(b)};
        \fill[scLead, even odd rule, rounded corners=10pt]
            (-2.5,-1.15) rectangle (2.5,1.25)
            [rounded corners=6pt] (-2.2,-0.85) rectangle (2.2,0.95);
        \begin{scope}
            \clip (-2.6,-1.3) rectangle (0,1.4);
            \fill[spinRed!25, even odd rule, rounded corners=10pt]
                (-2.5,-1.15) rectangle (2.5,1.25)
                [rounded corners=6pt] (-2.2,-0.85) rectangle (2.2,0.95);
        \end{scope}
            [rounded corners=6pt] (-2.2,-0.85) rectangle (2.2,0.95);
        \draw[thick, rounded corners=10pt] (-2.5,-1.15) rectangle (2.5,1.25);
        \draw[thick, rounded corners=6pt] (-2.2,-0.85) rectangle (2.2,0.95);
        \fill[white] (-0.55,-1.2) rectangle (0.55,-0.8);
        \draw[thick] (-0.55,-1.15) -- (-0.55,-0.85);
        \draw[thick] ( 0.55,-1.15) -- ( 0.55,-0.85);
        \draw[thick, fill=white] (0,-1.0) circle (0.27);
        \node at (0,-1.0) {\scriptsize QD};
        \fill[white] (-0.06,0.9) rectangle (0.06,1.3);
        \fill[black!55] (-0.06,0.95) rectangle (0.06,1.25);
        \draw[thick] (-0.06,0.95) -- (-0.06,1.25);
        \draw[thick] ( 0.06,0.95) -- ( 0.06,1.25);
        \node[above] at (0,1.25) {\scriptsize reference junction};
        \fill[spinRed!25] (-2.0,-1.75) rectangle (-0.8,-1.35);
        \draw[black, thick] (-2.0,-1.75) rectangle (-0.8,-1.35);
        \node[black] at (-1.4,-1.55) {\scriptsize $T+\Delta T$};
        \fill[spinBlue!20] (0.8,-1.75) rectangle (2.0,-1.35);
        \draw[black, thick] (0.8,-1.75) rectangle (2.0,-1.35);
        \node[black] at (1.4,-1.55) {\scriptsize $T$};
        \draw[<->] (-0.55,-0.55) -- (0.55,-0.55);
        \node[above] at (0,-0.55) {\small $\varphi$};
        \node[below] at (0,0.92) {\small $\phi_{\mathrm{ref}}$};
        \draw[->, thick, rounded corners=5pt]
            (-0.75,-0.62) -- (-1.95,-0.62) -- (-1.95,0.70) -- (-0.42,0.70);
        \node[right] at (-1.92,0.05) {\small $I_{\mathrm{circ}}$};
        \draw (-0.05,0.15) circle (0.13);
        \fill (-0.05,0.15) circle (0.035);
        \node[right] at (0.08,0.15) {\small $\Phi_{\mathrm{ext}}$};
        \begin{scope}[shift={(1.62,0.1)}]
            \draw[thick, fill=white] (-0.32,-0.32) rectangle (0.32,0.32);
            \draw[thick] (-0.07,0.25) -- (0.07,0.39);
            \draw[thick] (-0.07,0.39) -- (0.07,0.25);
            \draw[thick] (-0.07,-0.39) -- (0.07,-0.25);
            \draw[thick] (-0.07,-0.25) -- (0.07,-0.39);
            \node[below] at (0,-0.4) {\scriptsize SQUID};
        \end{scope}
    \end{tikzpicture}

    \vspace{0.3cm}

    \caption{(a) Sketch of the junction. A quantum dot with two orbital
    levels~$\epsilon_1$ and~$\epsilon_2$ is tunnel coupled to left~($L$)
    and right~($R$)~$s$-wave superconducting leads with order
    parameters~$\abs{\Delta_{\alpha}}e^{i\varphi_\alpha}$. A temperature
    difference~$\Delta T$ is applied across the junction, with the left
    lead at~$T_L=T+\Delta T$ (red shading) and the right lead
    at~$T_R=T$ (blue shading). Rashba spin-orbit interaction~$i\vec{a}\cdot\vec{\sigma}$
    couples the two orbitals. The coupling matrices~$\hat{\Gamma}_\alpha$
    include spin-conserving as well as spin-orbit-assisted tunneling. The
    inset shows how a Zeeman field~$\vec{b}$ lifts the spin degeneracy of
    orbital~$\epsilon_1$ into the levels~$\epsilon_{1\uparrow}$
    and~$\epsilon_{1\downarrow}$. The effects of intra- and inter-orbital Coulomb interactions at the
    mean-field level are discussed in Appendix~\ref{app:coulomb}.
    (b)~Measurement scheme. The quantum-dot junction and a conventional
    reference junction form a superconducting interferometer whose two
    arms act as the leads~$L$ and~$R$ of panel~(a).
    The circulating current~$I_{\mathrm{circ}}$ is detected by a SQUID that
    senses its magnetic flux. The thermophase follows from the external
    flux~$\Phi_{\mathrm{ext}}$ at which~$I_{\mathrm{circ}}$ vanishes
    (Sec.~\ref{sec:measuring}).}
    \label{fig:setup}
\end{figure}

We consider a Josephson junction consisting of a quantum dot with two orbital levels 
coupled to left ($L$) and right ($R$) $s$-wave superconducting leads, as illustrated 
in Fig.~\ref{fig:setup}. The quantum dot is subject to Rashba spin-orbit coupling and 
an external Zeeman field, and we allow the tunneling between the dot and the leads to be 
both spin-conserving and spin non-conserving. Coulomb interactions on the dot are introduced below and treated at the
mean-field level in Appendix~\ref{app:coulomb}. The full Hamiltonian reads

\begin{equation}\label{eq:total_hamiltonian}
    H = H_L + H_R + H_\mathrm{QD} + H_T + H_U,
\end{equation}
where~$H_\alpha$ ($\alpha = L, R$) describes the superconducting leads,~$H_\mathrm{QD}$ 
the quantum dot,~$H_T$ the tunneling, and~$H_U$ the Coulomb interaction. 
We now introduce each of these terms.
 
\textit{Superconducting leads.}---Each lead is modeled as a conventional~$s$-wave 
superconductor,
\begin{equation}\label{eq:lead_hamiltonian}
    H_\alpha = \sum_{k,\sigma} \epsilon_\alpha(k) 
    c_{\alpha k\sigma}^\dagger c_{\alpha k\sigma} + \big(
    \Delta_\alpha c_{\alpha k\uparrow}^\dagger c_{\alpha,-k,\downarrow}^\dagger + \mathrm{h.c.} \big),
\end{equation}
where~$\epsilon_\alpha(k)$ is the single-particle dispersion measured from the chemical 
potential and~$\Delta_\alpha = \abs{\Delta_\alpha}e^{i\phi_\alpha}$ is the 
superconducting order parameter. We adopt the phase convention~$\phi_L = -\phi_R = \phi/2$, 
so that~$\phi$ denotes the phase difference across the 
junction.
 
\textit{Quantum dot.}---The dot hosts two orbital levels with energies $\epsilon_1$ 
and~$\epsilon_2$. We include Rashba spin-orbit coupling on the dot by projecting the Rashba 
Hamiltonian~$H_{\rm so} = \alpha_R (\sigma_x p_y - \sigma_y p_x)$ onto the orbital 
eigenstates~$\chi_n(\vec{r}) = \braket{\vec r}{n}$ of the confined dot. The resulting matrix elements~\cite{dellannaJosephsonCurrentQuantum2007, zazunovAnomalousJosephsonCurrent2009}
\begin{equation}\label{eq:rashba_projection}
    \matrixel{n}{H_{\rm so}}{n'} = -i\,\vec{a}_{nn'} \cdot \vec{\sigma}
\end{equation}
are parametrized by a vector~$\vec{a}_{nn'}$ whose components are determined by 
the Rashba coupling strength~$\alpha_R$ and overlap integrals of the orbital wave 
functions. Time-reversal symmetry requires $\vec{a}_{nn'}$ to be real, and Hermiticity imposes $\vec a_{nn'} = -\vec a_{n'n}$, and thus~$\vec{a}_{nn} = 0$. At least two orbital levels are therefore 
needed for spin-orbit coupling to have any effect on the dot. For two orbitals, we 
define~$\vec{a} \equiv \vec{a}_{12}$. Together with an external Zeeman field~$\vec{b}$, 
the quantum dot Hamiltonian reads
\begin{align}\label{eq:dot_hamiltonian}
    \begin{split}
        H_\mathrm{QD} = \sum_{j,\sigma} &\epsilon_j n_{j\sigma} 
        + \sum_{j,\sigma\sigma'} (\vec{b} \cdot \vec{\sigma})_{\sigma\sigma'}
        d_{j\sigma}^\dagger d_{j\sigma'} \\
        &+ \sum_{\sigma\sigma'} \left[
            (-i\vec{a} \cdot \vec{\sigma})_{\sigma\sigma'}\, 
            d_{1\sigma}^\dagger d_{2\sigma'} + \mathrm{h.c.}
        \right],
    \end{split}
\end{align}
where~$n_{j\sigma} = d_{j\sigma}^\dagger d_{j\sigma}$ 
and~$\vec{\sigma} = (\sigma_x, \sigma_y, \sigma_z)$ is the vector of spin Pauli matrices.
We take the same Zeeman field for both orbitals. The spin-orbit mixing
between the two orbitals is treated explicitly through~$\vec{a}$,
while~$\vec{b}$ includes the renormalization of the~$g$ factor by remote
bands and orbitals~\cite{winklerSpinOrbitCoupling2003}.
 
\textit{Tunneling.}---The coupling between the dot and the leads allows for both spin 
conserving and spin-orbit-assisted 
hopping~\cite{danonPauliSpinBlockade2009},
\begin{equation}\label{eq:tunneling}
    H_T = \sum_{\substack{\alpha, k, j \\ \sigma\sigma'}}\left[
        t_{\alpha j} \delta_{\sigma\sigma'} + 
        i\eta_\alpha t_{\alpha j}^\mathrm{so}(\sigma_y)_{\sigma\sigma'}
    \right] c_{\alpha k\sigma}^\dagger d_{j\sigma'} + \mathrm{h.c.}
\end{equation}
Here,~$t_{\alpha j}$ and~$t_{\alpha j}^\mathrm{so}$ are real tunneling amplitudes for 
spin-conserving and spin-orbit-assisted tunneling processes to orbital~$j$ of lead~$\alpha$, 
respectively, and~$\eta_L = -\eta_R = -1$ encodes the reversal of the spin-orbit rotation 
with hopping direction.
The spin coordinate system is chosen such that the tunneling-induced spin rotation to the leads is around $\hat y$.
 
\textit{Coulomb interaction.}---We include an intra- and inter-orbital Coulomb repulsion,
\begin{equation}\label{eq:coulomb_interaction}
    H_U = U \sum_{i=1,2} n_{i\uparrow}\, n_{i\downarrow} + U' n_1 n_2,
\end{equation}
where~$n_i = n_{i\uparrow} + n_{i\downarrow}$ is the occupation of
orbital~$i$. In semiconductor quantum dots the inter-orbital
repulsion~$U'$ is typically comparable to the intra-orbital
repulsion~$U$~\cite{kouwenhovenFewelectronQuantumDots2001}. At the
mean-field level~\cite{vecinoJosephsonCurrentCorrelated2003,mengSelfconsistentDescriptionAndreev2009,martinroderoAndreevStatesSuperconducting2012}, both terms act as renormalizations of dot parameters. 
At the mean-field level, both terms mainly renormalize the orbital energies, the Zeeman field and the inter-orbital coupling, which are already independent control parameters of our model. In spin-polarized states, the interaction additionally generates exchange fields that differ between the orbitals or couple them. Appendix~\ref{app:coulomb} shows that these act like orbital-dependent Zeeman fields. We therefore set~$U=U'=0$ in the main text.
 
\textit{Methodology.}---We compute the current through the junction using the 
Keldysh nonequilibrium Green's function formalism. 
The total current is given by~\cite{kleeorinLargeTunableThermophase2016}
\begin{equation}\label{eq:current}
    J = -\frac{e}{h} \int \dif\omega\, \Re \tr_p \left[
        \left(\hat{\Sigma}_L - \hat{\Sigma}_R\right) \hat{G} 
    \right]^{<},
\end{equation}
where the square bracket is evaluated using the Langreth relation~$[AB]^{<}=A^rB^{<}+A^{<}B^a$.
In this context,~$\hat{G}^{r,a,<}$ are the retarded, advanced, and lesser Green's functions of 
the dot and~$\hat{\Sigma}_\alpha^{r,a,<}$ the corresponding self-energies due to lead $\alpha$, all of which are~$8\times 8$ matrices in the combined Nambu, orbital, and
spin space. The subscript~$p$ denotes a projection onto the particle block of
Nambu space, so that no explicit Nambu charge matrix $\tau_3$ is required, followed by a trace over the remaining orbital and spin 
indices,~$\tr_p[\hat{X}] = \sum_{j \sigma} X_{j\sigma,j\sigma},$
with both indices restricted to the particle sector.
The
self-energies are obtained by integrating out the leads within the wide-band
approximation~\cite{sunElectronTransportMesoscopic2000}. Explicit expressions for these quantities
are given in Appendix~\ref{app:greens_function}.

\textit{Thermoelectric response.}---We apply a temperature bias~$\Delta T$ across the
junction by setting~$T_L = T + \Delta T$ and~$T_R = T$, as indicated in
Fig.~\ref{fig:setup}. In an open superconducting
junction, the thermally induced quasiparticle current can be canceled by a
counterflowing supercurrent. The phase difference that enforces zero total current
is called the thermophase and is denoted by~$\phi_\mathrm{th}$~\cite{kleeorinLargeTunableThermophase2016},
\begin{equation}\label{eq:thermophase}
    J\left(\phi_\mathrm{th},\Delta T\right) = 0.
\end{equation}
Such a stationary solution exists only as long as the supercurrent can
compensate the thermally induced quasiparticle current,
i.e., for~$\abs{J_\mathrm{qp}}<I_c$ with~$I_c$ the critical current of the
junction. For larger temperature differences, a thermovoltage develops
across the junction. The phase then winds in time according to the ac
Josephson relation~\cite{mansikkamakiThermoelectricACJosephson2025}, and the junction no longer has a stationary
thermophase. We restrict ourselves to the stationary regime throughout.
The thermophase Seebeck coefficient is then defined as the differential response
\begin{equation}\label{eq:seebeck}
    S_\phi = \odv{\phi_\mathrm{th}}{(\Delta T)},
\end{equation}
which quantifies how sensitively the stationary phase responds to a temperature
difference. This definition does not assume linear response and captures the full
nonlinear dependence of~$\phi_\mathrm{th}$ on~$\Delta T$.
In the linear-response regime, the current decomposes
as~$J(\phi) = J_\mathrm{sc}(\phi) + \mathcal{L}(\phi)\,\Delta T$,
where~$J_\mathrm{sc}$ is the equilibrium supercurrent and~$\mathcal{L}$ is a
quasiparticle transport coefficient (explicit expressions are given in
Appendix~\ref{app:current}). Near the equilibrium phase~$\phi_0$,
where~$J_\mathrm{sc}(\phi_0)=0$, we write~$\phi_\mathrm{th}=\phi_0+\delta\phi$
and expand this expression to first order. We then find
\begin{equation}\label{eq:seebeck_linear}
    S_\phi = -\frac{\mathcal{L}(\phi_0)}{K_\phi},
    \qquad
    K_\phi \equiv \partial_\phi J_\mathrm{sc}\big|_{\phi_0}.
\end{equation}
Here,~$K_\phi$ is the local current-phase slope. The equilibrium
phase~$\phi_0$ is the stable zero of the equilibrium supercurrent and need
not vanish in general. A singly occupied dot level, for example induced by
a Zeeman field, can reverse the sign of the Josephson coupling and turn
the junction into a~$\pi$ junction with~$\phi_0=\pi$~\cite{glazmanResonantJosephsonCurrent1989,
spivakNegativeLocalSuperfluid1991,zhuAndreevBoundStates2001,vecinoJosephsonCurrentCorrelated2003,
vandamSupercurrentReversalQuantum2006,cleuziouCarbonNanotubeSuperconducting2006,
jorgensenCriticalCurrentTransition2007,martin-roderoJosephsonAndreevTransport2011}.
Spin-orbit coupling combined with a Zeeman field can further produce
a~$\phi_0$ junction with an arbitrary equilibrium
phase~\cite{buzdinDirectCouplingMagnetism2008,
zazunovAnomalousJosephsonCurrent2009,brunettiAnomalousJosephsonCurrent2013,
yokoyamaAnomalousJosephsonEffect2014,szombatiJosephsonF0junctionNanowire2016,
assoulineSpinorbitInducedPhaseshift2019}. In our model, such an anomalous phase additionally requires two orbitals
with different coupling matrices for the two
leads~\cite{zazunovAnomalousJosephsonCurrent2009}. Spin-orbit-assisted
tunneling provides such a difference even for equal tunneling strengths,
since the spin rotation it induces reverses with the hopping direction. For the identical coupling matrices considered here,
exchanging the leads maps~$\phi\to-\phi$. The equilibrium supercurrent is
then odd in~$\phi$, which restricts~$\phi_0$ to~$0$ or~$\pi$. For all
parameters studied in Sec.~\ref{sec:results}, the sign of~$K_\phi$
at~$\phi=0$ corresponds to a stable~$0$ junction over the whole gate
range, even where a Zeeman field brings the junction close to a~$0$-$\pi$
transition. We therefore evaluate~$\mathcal{L}(\phi_0)$ and~$K_\phi$
at~$\phi_0=0$. In a~$\pi$ junction, Eq.~\eqref{eq:seebeck_linear} applies
unchanged with both coefficients evaluated at~$\phi_0=\pi$.
Unless stated otherwise, the numerical results in Sec.~\ref{sec:results}
are obtained in linear response from Eq.~\eqref{eq:seebeck_linear}. The nonlinear
definition in Eqs.~\eqref{eq:thermophase} and~\eqref{eq:seebeck} is used when
we explicitly consider finite temperature differences, in particular when
determining the linear operating range in
Sec.~\ref{sec:sensing_linear_range}.

\section{Results}\label{sec:results}

We now present our numerical results for the thermoelectric response of the
junction. Throughout, energies are measured in units of the superconducting
gap~$\abs{\Delta}$, and temperatures in units of~$\abs{\Delta}/k_{\mathrm{B}}$. We
assume symmetric leads with equal gaps and coupling strengths.
We first validate our formalism against previous results for a single
spin-degenerate level and then systematically investigate the roles of the
second orbital, spin-orbit coupling, and the Zeeman field. Section~\ref{sec:sensing} then 
assesses the potential of the junction as a quantum sensor for small
temperature differences.

\begin{figure*}[t!]
    \centering
    \includegraphics[width=.95\textwidth]{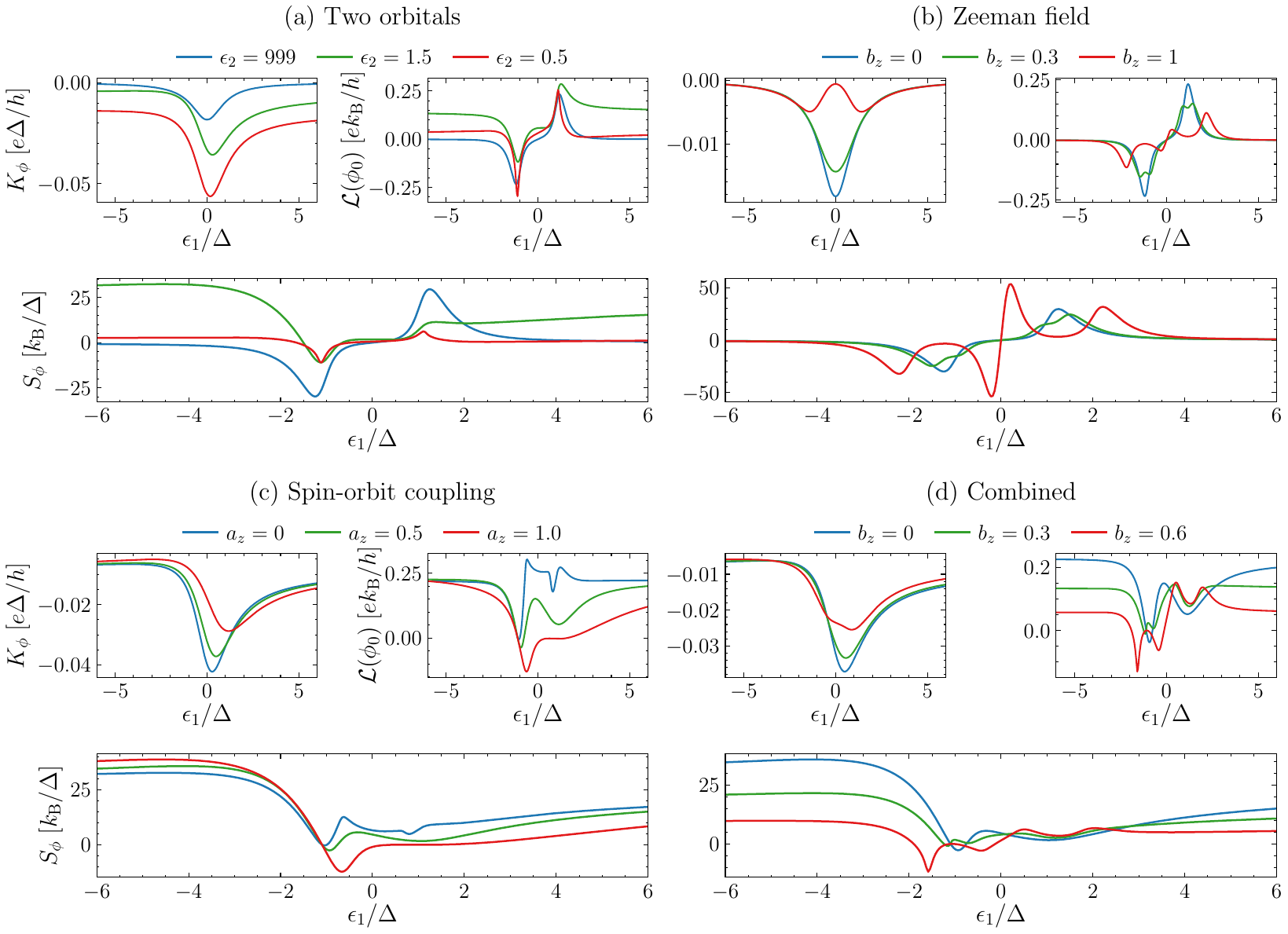}
    \caption{Thermoelectric response of the junction as a function of the
    level position~$\epsilon_1$, shown for four scenarios that successively introduce
    the ingredients of the model. Each panel displays the local current-phase
    slope~$K_\phi$
    (top left), the quasiparticle
    transport coefficient~$\mathcal{L}(\phi_0)$
    (top right), and the resulting Seebeck coefficient~$S_\phi$ (bottom).
    All three subpanels of a given scenario share the same horizontal
    axis~$\epsilon_1$ and the same color coding.
    Here,~$\epsilon_1$ and~$\epsilon_2$ denote the bare orbital energies
    of the dot, before spin-orbit hybridization and Zeeman splitting.
    (a)~Effect of the second orbital, where $\epsilon_2$ is lowered toward the gap edge
    while spin-orbit coupling and the Zeeman field are switched off. The
    curve with~$\epsilon_2 = 999$ is the single-orbital reference.
    (b)~Effect of the Zeeman field~$\vec{b}=b_z\hat{z}$ for a single active
    orbital ($\epsilon_2 = 999$) and no spin-orbit coupling.
    (c)~Effect of spin-orbit coupling~$\vec{a}=a_z\hat{z}$ with both orbitals
    active ($\epsilon_2 = 1.1$) and no Zeeman field.
    (d)~Combined effect of spin-orbit coupling~$\vec{a} = (0.3, 0.2, 0.35)^{\top}$ and a Zeeman
    field~$b_z$ with both orbitals active ($\epsilon_2 = 1.1$). 
    All panels share the parameters
    $T = 0.3$, $\abs{\Delta} = 1$, $U=U'=0$,~$t_{\alpha j} =\sqrt{0.1}$, and
    $t_{\alpha j}^{\,\mathrm{so}} = 0$ for $\alpha = {\rm L},{\rm R}$ and $j=1,2$.}
    \label{fig:seebeck}
\end{figure*}

Throughout, we  guide our interpretation by Eq.~\eqref{eq:seebeck_linear}, where
$S_\phi$ is the ratio of the quasiparticle thermal
response~$\mathcal{L}$ to the phase sensitivity of the supercurrent. The
numerator requires particle-hole asymmetry and thermally excited
quasiparticles, and is therefore maximal when a dot level aligns with the BCS
coherence peaks just above the gap
edge~\cite{kleeorinLargeTunableThermophase2016}. The denominator is governed
by the subgap Andreev bound states (ABS), whose contribution to the
supercurrent takes the form~\cite{beenakkerUniversalLimitCriticalcurrent1991a,
golubovCurrentphaseRelationJosephson2004}
\begin{equation}\label{eq:abs_current}
    J_\mathrm{ABS} = -\frac{e}{\hbar} \sum_n f(\omega_n)\,
    \pdv{\omega_n}{\phi},
\end{equation}
where~$\omega_n(\phi)$ are the ABS energies and~$f$ is the Fermi function. The sum runs over the complete Bogoliubov–de Gennes (BdG) spectrum, including positive and
negative ABS energies and both spin sectors.
The supercurrent is thus controlled by the phase dispersion
and the thermal occupation of the bound states. Since~$S_\phi = -\mathcal{L}/K_\phi$, a large thermoelectric response
therefore requires suppressing the pair transport without simultaneously suppressing
the quasiparticle channel. 

We will mainly present our results by sweeping $\epsilon_1$ at fixed $\epsilon_2$ while varying the other system parameters. This should be interpreted as a map of the underlying parameter space rather than necessarily as the trajectory generated by a single experimental gate, which will generally affect both levels simultaneously. In this way, we identify the relative orbital configurations that an experiment should aim to realize through device design and the available gate control.

\subsection{Effect of the second orbital}\label{sec:second_orbital}

We first recover the spin-degenerate single-orbital results of
Ref.~\cite{kleeorinLargeTunableThermophase2016} by placing the second
orbital far off resonance and switching off all spin-dependent
couplings. This gives the reference in Fig.~\ref{fig:seebeck}(a).
Its thermophase Seebeck coefficient is odd in~$\epsilon_1$, since
changing the sign of the orbital energy reverses~$\mathcal{L}$ while
leaving~$K_\phi$ unchanged.

Bringing the second orbital into the relevant energy range [green and red traces in Fig.~\ref{fig:seebeck}(a)] opens a second tunneling path between
the leads. For equal tunneling amplitudes, the leads couple only to the symmetric
combination~$(d_1+d_2)/\sqrt{2}$ of the two orbitals. The antisymmetric
combination has no direct coupling to the leads. The transport coefficients therefore depend on the dot only through~$g(\omega)=\sum_j t_j^2/(\omega-\epsilon_j)$, the Green's function of the
isolated dot projected onto the coupled combination and weighted by the
tunneling amplitudes. The contributions of the two orbitals add in~$g$ before the transport
coefficients are formed. They reinforce each other when~$\omega$ lies
above or below both levels and partially cancel when~$\omega$ lies
between them. Which case applies depends on the energy window that
dominates each response. The pair response is governed by~$g$ in the subgap spectrum, while the quasiparticle thermoelectric response is governed by~$g$ at above-gap energies near
the coherence peaks.

In the weak-tunneling regime, the additional path benefits pair transport more than thermoelectric
transport. A Cooper pair transfers two electrons through each barrier, so
$K_\phi$ grows roughly quadratically with the tunneling rate. The thermoelectric coefficient instead depends on the electron-hole
asymmetry of the transmission just above the gap. A level near the coherence peak carries the quasiparticle current through sequential tunneling, whose rate is of first order in~$\Gamma \propto t^2$. Off resonance, the current instead flows by cotunneling, which is of second order in~$\Gamma$. Within the
gap, the same level is off-resonant and transfers Cooper pairs only through
virtual processes of second order in $\Gamma$. A stronger coupling therefore provides 
$K_\phi$ with a much larger relative increase than~$\mathcal{L}$. When the two levels coincide, the symmetric combination is itself an eigenstate of the dot, so the junction sees a single level whose tunneling amplitude is enhanced by~$\sqrt{2}$. For split levels the same enhancement survives at energies far from both levels compared with their splitting, where~$g(\omega) \approx 2t^2/(\omega - \bar\epsilon)$ with~$\bar\epsilon = (\epsilon_1 + \epsilon_2)/2$.
The junction then behaves like the reference with a doubled tunneling rate. Since this enhances $K_\phi$ relatively more than $\mathcal{L}$, the resulting~$S_\phi$ is smaller. This is why~$S_{\phi}$ stays
well below the reference near~$\epsilon_1\simeq1.2$, where both orbitals
lie close to the coherence peak.

For~$\abs{\epsilon_1}\gg\abs{\Delta}$, the swept orbital adds only the
small term~$-t_1^2/\epsilon_1$ to~$g$.
This contribution decays only algebraically, so the response approaches
that of the fixed orbital alone as~$1/\epsilon_1$. Its sign follows the
position of the swept orbital relative to the Fermi energy. For
$\epsilon_1>0$ the pair-transport amplitudes of both orbitals add, while
for~$\epsilon_1<0$ they partially cancel. In the green trace, the pair
response is therefore markedly weaker on the negative side of the sweep,
while the quasiparticle response is still carried by the fixed orbital near
the coherence peak. As a result,~$S_\phi$ remains large over the whole
negative side and even slightly exceeds the single-orbital maximum. An
orbital inside the gap (red,~$\epsilon_2=0.5$) instead mainly strengthens
pair transport and suppresses~$S_\phi$ over most of the sweep.

The fixed orbital also removes the antisymmetry of the reference
under~$\epsilon_1\to-\epsilon_1$. Particle-hole symmetry instead
maps~$(\epsilon_1,\epsilon_2)$ onto~$(-\epsilon_1,-\epsilon_2)$ and
reverses~$\mathcal{L}$. For the equal couplings considered here,
$\mathcal{L}$ and~$S_\phi$ therefore vanish at~$\epsilon_1=-\epsilon_2$,
where the two levels lie symmetrically around the Fermi energy.

\subsection{Effect of the Zeeman field}\label{sec:zeeman}

Figure~\ref{fig:seebeck}(b) shows the effect of a Zeeman field~$b_z$ for
a single active orbital ($\epsilon_2=999$) without spin-orbit coupling.
The field splits the bare spin levels into~$\epsilon_1\pm b_z$. We
consider Zeeman energies up to~$b_z=\abs{\Delta}$, corresponding
to~$200\,\mu\mathrm{eV}$ for a representative Al gap. 
We assume that the field acts only on the dot, which is a good approximation since the~$g$
factor of semiconductor dots exceeds that of Al by about an order of
magnitude~\cite{muMeasurementsAnisotropicGfactors2020}.

The quasiparticle response splits into two peaks as the spin levels
separate [green and red traces in Fig.~\ref{fig:seebeck}(b)]. Each peak
appears when one spin level reaches the optimal
alignment~$\epsilon^*\simeq1.2\abs{\Delta}$ of the spin-degenerate
reference, so the peaks lie near~$\epsilon_1\simeq\epsilon^*\mp b_z$.
Each peak is weaker than the reference, since only one spin level is
aligned at a time.

The field also suppresses the pair response, as generally expected for
spin-singlet Cooper pairs passing through a spin-split
region~\cite{buzdinProximityEffectsSuperconductorferromagnet2005,
zhuAndreevBoundStates2001}. It shifts the two spin components of each Andreev level in
opposite directions, by nearly~$b_z$ for states localized mainly on the
dot. The shift hardly has any effect on the phase dependence. The lower component originating from the positive-energy branch therefore retains a phase dispersion opposite to that of the occupied negative-energy branches. As this component approaches zero energy from above, it becomes thermally occupied and its contribution counteracts that of the occupied negative-energy branches. The suppression is complete only once the component lies several~$k_BT$ below zero. At~$k_BT=0.3\abs{\Delta}$ used in Fig.~\ref{fig:seebeck}, a field~$b_z=0.3\abs{\Delta}$ shifts it by only about~$k_BT$. It therefore remains partially occupied, so~$\abs{K_\phi}$ is only moderately reduced. For~$b_z=\abs{\Delta}$ the component lies well below zero near~$\epsilon_1=0$, where~$\abs{K_\phi}$ almost vanishes. At even stronger fields,
the net Josephson coupling changes sign and the equilibrium phase moves
from~$0$ to~$\pi$, which marks a~$0$-$\pi$
transition~\cite{zhuAndreevBoundStates2001,
martin-roderoJosephsonAndreevTransport2011}. For the fields shown
here,~$K_\phi$ keeps its sign and the junction remains in the~$0$ state.
For~$b_z=0.3\abs{\Delta}$, the reduced pair response does not compensate
for the weaker quasiparticle peaks, and~$S_\phi$ stays slightly below the
reference. For~$b_z=\abs{\Delta}$, instead,~$S_\phi$ develops two
pronounced peaks at positive~$\epsilon_1$, with mirror images of opposite
sign at negative~$\epsilon_1$. The lower peak appears where the
upper spin level is aligned with the coherence peak while the lower one
is occupied. The occupied low-energy level contributes to the pair response with the opposite sign to the other occupied levels, so that $K_\phi$ is nearly cancelled. This yields the
largest thermophase response in Fig.~\ref{fig:seebeck}. The upper peak appears where the lower spin level is aligned with the coherence peak and the dot is empty. In this regime, the pair transport is weak as well. Both peaks therefore exceed the
spin-degenerate reference, although the quasiparticle response alone is
smaller, since~$S_\phi=-\mathcal{L}/K_\phi$.

\subsection{Effect of spin-orbit coupling}\label{sec:soc}

Figure~\ref{fig:seebeck}(c) shows the effect of Rashba spin-orbit
coupling on the dot with both orbitals active and no Zeeman field.
For a representative gap~$\abs{\Delta}=200\,\mu\mathrm{eV}$,
$a_z=0.5$ and~$1$ correspond to orbital mixing energies of~$100$ and
$200\,\mu\mathrm{eV}$. These are comparable to the SOC-induced level
mixing of about~$180\,\mu\mathrm{eV}$ reported in few-electron InSb
nanowire dots~\cite{muMeasurementsAnisotropicGfactors2020}.

The Rashba interaction hybridizes the two orbitals and moves their
energies apart. The isolated dot then has the orbital eigenenergies
\begin{equation}\label{eq:level_repulsion}
    \epsilon_\pm = \frac{\epsilon_1 + \epsilon_2}{2} \pm
    \sqrt{\left(\frac{\epsilon_1 - \epsilon_2}{2}\right)^2 + \abs{\vec{a}}^2}.
\end{equation}

For~$a_z=0.5$, the curves retain the overall shape of the result without
spin-orbit coupling. For~$a_z=1$, the change is more pronounced.
At~$\epsilon_1=-5$, the upper hybridized level lies at~$\epsilon_+\simeq1.26$,
close to the optimal quasiparticle alignment~$\epsilon^*$. This supports
substantial quasiparticle transport on the negative side of the sweep.
As~$\epsilon_1$ increases, the upper level moves away from this alignment.
At~$\epsilon_1=0$, for example, it lies at~$\epsilon_+\simeq1.69$,
while the lower level lies inside the gap at~$\epsilon_-\simeq-0.59$.
The quasiparticle response is therefore reduced in this region.
The resulting~$S_\phi$ is negative between~$\epsilon_1=-1.1$
and~$\epsilon_1\simeq1.6$. At large positive~$\epsilon_1$, the lower
hybridized level approaches~$\epsilon_2$ from below, and the
quasiparticle response recovers.

Pair transport instead favors a level close to the Fermi energy.
On the negative side of the sweep, stronger hybridization moves both
levels farther from zero and reduces~$\abs{K_\phi}$. As~$\epsilon_1$
increases, the lower level approaches and crosses the Fermi energy.
Setting~$\epsilon_-=0$ in Eq.~\eqref{eq:level_repulsion} gives
$\epsilon_1=a_z^2/\epsilon_2\simeq0.23$ and~$0.91$ for~$a_z=0.5$
and~$1$, respectively. The numerical maxima of~$\abs{K_\phi}$ lie
at slightly higher gate values because the dot is also coupled to
the leads. Beyond the maximum, stronger SOC keeps the lower
hybridized level closer to the Fermi energy, so the pair response
decreases more slowly. The~$a_z=1$ curve consequently crosses the
weaker-SOC curves near~$\epsilon_1=1.5$ and has the larger magnitude
at higher~$\epsilon_1$.

We also investigated spin-orbit-assisted tunneling with~$t^{\mathrm{so}}\neq0$. For a single orbital, a rotation of the lead spins removes it and leaves the effective amplitude~$\sqrt{t^2+(t^{\mathrm{so}})^2}$~\cite{padurariuTheoreticalProposalSuperconducting2010}. Spin-orbit-assisted tunneling therefore redistributes the coupling between the two tunneling channels rather than increasing it. The same holds for two orbitals if~$t^{\mathrm{so}}_{\alpha j}$ is proportional to~$t_{\alpha j}$, in which case~$\Lambda_\alpha=0$ in Eq.~\eqref{eq:gamma_eff}. Otherwise spin-orbit-assisted tunneling changes which orbital combinations couple to the leads. At fixed~$t^2+(t^{\mathrm{so}})^2$, this modifies the interference between the orbitals discussed above rather than the overall coupling strength. These results are not shown.

\begin{figure*}[t!]
    \centering
    \includegraphics[width=.95\textwidth]{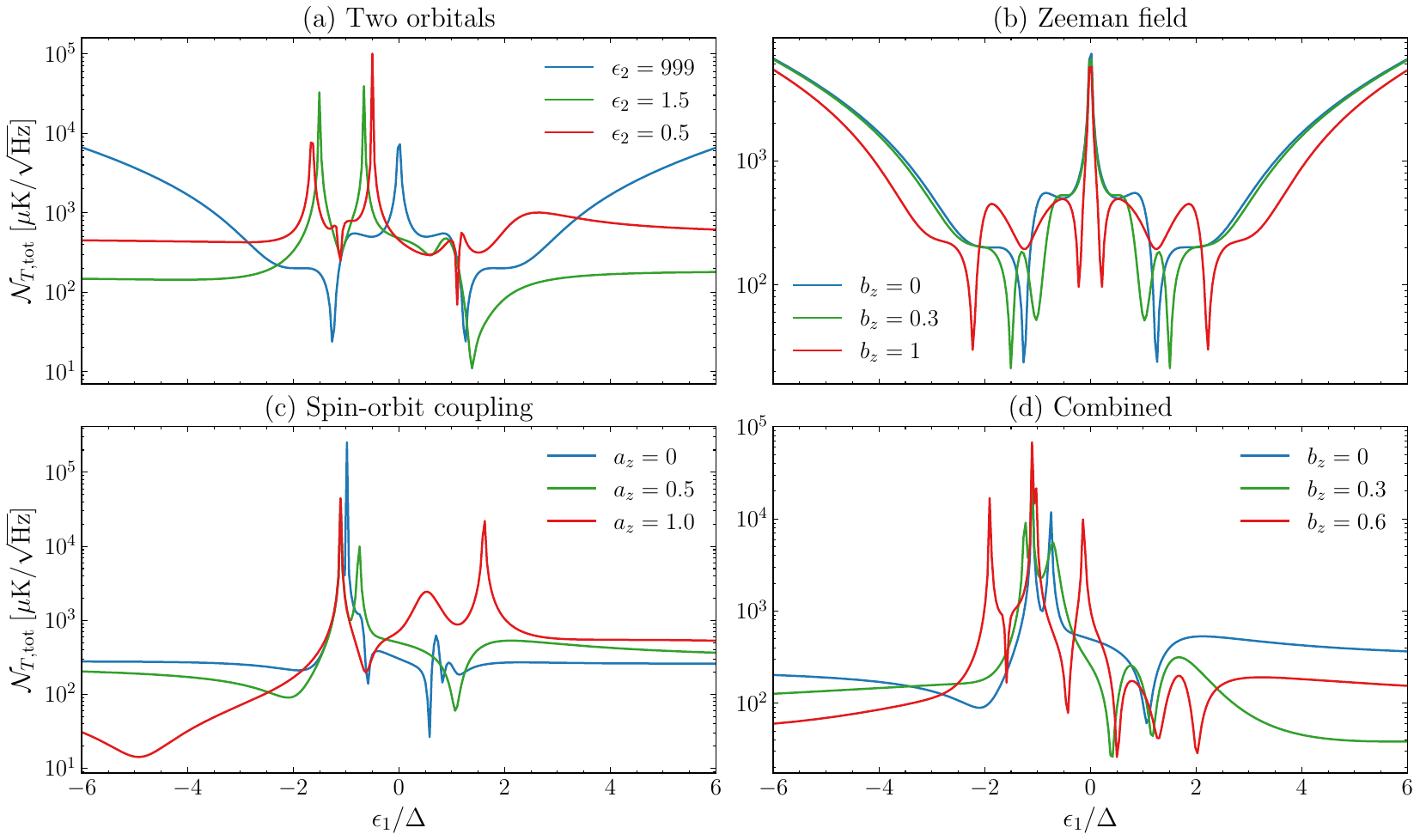}
    \caption{Total noise-equivalent temperature for the parameter families of
    Fig.~\ref{fig:seebeck}. The panels show the effects of a second orbital,
    a Zeeman field, spin-orbit coupling, and their combination. A smaller value of
    $\mathcal{N}_{T,\mathrm{tot}}$ corresponds to a more sensitive thermometer.
    All noise amplitudes and~$\Delta T_{\mathrm{op}}$ are listed in Table~\ref{tab:noise_budget}.}
    \label{fig:total_net}
\end{figure*}

\begin{figure*}[t!]
    \centering
    \includegraphics[width=.95\textwidth]{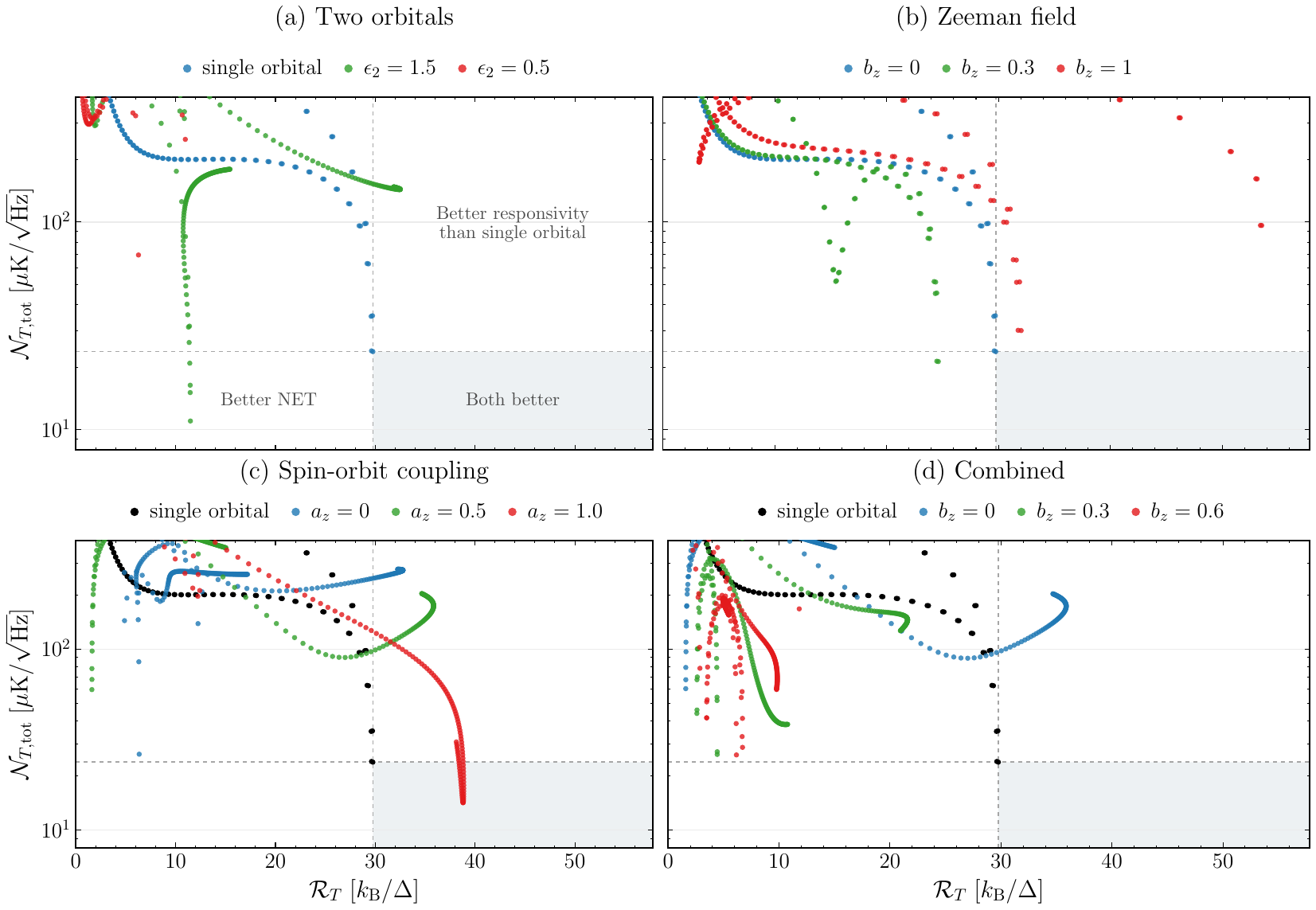}
    \caption{Thermophase responsivity and projected noise-equivalent
    temperature for the parameter families of Fig.~\ref{fig:seebeck}.
    Each point corresponds to one sampled gate value. The black points in
    panels (c) and (d) show the spin-degenerate single-orbital reference.
    The vertical and horizontal dashed lines mark its maximum responsivity
    and minimum NET, respectively. The shaded region combines a larger
    responsivity with a lower NET than these reference limits.
    Parameters and noise assumptions are the same as in
    Fig.~\ref{fig:total_net}.}
    \label{fig:tradeoff}
\end{figure*}

\subsection{Combined effects}\label{sec:combined}

Figure~\ref{fig:seebeck}(d) shows the combined effect of spin-orbit
coupling and a Zeeman field~$\vec{b}=b_z\hat{z}$ for~$\epsilon_2=1.1$.
We choose~$\vec{a}=(0.3,0.2,0.35)$, which has nearly the same magnitude
as~$a_z=0.5$ in Fig.~\ref{fig:seebeck}(c) but is not parallel to the
field. Without a field only~$\abs{\vec{a}}$ matters, so the~$b_z=0$ curve
coincides with the~$a_z=0.5$ curve of Fig.~\ref{fig:seebeck}(c). 

For a field parallel to~$\vec{a}$, the spin along~$\hat{a}$ is conserved and the dot levels are~$\epsilon_\pm\pm b_z$. Spin-orbit coupling mixes the orbitals with a relative phase~$\pm i$. The tunneling amplitudes into the two orbital components of a hybridized level then do not interfere, so each level couples to the leads with the same strength as a single orbital for~$t_1=t_2$. With both the level positions and their couplings unchanged, the junction behaves exactly like two independent orbitals at these energies.

The component of~$\vec{a}$ perpendicular to~$\vec{b}$ instead couples
opposite spin states of the two hybridized levels. Where the
levels~$\epsilon_+-b_z$ and~$\epsilon_-+b_z$ would cross, which
requires~$b_z>\abs{\vec{a}}$, they anticross and acquire mixed spin
character. The eigenstates then no longer share a common spin axis, so
no model of independent orbitals in a Zeeman field reproduces the
response, even with matched level positions. For~$b_z=0.6$ the anticrossing occurs
for~$0.4\lesssim\epsilon_1\lesssim1.8$. Near~$\epsilon_1=1.2$, the
anticrossing pushes both levels away from the optimal alignment, which
reduces~$\mathcal{L}$ and~$S_\phi$ compared with a field parallel
to~$\vec{a}$. Near~$\epsilon_1=0.4$, it instead moves one level toward
this alignment and enhances the response. The pair response is hardly
affected. For~$b_z=0.3$ the levels do not cross, and the orientation
of~$\vec{a}$ has only a small effect. The angle between~$\vec{a}$
and~$\vec{b}$ thus provides an additional control parameter, which is
absent for two orbitals without spin-orbit coupling.

\section{Sensing applications}\label{sec:sensing}

We now assess the junction as a phase-based thermometer for the
temperature difference~$\Delta T=T_L-T_R$. We shall use the orbital energies, spin-orbit coupling,
and Zeeman field as control parameters to select the operating point of the thermometer.

\subsection{Measuring the thermophase}\label{sec:measuring}

Experimentally, the thermophase~$\phi_{\mathrm{th}}$ may be detected as follows. The quantum-dot Josephson junction is embedded in a superconducting loop and the resulting circulating current is detected magnetically with a nearby SQUID. Such phase-readout geometries are well-established experimentally \cite{nicheleRelatingAndreevBound2020}, while temperature-biased superconducting loops with local heating and thermometry have been demonstrated for phase-coherent heat transport \cite{giazottoJosephsonHeatInterferometer2012}.

To determine the intrinsic open-circuit thermophase, we consider a superconducting interferometer containing the thermoelectric quantum-dot junction in one arm and a separate conventional reference Josephson junction in the other. The reference junction has suppressed thermoelectric response and a standard sinusoidal current-phase relation. When the quantum-dot junction is embedded in this superconducting loop, the phase difference across it is fixed by the current-phase relations of both junctions together with current conservation and fluxoid quantization. Consequently, the actual phase across the quantum-dot junction need not coincide with its isolated zero-current thermophase. The latter can nevertheless be determined through an interferometric null measurement. An external magnetic flux controls the phase difference between the two arms and is adjusted until the circulating current vanishes. At this null point, both junctions carry zero current, and the phase across the quantum-dot junction equals its intrinsic thermophase. The required compensation flux therefore provides a measure of $\phi_{\mathrm{th}}$.

To see this quantitatively, recall that the intrinsic thermophase is defined by the open-circuit condition~$J\left(\phi_{\mathrm{th}},\Delta T\right)= 0.$ When the junction is embedded in a superconducting interferometer, its actual phase~$\phi$ is generally modified by the circulating
current. The junction currents satisfy
\begin{equation}
    J
    \left(
        \phi,
        \Delta T
    \right)
    = I_{\mathrm{circ}},
    \qquad
    J_{\mathrm{ref}}
    \left(
        \phi_{\mathrm{ref}}
    \right)
    = -I_{\mathrm{circ}},
    \label{eq:loop_current_conservation}
\end{equation}
while fluxoid quantization imposes
\begin{equation}
    \phi
    - \phi_{\mathrm{ref}}
    + \frac{2\pi}{\Phi_0}
    \left(
        \Phi_{\mathrm{ext}}
        + L I_{\mathrm{circ}}
    \right)
    = 2\pi n.
    \label{eq:loop_fluxoid_quantization}
\end{equation}
Here, $J_{\mathrm{ref}}$ denotes the current through a separate conventional reference Josephson junction in the second arm of the interferometer. We assume this reference junction to have a suppressed thermoelectric response.  Moreover, $L$ is the loop inductance, $\Phi_{\mathrm{ext}}$ is the externally applied flux, and $n$ is the fluxoid number. 

To determine the open-circuit thermophase, the external flux is adjusted to a value $\Phi_\text{comp}$
at which the circulating current vanishes, $I_{\mathrm{circ}}=0.$ At this flux value, the quantum-dot junction carries no current and
therefore $\phi = \phi_{\mathrm{th}}.$ The reference junction also carries no current due to current conservation. For a conventional
$0$-junction with a known current-phase relation, its stable zero-current
phase is $\phi_{\mathrm{ref}}=0$. Moreover, the self-flux $L I_{\mathrm{circ}}$ vanishes at the null point.
Equation~\eqref{eq:loop_fluxoid_quantization} then gives, for a fixed temperature difference, 
\begin{equation}
    \phi_{\mathrm{th}}
    = -\frac{2\pi}{\Phi_0}
    \Phi_{\mathrm{comp}}
    + 2\pi n.
    \label{eq:thermophase_compensation_flux}
\end{equation}
Experimentally, the relevant observable is the shift of the compensation flux relative to its value at zero temperature bias. We first determine $\Phi_{\mathrm{comp}}(0)$ and then apply $\Delta T$ and adjust the external flux until $I_{\mathrm{circ}}=0$ is recovered at $\Phi_{\mathrm{comp}}(\Delta T)$. Subtracting Eq. (\ref{eq:thermophase_compensation_flux}) at the two temperature biases eliminates the fluxoid number and any static flux offset, 
\begin{equation} 
    \phi_{\mathrm{th}}(\Delta T) = -\frac{2\pi}{\Phi_0} \left[ \Phi_{\mathrm{comp}}(\Delta T) - \Phi_{\mathrm{comp}}(0) \right],
    \label{eq:thermophase_null_measurement} 
\end{equation} 
since $\phi_\text{th}(0)=0$. The open-circuit thermophase is therefore measured as the temperature-induced displacement of the external flux at which the circulating current vanishes, even when the quantum-dot Josephson junction is part of a closed loop.
In practice, however, the external flux also couples directly to the SQUID, so identifying the null point requires a careful calibration of this cross-coupling.

\subsection{Noise-equivalent temperature}
\label{sec:sensing_responsivity}

\begin{figure}[t]
    \centering
    \includegraphics[width=\columnwidth]{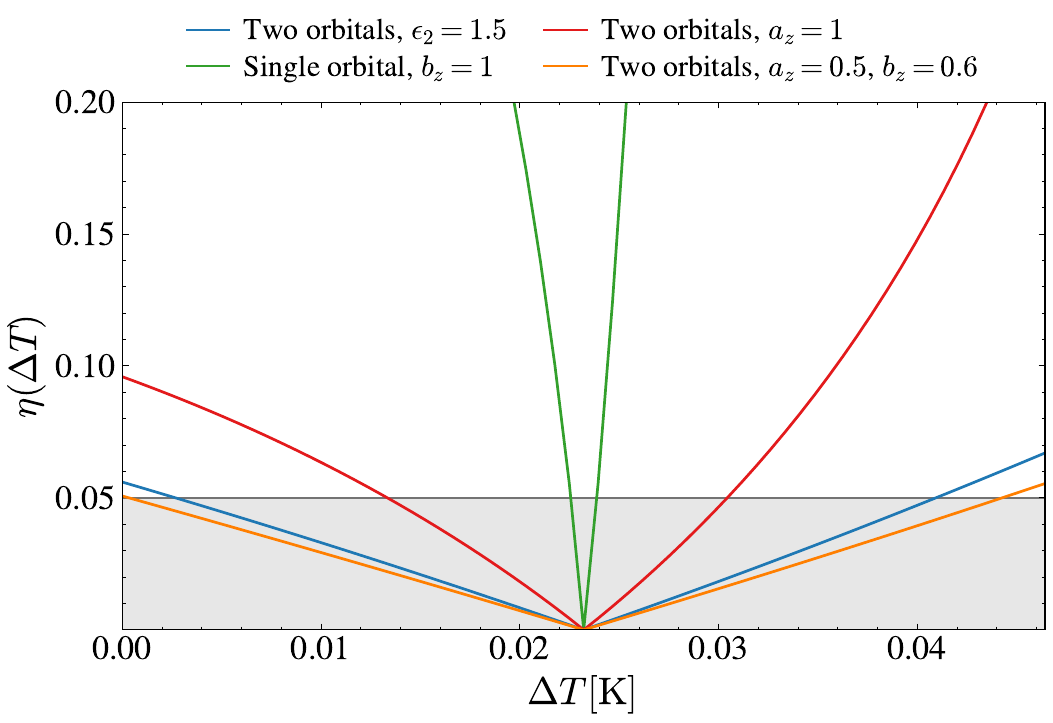}
    \caption{Relative deviation from the local linear calibration at
    $\Delta T_{\mathrm{op}}=23.2\,\mathrm{mK}$ for four representative
    operating points. The gray area marks the range~$\eta<0.05$.}
    \label{fig:thermophase_nonlinearity}
\end{figure}

We write the temperature difference as~$\Delta T = \Delta T_{\mathrm{op}} + \delta T$, where~$\Delta T_{\mathrm{op}}$ is a fixed operating point and~$\delta T$ is a small variation around it. We take~$\Delta T_{\mathrm{op}}$ within the linear-response regime of Fig.~\ref{fig:thermophase_nonlinearity}, so the slope at the operating point equals the linear-response value. The phase signal is then~$\delta\phi_{\mathrm{th}} = S_\phi\,\delta (\Delta T)$. We define the temperature responsivity as
\begin{equation}
    \mathcal{R}_T \equiv \abs{S_\phi}.
    \label{eq:temperature_responsivity}
\end{equation}
A large~$\mathcal{R}_T$ converts a small temperature change into a large
phase signal, but it does not by itself guarantee a precise thermometer. For example, a nearly vanishing~$K_\phi$ close to a~$0$-$\pi$ transition greatly increases the responsivity but also makes the device more susceptible to gate noise, as we will see below.

For this reason, we use the noise-equivalent temperature (NET) as a figure of merit to compare the useful
temperature response with the relevant noise sources. Below,~$\mathcal{N}_x$ denotes the amplitude spectral density of a noise source~$x$, i.e., the rms noise amplitude per square root of bandwidth. The NET expresses the combined output noise as the amplitude spectral density of an equivalent, time-dependent temperature fluctuation that would produce the same thermophase noise. It therefore represents an effective uncertainty in the inferred temperature difference per square root of bandwidth. For a symmetric
junction with~$\phi_0=0$, the total NET in the small signal regime $\mathcal{N}_{T,\mathrm{tot}}$ follows from
\begin{align}
    \mathcal{N}_{T,\mathrm{tot}}^2
    =\left(\frac{\mathcal{N}_{\phi,\mathrm{ro}}}
    {\mathcal{R}_T}\right)^2
    \!\!+\left(\frac{\mathcal{N}_I}{\abs{\mathcal{L}}}\right)^2
    \!\!+\Delta T_{\mathrm{op}}^2
    \hspace{-1em}\sum_{x\in\{\epsilon_1,\epsilon_2,b_z\}}\hspace{-1em}
    \chi_x^2\mathcal{N}_x^2,
    \label{eq:total_net}
\end{align}
where~$\mathcal{N}_{\phi,\mathrm{ro}}$ is the phase-noise amplitude of the
readout,~$\mathcal{N}_I$ is the amplitude of physical current fluctuations in the
junction and its circuit, and~$\mathcal{N}_x$ denotes the noise amplitude of a
control parameter~$x$. The logarithmic sensitivity
\begin{equation}
    \chi_x\equiv\frac{\partial\ln\mathcal{R}_T}{\partial x}
    \label{eq:logarithmic_sensitivity}
\end{equation}
measures how strongly such a fluctuation changes the thermometer
calibration. Independent noise sources are added in Eq.~\eqref{eq:total_net}.
Correlated shifts of the two orbital energies can be included through the
corresponding cross spectral density.

The first term of Eq.~\eqref{eq:total_net} is the phase-readout limit. If
the SQUID noise is quoted as a flux-noise amplitude, it is converted through
$\mathcal{N}_{\phi,\mathrm{ro}}=2\pi \mathcal{N}_{\Phi,\mathrm{ro}}/\Phi_0$. The
second term follows because a current fluctuation changes the phase by
$\delta\phi=-\delta I/K_\phi$, while
$\mathcal{R}_T=\abs{\mathcal{L}/K_\phi}$. The factors of~$K_\phi$ therefore
cancel when the same fluctuation is referred back to an equivalent
temperature. The last term describes noise arising from fluctuations of the 
thermometer responsivity, sometimes referred to as gain noise. It is proportional to the
finite bias~$\Delta T_{\mathrm{op}}$ because gate and field fluctuations then
modulate an existing thermophase signal.

\begin{table}[t]
    \begin{ruledtabular}
    \begin{tabular}{llll}
        Source & Amplitude~$\mathcal{N}_x$ & NET term & Reduced by \\
        \hline
        Readout & $10^{-5}\,\mathrm{rad}/\sqrt{\mathrm{Hz}}$~\cite{jabdaraghiNoiseSuperconductingMagnetic2017}
            & $\mathcal{N}_{\phi,\mathrm{ro}}/\mathcal{R}_T$
            & large $\mathcal{R}_T$ \\
        Current & $5-10\,\mathrm{fA}/\sqrt{\mathrm{Hz}}$
            & $\mathcal{N}_I/\abs{\mathcal{L}}$
            & large $\abs{\mathcal{L}}$ \\
        Gate & $1\,\mu\mathrm{eV}/\sqrt{\mathrm{Hz}}$~\cite{jekatExfoliatedHexagonalBN2020,hendrickxGatecontrolledQuantumDots2018}
            & $\Delta T_{\mathrm{op}}\abs{\chi_{\epsilon_i}}\mathcal{N}_{\epsilon_i}$
            & small $\abs{\chi_{\epsilon_i}}$ \\
        Field & $10\,\mathrm{nT}/\sqrt{\mathrm{Hz}}$~\cite{yonedaQuantumdotSpinQubit2018}
            & $\Delta T_{\mathrm{op}}\abs{\chi_{b_z}}\mathcal{N}_{b_z}$
            & small $\abs{\chi_{b_z}}$ \\
    \end{tabular}
    \end{ruledtabular}
    \caption{Illustrative low-frequency noise budget entering
    Eq.~\eqref{eq:total_net}, together with the combination of transport
    coefficients that controls each contribution. The gate noise refers to each orbital energy~$\epsilon_i$ and the field noise to~$b_z$. The values are quoted
    near~$1\,\mathrm{Hz}$ for~$\abs{\Delta}=200\,\mu\mathrm{eV}$, with~$g=15$~\cite{bjorkTunableEffectiveFactor2005,csonkaGiantFluctuationsGate2008}.
    Together with the operating point~$\Delta T_{\mathrm{op}}=0.01\abs{\Delta}/k_{\mathrm B}=23.2\,\mathrm{mK}$, these values fully specify the noise-equivalent temperatures shown in Figs.~\ref{fig:total_net} and~\ref{fig:tradeoff}.
    They depend on the materials, device geometry, readout circuit and measurement bandwidth. They mainly serve to compare the different operating points.}
    \label{tab:noise_budget}
\end{table}

The noise amplitudes used below are summarized in
Table~\ref{tab:noise_budget}. The readout value is of the order of the flux noise reported for a
proximity-junction interferometer~\cite{jabdaraghiNoiseSuperconductingMagnetic2017}.
Nanoscale SQUIDs reach even lower flux noise~\cite{vasyukovScanningSuperconductingQuantum2013}. The current noise is bounded from below by the Johnson noise~$\sqrt{4k_BTG}$ of the junction, where~$G=(e^2/h)\int d\omega\,(-\partial f/\partial\omega)\,\mathcal{T}(\omega)$ is the quasiparticle conductance and~$\mathcal{T}$ the transmission trace entering~$\mathcal{L}$. At the NET minima of Fig.~\ref{fig:total_net} this gives~$5$ to~$10\,\mathrm{fA}/\sqrt{\mathrm{Hz}}$. The orbital-energy noise
corresponds to charge-noise measurements in semiconductor quantum
dots~\cite{jekatExfoliatedHexagonalBN2020,hendrickxGatecontrolledQuantumDots2018,
lodariLowPercolationDensity2021,paqueletwuetzReducingChargeNoise2023}, and we treat the two orbital energies
as independent. 
The field noise is a conservative estimate for superconducting magnets~\cite{yonedaQuantumdotSpinQubit2018} and enters through~$\mathcal{N}_{b_z}=g\mu_{\mathrm B}\mathcal{N}_{B_z}/2$.

The last column of Table~\ref{tab:noise_budget} indicates how each
contribution can be reduced. Only the readout contribution benefits from a small~$K_\phi$,
because Josephson softening amplifies the temperature-induced phase signal relative to a fixed phase-readout imprecision. The current contribution depends
on the quasiparticle coefficient~$\mathcal{L}$ alone, since a current
fluctuation and the thermoelectric signal act on the junction phase in the
same way. The gain-noise contributions instead favor operating points where
the calibration~$S_\phi$ is insensitive to the control parameters. Josephson
softening therefore improves only the first of these contributions, whereas
a large quasiparticle response and a flat calibration improve the others.

\subsection{Projected sensitivity}
\label{sec:numerical_sensing_results}

Figure~\ref{fig:total_net} shows that the largest~$\abs{S_\phi}$ does not
necessarily give the lowest NET. A favorable point must combine a sizable
thermoelectric coefficient~$\abs{\mathcal{L}}$, a phase response that can be
resolved above the readout noise, and weak dependence of the calibration on
the control parameters. Sharp maxima occur when~$\mathcal{L}$ or~$S_\phi$
vanishes, or when~$S_\phi$ varies rapidly with gate voltage. 

For the noise budget and gate grid shown, the spin-degenerate
single-orbital reference reaches a minimum NET of approximately
$24\,\mu\mathrm{K}/\sqrt{\mathrm{Hz}}$. In
Fig.~\ref{fig:total_net}(a), a second orbital at
$\epsilon_2=1.5\abs{\Delta}$ lowers this value to approximately
$11\,\mu\mathrm{K}/\sqrt{\mathrm{Hz}}$. The minimum occurs where the two
orbitals are nearly degenerate and act as a single level with tunneling
amplitude~$\sqrt{2}\,t$. A single orbital with this amplitude reaches the
same NET. The second orbital thus enhances the effective coupling by
tuning the orbital energies alone, without modifying the tunnel barriers.
The near degeneracy requires the two orbitals to lie within a fraction
of~$\abs{\Delta}$ of each other. Coulomb
repulsion between the orbitals, neglected here, would modify this regime.
The case $\epsilon_2=0.5\abs{\Delta}$ performs worse for the noise budget
considered. The reduction in NET does not result from a thermophase response exceeding the single-orbital maximum, but from the enhanced effective coupling and the resulting improvement of the noise contributions. Combining a low NET with a large response is
instead enabled by spin-orbit coupling, as we will discuss further below. 

In Fig.~\ref{fig:total_net}(b), the moderate field
$b_z=0.3\abs{\Delta}$ slightly lowers the minimum NET to
$21.3\,\mu\mathrm{K}/\sqrt{\mathrm{Hz}}$. Although the current-noise
contribution is larger at this operating point, the gain-noise
contribution is smaller. For $b_z=\abs{\Delta}$, the minimum instead
rises to approximately $30\,\mu\mathrm{K}/\sqrt{\mathrm{Hz}}$.
The large thermophase response at strong field therefore does not
give the lowest temperature noise.

Spin-orbit coupling shifts the favorable operating regions through
orbital level repulsion. In Fig.~\ref{fig:total_net}(c), the case
$a_z=1$ reaches approximately
$14.2\,\mu\mathrm{K}/\sqrt{\mathrm{Hz}}$ at negative $\epsilon_1$.
Here, a large thermophase response coincides with a small
gain-noise contribution. This case performs better than the
two weaker SOC couplings shown.

In the combined case of Fig.~\ref{fig:total_net}(d), the fields
$b_z=0.3\abs{\Delta}$ and $b_z=0.6\abs{\Delta}$ give similar minimum
NET values of approximately
$26.2\,\mu\mathrm{K}/\sqrt{\mathrm{Hz}}$ and
$26.1\,\mu\mathrm{K}/\sqrt{\mathrm{Hz}}$, respectively.
Multiorbital and spin-dependent control therefore allow different
compromises between response and noise, rather than a universal
improvement of the sensitivity.
These values are comparable to the noise-equivalent temperatures of state-of-the-art electron thermometers based on superconducting junctions, which reach~$10$ to~$90\,\mu\mathrm{K}/\sqrt{\mathrm{Hz}}$~\cite{gasparinettiFastElectronThermometry2015,sairaDispersiveThermometryJosephson2016,karimiNoninvasiveThermometerBased2018}. Those devices measure an absolute electron temperature with megahertz bandwidth, whereas the thermophase junction directly measures a temperature difference.

Figure~\ref{fig:tradeoff} compares the response and projected NET
directly, with each point corresponding to one sampled gate value. The shaded
region requires both a response above the largest single-orbital
response and a NET below its lowest value. The strong-SOC case
enters this region for approximately
$-5.6\lesssim\epsilon_1/\abs{\Delta}\lesssim-4.3$.
It therefore improves both quantities simultaneously over the
chosen spin-degenerate single-orbital reference.

The NET is approximately even in gate voltage for the single-orbital
curves. The thermophase reverses sign between electron-like and hole-like
transport, while the NET depends on magnitudes and noise powers. A fixed
second orbital breaks this simple gate symmetry and can leave a finite
response after the swept orbital is detuned. This explains why useful
operating regions can also occur at negative gate energies and at large
$\abs{\epsilon_1}$.

\subsection{Linear and temperature-insensitive operation}
\label{sec:sensing_linear_range}

The NET describes small temperature variations around a calibrated
operating point. To determine the range over which one constant responsivity
remains accurate, we compare the full thermophase with the local tangent
\begin{equation}
    \phi_{\mathrm{th}}^{\mathrm{lin}}(\Delta T)
    =\phi_{\mathrm{th}}(\Delta T_{\mathrm{op}})
    +S_\phi(\Delta T_{\mathrm{op}})
    \left(\Delta T-\Delta T_{\mathrm{op}}\right).
    \label{eq:linear_thermophase}
\end{equation}
The relative deviation is defined as
\begin{equation}
    \eta(\Delta T)=
    \frac{\left|\phi_{\mathrm{th}}(\Delta T)
    -\phi_{\mathrm{th}}^{\mathrm{lin}}(\Delta T)\right|}
    {\left|\phi_{\mathrm{th}}^{\mathrm{lin}}(\Delta T)
    -\phi_{\mathrm{th}}(\Delta T_{\mathrm{op}})\right|}.
    \label{eq:thermophase_nonlinearity}
\end{equation}
At $\Delta T = \Delta T_\text{op}$, $\eta=0$ by continuity. We define the linear operating interval by~$\eta<0.05$. We note that this threshold
is only used as a practical calibration criterion and should not be seen as a fundamental boundary.

For the two-orbital point with~$\epsilon_2=1.5\abs{\Delta}$, the five-percent interval extends from~$2.7$ to~$41.0\,\mathrm{mK}$. The spin-orbit-coupled point remains linear from~$13$ to~$30\,\mathrm{mK}$, while the combined point stays within the criterion over the entire range shown, from~$0$ to~$46\,\mathrm{mK}$. By contrast, the single-orbital point at~$b_z=\abs{\Delta}$ meets the same criterion only between~$22.6$ and~$23.9\,\mathrm{mK}$.
Its enhanced local response is therefore accompanied
by a very narrow calibration range. 
The combined point thus pairs a competitive noise-equivalent temperature with the widest calibration range.
Outside the reported intervals the
thermophase can still be used, but the temperature must be inferred from the
full nonlinear calibration curve rather than one constant~$S_\phi$.

The preceding analysis concerns the temperature range around operating points with a large thermophase response. The same decomposition $S_\phi=-\mathcal{L}/K_\phi$ also identifies the opposite regime, in which the thermophase is insensitive to a temperature difference to first order. In particular, $S_\phi$ can change sign in two physically distinct ways. Close to a $0$-$\pi$ transition, $K_\phi$ approaches zero and $S_\phi$ grows strongly, as discussed in Sec.~III B. By contrast, a zero of $\mathcal{L}$ at finite $K_\phi$ reverses the thermoelectric response while preserving a well-defined Josephson coupling. The separate display of $\mathcal{L}$ and $K_\phi$ in Fig.~2 allows these two mechanisms to be distinguished, which is not possible from $S_\phi$ alone. In all two-orbital configurations of Fig.~2, $S_\phi$ vanishes at $\epsilon_1=-\epsilon_2$, where the two dot levels are arranged symmetrically around the Fermi energy, while $K_\phi$ remains finite. This generalizes the zero at $\epsilon_1=0$ in the single-level junction and occurs independently of the spin-orbit coupling and Zeeman field considered here. At this point, the linear thermophase response vanishes because 
\begin{equation} 
    S_\phi = \left. \frac{\rm d\phi_{\mathrm{th}}} {\rm d(\Delta T)} \right|_{\mathrm{op}} = 0. 
\end{equation} 
For a small change $\delta(\Delta T)$ around the operating temperature difference $\Delta T_{\mathrm{op}}$, the thermophase therefore becomes 
\begin{align} 
    &\phi_{\mathrm{th}}\left[ \Delta T_{\mathrm{op}}+\delta(\Delta T) \right] - \phi_{\mathrm{th}}(\Delta T_{\mathrm{op}}) \notag\\
    &\quad = \frac{1}{2} \left. \frac{\partial^2\phi_{\mathrm{th}}} {\partial(\Delta T)^2} \right|_{\mathrm{op}} \left[\delta(\Delta T)\right]^2 + \mathcal{O}\left( \left[\delta(\Delta T)\right]^3 \right). 
\end{align} 
The thermophase is consequently insensitive to temperature variations to first order, while the finite~$K_\phi$ shows that the junction retains a well-defined Josephson coupling. The same gate-tunable junction can therefore operate as a thermoelectric sensor near a high-responsivity point and as a temperature-insensitive phase element near~$\epsilon_1=-\epsilon_2$. Although the latter function is secondary to the thermometry application considered here, it illustrates the broad tunability of the temperature-to-phase conversion.

\section{Conclusion}\label{sec:conclusion}

Quantum-dot Josephson junctions convert a temperature difference into a phase signal whose size and sign can be set electrically and magnetically. We have studied this thermophase in a two-orbital dot with spin-orbit coupling and a Zeeman field. Writing the thermophase Seebeck coefficient as~$S_\phi=-\mathcal{L}/K_\phi$ separates the quasiparticle response from the stiffness of the junction, so the effect of each control parameter can be traced to one of the two. A second orbital mainly strengthens the pair transport. When it lies close to the optimal quasiparticle alignment, however, it turns the sharp single-level maximum of~$S_\phi$ into a broad region of large response. The Zeeman field enhances~$S_\phi$ near the~$0$-$\pi$ transition by softening the junction rather than by increasing~$\mathcal{L}$. Spin-orbit coupling shifts the orbital levels through level repulsion. Depending on the gate position it either suppresses the response or raises it beyond the single-orbital maximum. For a field that is not parallel to the spin-orbit vector, the hybridized levels acquire mixed spin character near their anticrossing, which makes the field direction a further control parameter.

The junction thus offers an unusually rich set of knobs that can be adjusted \textit{in situ}. Gate voltages set the level positions, while the magnitude and direction of the field tune the spin splitting and its interplay with spin-orbit coupling. Each knob acts differently on the quasiparticle response and the stiffness. The same device can therefore be tuned between a large thermophase and a robust calibration without any change in its fabrication. This tunability matters because a large thermophase alone does not make a good thermometer. Josephson softening improves only the readout contribution to the noise-equivalent temperature and narrows the range over which a linear calibration holds. The lowest noise-equivalent temperatures are instead reached where a large quasiparticle response coincides with a calibration that is insensitive to gate fluctuations. For the noise levels assumed here, they are of order~$10\,\mu\mathrm{K}/\sqrt{\mathrm{Hz}}$. Strong spin-orbit coupling provides a broad gate range in which such operating points exist.

These predictions rest on several simplifications. Coulomb interactions enter only at the mean-field level, where they renormalize the model parameters and add polarization-induced exchange fields. Kondo correlations, which become relevant when the Kondo temperature approaches the superconducting gap, lie beyond this treatment. The noise budget relies on representative noise amplitudes rather than on a model of a specific readout circuit. We have also considered couplings that preserve the symmetry pinning the equilibrium phase. An anomalous phase would let gate and field fluctuations shift the phase directly, adding a noise contribution that does not vanish for small~$\Delta T_{\mathrm{op}}$. Natural extensions include a treatment of interactions beyond mean field and an explicit model of the readout circuit.

Nanowire and planar quantum dots with superconducting contacts already combine all the required ingredients. We therefore expect gate-tunable quantum-dot junctions to provide a viable phase-based thermometer. Such devices could help track dissipation and heat flow in cryogenic quantum technologies, where managing small heat loads is becoming an increasingly important challenge.

\begin{acknowledgments}
We thank C. Sun for useful discussions. This work was
supported by the Research Council of Norway through Grant
No.~353894 and its Centres of Excellence funding scheme
Grant No.~262633 “QuSpin.”
\end{acknowledgments}

\appendix
 
\section{Green's function formalism}\label{app:greens_function}
 
In this appendix, we recast the model of Sec.~\ref{sec:model} in matrix Nambu form and 
derive the retarded self-energies of the superconducting leads, the Green's
function of the dot, and the current in the linear-response regime. The
mean-field treatment of the Coulomb interaction is described in
Appendix~\ref{app:coulomb}.
 
\subsection{Nambu representation}\label{app:nambu}
 
To treat superconducting correlations, we introduce Nambu spinors that combine 
particle and hole degrees of freedom. For the quantum dot, the eight-component spinor
\begin{equation}\label{eq:nambu_dot}
    \Psi_\mathrm{QD} = \left(
        d_{1\uparrow}, d_{1\downarrow}, d_{2\uparrow}, d_{2\downarrow},
        d_{1\downarrow}^\dagger, -d_{1\uparrow}^\dagger, 
        d_{2\downarrow}^\dagger, -d_{2\uparrow}^\dagger
    \right)^\top
\end{equation}
lives in the tensor product of Nambu ($\tau$), orbital ($\nu$), and spin ($\sigma$) 
spaces, with corresponding Pauli matrices $\tau_i$, $\nu_i$, and $\sigma_i$. We denote 
identity matrices in each space by $\mathbbm{1}_\tau$, $\mathbbm{1}_\nu$, and 
$\mathbbm{1}_\sigma$. For the leads, we use the four-component spinor
\begin{equation}\label{eq:nambu_lead}
    \Psi_{\alpha k} = \left(
        c_{\alpha k\uparrow}, c_{\alpha \bar{k}\downarrow},
        c_{\alpha \bar{k}\downarrow}^\dagger, -c_{\alpha k\uparrow}^\dagger
    \right)^\top,
\end{equation}
with $\bar{k} \equiv -k$, spanning Nambu~$\otimes$~spin space.
 
In this basis, the lead Hamiltonian, Eq.~\eqref{eq:lead_hamiltonian}, takes the 
Bogoliubov-de~Gennes form
\begin{equation}\label{eq:bdg}
    H_\alpha = \frac{1}{2}\sum_k \Psi_{\alpha k}^\dagger\,
    \begin{pmatrix}
        \epsilon_\alpha(k)\,\mathbbm{1}_\sigma & \Delta_\alpha\mathbbm{1}_\sigma \\
        \Delta_\alpha^*\,\mathbbm{1}_\sigma & -\epsilon_\alpha(k)\mathbbm{1}_\sigma
    \end{pmatrix}
    \Psi_{\alpha k}.
\end{equation}
The quantum dot Hamiltonian, Eq.~\eqref{eq:dot_hamiltonian}, can be written as
\begin{equation}\label{eq:hqd_nambu}
    H_\mathrm{QD} = \frac{1}{2}\Psi_\mathrm{QD}^\dagger\,
    \mathcal{H}_\mathrm{QD}\Psi_\mathrm{QD},
\end{equation}
with the $8\times 8$ matrix
\begin{equation}\label{eq:hqd_matrix}
    \mathcal{H}_\mathrm{QD} = 
    \tau_z \otimes \left[
        \mathcal{E}_\nu \otimes \mathbbm{1}_\sigma + 
        \nu_y \otimes \vec{a} \cdot \vec{\sigma}
    \right] +
    \mathbbm{1}_\tau \otimes \mathbbm{1}_\nu \otimes 
    \vec{b} \cdot \vec{\sigma}.
\end{equation}
Here,~$\mathcal{E}_\nu=\diag(\epsilon_1,\epsilon_2)$ contains the
orbital energies.
The structure of this matrix follows from the particle-hole conjugation of the hole 
block~\cite{zazunovAnomalousJosephsonCurrent2009}. The single-particle terms 
(orbital energies and spin-orbit coupling) acquire a $\tau_z$ prefactor reflecting 
their sign change under conjugation, while the Zeeman field, which breaks time 
reversal symmetry, couples to $\mathbbm{1}_\tau$ because 
$\sigma_y\,\vec{\sigma}^{\,*}\sigma_y = -\vec{\sigma}$ compensates the overall 
minus sign from the hole block.
 
The tunneling Hamiltonian, Eq.~\eqref{eq:tunneling}, takes the Nambu form
\begin{equation}\label{eq:ht_nambu}
    H_T = \sum_{\alpha, k}
    \Psi_{\alpha k}^\dagger\,\hat{V}_\alpha\,\Psi_\mathrm{QD},
\end{equation}
where the $4\times 8$ coupling matrix factorizes as
\begin{equation}\label{eq:coupling_nambu}
    \hat{V}_\alpha = \tau_z \otimes W_\alpha,
\end{equation}
with $\tau_z$ acting in Nambu space and $W_\alpha$ being a $2\times 4$ matrix in 
orbital $\otimes$ spin space,
\begin{equation}\label{eq:W_matrix}
    W_\alpha = 
    (t_{\alpha 1},t_{\alpha 2}) \otimes \mathbbm{1}_\sigma + 
    i\eta_\alpha(t_{\alpha 1}^{\mathrm{so}},t_{\alpha 2}^{\mathrm{so}}) \otimes \sigma_y.
\end{equation}

\subsection{Retarded self-energies}\label{app:self_energy}
 
The retarded self-energy due to lead $\alpha$ is obtained by integrating out the lead 
degrees of freedom,
\begin{equation}\label{eq:sigma_def}
    \hat{\Sigma}_\alpha^r(\omega) = 
    \sum_k \hat{V}_\alpha^\dagger\,
    \hat{g}_\alpha^r(k,\omega)\,\hat{V}_\alpha,
\end{equation}
where $\hat{g}_\alpha^r(k,\omega)$ is the bare retarded Green's function of 
lead~$\alpha$. From the BdG Hamiltonian, Eq.~\eqref{eq:bdg}, one finds
\begin{align}\label{eq:g_lead_k}
    \begin{split}
        \hat{g}_\alpha^r(k,\omega) = 
        {} & {} \frac{1}{(\omega + i\eta)^2 - \epsilon_\alpha^2(k) - \abs{\Delta_\alpha}^2}\\
        {} & {} \times \begin{pmatrix}
            \omega + \epsilon_\alpha(k) & \Delta_\alpha \\
            \Delta_\alpha^* & \omega - \epsilon_\alpha(k)
        \end{pmatrix}
        \otimes \mathbbm{1}_\sigma,
    \end{split}
\end{align}
where $\eta \to 0^+$.
 
We convert the momentum sum into an energy integral, 
$\sum_k \to \int d\epsilon\, N(\epsilon)$, and employ the wide bandwidth 
approximation~\cite{sunElectronTransportMesoscopic2000}, replacing the density of 
states by its value at the Fermi level, $N(\epsilon) \approx N(0)$, which yields
\begin{equation}\label{eq:g_lead_summed}
    \hat{g}_\alpha^r(\omega) = 
    -i\pi N(0)\,\rho_\alpha(\omega)
    \begin{pmatrix}
        1 & \Delta_\alpha/\omega \\
        \Delta_\alpha^{\ast}/\omega & 1
    \end{pmatrix}
    \otimes \mathbbm{1}_\sigma.
\end{equation}
Here, we introduced the dimensionless BCS density of states
\begin{equation}\label{eq:bcs_dos}
    \rho_\alpha(\omega)=\frac{\abs{\omega}\,\theta(\abs{\omega}-\abs{\Delta_\alpha})}{\sqrt{\omega^2-\abs{\Delta_\alpha}^2}}+\frac{\omega\,\theta(\abs{\Delta_\alpha}-\abs{\omega})}{i\sqrt{\abs{\Delta_\alpha}^2-\omega^2}},
\end{equation}
which encodes both the above-gap quasiparticle continuum and the subgap analytic 
continuation responsible for the Andreev bound states.
 
Inserting Eqs.~\eqref{eq:coupling_nambu} and~\eqref{eq:g_lead_summed} 
into Eq.~\eqref{eq:sigma_def}, we use the factorization 
$\hat{V}_\alpha^\dagger(\hat{B}\otimes\mathbbm{1}_\sigma)\hat{V}_\alpha = 
(\tau_z\hat{B}\,\tau_z) \otimes (W_\alpha^\dagger W_\alpha)$ to obtain
\begin{equation}\label{eq:self_energy_result}
    \hat{\Sigma}_\alpha^r(\omega) = 
    -\frac{i}{2}\,\rho_\alpha(\omega)
    \begin{pmatrix}
        1 & -\Delta_\alpha / \omega \\
        -\Delta_\alpha^* / \omega & 1
    \end{pmatrix}
    \otimes\,\hat{\Gamma}_\alpha^\mathrm{eff},
\end{equation}
where the sign flip in the off-diagonal entries arises from the conjugation by $\tau_z$, 
and we defined the $4\times 4$ effective coupling matrix
\begin{equation}\label{eq:gamma_eff}
    \hat{\Gamma}_\alpha^\mathrm{eff} = 
    2\pi N(0)W_\alpha^\dagger W_\alpha = 
    \hat{\Gamma}_\alpha \otimes \mathbbm{1}_\sigma -
    \eta_\alpha \Lambda_\alpha\nu_y \otimes \sigma_y.
\end{equation}
Here,
the $2\times 2$ orbital coupling matrix $\hat{\Gamma}_\alpha$ is defined through
\begin{equation}\label{eq:gamma_orbital}
    \big(\hat{\Gamma}_\alpha\big)_{nm} = 
    2\pi N(0)\left[
        t_{\alpha n}t_{\alpha m} + 
        t_{\alpha n}^{\mathrm{so}}t_{\alpha m}^{\mathrm{so}}
    \right]
\end{equation}
and~$\Lambda_\alpha = 2\pi N(0)(t_{\alpha 1}t_{\alpha 2}^\mathrm{so} - 
t_{\alpha 1}^\mathrm{so} t_{\alpha 2})$. We note that the second term in 
Eq.~\eqref{eq:gamma_eff}, which couples orbital ($\nu_y$) and spin ($\sigma_y$) 
degrees of freedom, 
is absent in a single-orbital model and is only
nonzero for two orbitals when the ratio 
$t_{\alpha n}^\mathrm{so}/t_{\alpha n}$ differs between the two orbitals.

\subsection{Green's function}\label{app:green_function}

The retarded Green's function of the dot is determined by the Dyson equation
\begin{equation}\label{eq:dyson_app}
    \hat{G}^r(\omega) = \left[
        (\hat{g}_0^r)^{-1}(\omega)
        - \hat{\Sigma}_L^r(\omega) - \hat{\Sigma}_R^r(\omega)
    \right]^{-1},
\end{equation}
where the inverse bare Green's function follows directly from the Nambu dot
Hamiltonian, Eq.~\eqref{eq:hqd_matrix},
\begin{equation}\label{eq:g0_inv}
    (\hat{g}_0^r)^{-1}(\omega) =
    \omega\,\mathbbm{1}_8 - \mathcal{H}_\mathrm{QD}.
\end{equation}
The lesser Green's function is obtained from the Keldysh
equation~\cite{stefanucciNonequilibriumManyBodyTheory2013}
\begin{equation}\label{eq:keldysh_lesser}
    \hat{G}^{<}(\omega) =
    \hat{G}^r(\omega)\,\hat{\Sigma}^{<}(\omega)\,\hat{G}^a(\omega).
\end{equation}
Here, the advanced Green's function is~$\hat{G}^a = (\hat{G}^r)^\dagger$,
the total lesser self-energy
is~$\hat{\Sigma}^{<} = \hat{\Sigma}_L^{<} + \hat{\Sigma}_R^{<}$ and the
lesser self-energy of each lead follows from the fluctuation-dissipation
theorem~\cite{stefanucciNonequilibriumManyBodyTheory2013}
\begin{equation}\label{eq:fdt_sigma}
    \hat{\Sigma}_\alpha^{<}(\omega) =
    f_\alpha(\omega)\left(
        \hat{\Sigma}_\alpha^a(\omega) - \hat{\Sigma}_\alpha^r(\omega)
    \right),
\end{equation}
with~$f_\alpha(\omega) = [e^{\omega/k_BT_\alpha} + 1]^{-1}$ the Fermi
function of lead~$\alpha$.

\subsection{Current in the linear-response regime}\label{app:current}
 
The charge current through the junction is given 
by~\cite{QuantumKineticsTransport2008}
\begin{equation}\label{eq:current_general}
    J = -\frac{e}{h} \int d\omega\,
    \Re\,\tr_p\!\left[
        (\hat{\Sigma}_L - \hat{\Sigma}_R)\,\hat{G}(\omega)
    \right]^{<}\!,
\end{equation}
where $\tr_p$ denotes the trace restricted to the particle block of Nambu
space, meaning the sum of the diagonal elements with indices $1$-$4$ in the basis of
Eq.~\eqref{eq:nambu_dot}. Those four indices label the orbital and spin degrees of
freedom of the particle sector.
Applying the Langreth
rule~\cite{langrethLinearNonlinearResponse1976, stefanucciNonequilibriumManyBodyTheory2013} 
$[AB]^{<} = A^r B^{<} + A^{<} B^a$ gives the expression quoted in the main text, 
Eq.~\eqref{eq:current}.
 
We now specialize to the linear-response regime. A temperature difference 
$\Delta T$ is applied across the junction such that $T_L = T + \Delta T$ and $T_R = T$. 
Expanding the Fermi function of the left lead to first order,
\begin{equation}\label{eq:f_linearized}
    f_L(\omega) = f(\omega) + \frac{\partial f}{\partial T}\Delta T,
    \qquad
    f_R(\omega) = f(\omega),
\end{equation}
where $f(\omega)$ is the equilibrium distribution at temperature $T$, the current 
separates into an equilibrium contribution (the supercurrent $J_\mathrm{sc}$, 
independent of $\Delta T$) and a non-equilibrium contribution (the quasiparticle 
current $J_\mathrm{qp}$, linear in $\Delta T$).
 
\textit{Supercurrent.}---Inserting the equilibrium part of 
Eq.~\eqref{eq:f_linearized} into the current and using the Keldysh equation, 
Eq.~\eqref{eq:keldysh_lesser}, together with the identity
\begin{equation}\label{eq:ga_gr_identity}
    \hat{G}^r(\omega)\left[
        \hat{\Sigma}^a(\omega) - \hat{\Sigma}^r(\omega)
    \right]\hat{G}^a(\omega) = 
    \hat{G}^a(\omega) - \hat{G}^r(\omega),
\end{equation}
one arrives at the supercurrent
\begin{equation}\label{eq:supercurrent}
    J_\mathrm{sc} = -\frac{e}{h} \int \dif{\omega} f(\omega)
    \Re\tr_p\left[
        \hat{D}^a\hat{G}^a - \hat{D}^r\hat{G}^r
    \right],
\end{equation}
where $\hat{D}^r \equiv \hat{\Sigma}_L^r - \hat{\Sigma}_R^r$ and 
$\hat{D}^a = (\hat{D}^r)^\dagger$. This contribution is physically identified as a supercurrent since it does not depend on~$\Delta T$ in~$f(\omega)$, and thus exists even in equilibrium.
 
\textit{Quasiparticle current.}---The terms proportional to $\Delta T$ in 
Eqs.~\eqref{eq:f_linearized} give rise to a quasiparticle current
\begin{align}\label{eq:qp_current}
    \begin{split}
        \mathcal{L}(\phi)\Delta T = 
        -\frac{e}{h} \int d\omega \frac{\partial f}{\partial T}\,
        &\Re\tr_p\Big[
            \hat{D}^r\hat{G}^r\left(
                \hat{\Sigma}_L^a - \hat{\Sigma}_L^r
            \right)\hat{G}^a \\&+
            \left(
                \hat{\Sigma}_L^a - \hat{\Sigma}_L^r
            \right)\hat{G}^a
        \Big]\Delta T.
    \end{split}
\end{align}
Note that both Eqs.~\eqref{eq:supercurrent} and~\eqref{eq:qp_current} involve traces
of products of $8\times 8$ matrices, which we evaluate numerically. In each case the upper-left $4\times 4$ particle block of the product is
retained and traced, following the definition below
Eq.~\eqref{eq:current_general}.

\section{Coulomb interaction at the mean-field level}\label{app:coulomb}

In this appendix, we treat the intra- and inter-orbital Coulomb repulsion of
Eq.~\eqref{eq:coulomb_interaction} in the Hartree-Fock approximation. We show that it renormalizes the orbital energies, the Zeeman field and the inter-orbital coupling, which are independent control parameters in
the main text. In spin-polarized states, it additionally generates
orbital-dependent exchange fields, whose effect we discuss in
Sec.~\ref{app:hf_results}.

\subsection{Hartree-Fock decoupling}\label{app:hf_decoupling}

We label the dot operators by~$a=(\nu,\sigma)$ and define the
single-particle density matrix~$\rho_{ab}=\avg{d_b^\dagger d_a}$. Its
$2\times2$ spin blocks~$\rho_{\nu\nu'}$ have the
elements~$(\rho_{\nu\nu'})_{\sigma\sigma'}=\avg{d_{\nu'\sigma'}^\dagger
d_{\nu\sigma}}$. The diagonal block~$\rho_{\nu\nu}$ contains the
occupation~$n_\nu=\tr\rho_{\nu\nu}$ and the spin
polarization~$\vec{m}_\nu=\tr(\rho_{\nu\nu}\vec{\sigma})$ of
orbital~$\nu$. The off-diagonal block~$\rho_{\nu\bar{\nu}}$, with~$\bar{\nu}$
the other orbital, describes the coherence between the two orbitals.
Both terms of Eq.~\eqref{eq:coulomb_interaction} are invariant under spin rotations. We
therefore use a normal Hartree-Fock decoupling that keeps all spin and
orbital coherences, 
\begin{equation}\label{eq:hf_mean_field}
    H_U \approx \sum_{ab} V_{ab}\, d_a^\dagger d_b
    - \frac{1}{2}\tr\left(V\rho\right),
\end{equation}
with the~$2\times2$ spin blocks of the mean-field potential
\begin{align}\label{eq:hf_potential}
    \begin{split}
        V_{\nu\nu} &= U\left(n_\nu\mathbbm{1}_\sigma-\rho_{\nu\nu}\right) + U' n_{\bar{\nu}}\,\mathbbm{1}_\sigma, \\
        V_{\nu\bar{\nu}} &= -U'\rho_{\nu\bar{\nu}},
    \end{split}
\end{align}
where~$\bar{\nu}$ denotes the other orbital. An interaction-generated pairing field is not included,
as in Ref.~\cite{kleeorinLargeTunableThermophase2016}.

Adding Eq.~\eqref{eq:hf_potential} to the dot Hamiltonian of
Eq.~\eqref{eq:dot_hamiltonian} gives a mean-field Hamiltonian of the
same form,
\begin{align}\label{eq:hf_effective}
    \begin{split}
        H_\mathrm{QD}+\sum_{ab}&V_{ab}d_a^\dagger d_b
        = \sum_{\nu} d_\nu^\dagger\left(\tilde{\epsilon}_\nu\mathbbm{1}_\sigma
        +\tilde{\vec{b}}_\nu\cdot\vec{\sigma}\right)d_\nu \\
        &+\left[d_1^\dagger\left(\tilde{t}\,\mathbbm{1}_\sigma
        -i\tilde{\vec{a}}\cdot\vec{\sigma}\right)d_2+\mathrm{h.c.}\right],
    \end{split}
\end{align}
where~$d_\nu=(d_{\nu\uparrow},d_{\nu\downarrow})^T$. With~$\rho_{\nu\nu}=(n_\nu\mathbbm{1}_\sigma+\vec{m}_\nu\cdot\vec{\sigma})/2$,
the renormalized parameters read
\begin{subequations}\label{eq:hf_parameters}
\begin{align}
    \tilde{\epsilon}_\nu &= \epsilon_\nu+\frac{U}{2}n_\nu+U'n_{\bar{\nu}}, \\
    \tilde{\vec{b}}_\nu &= \vec{b}-\frac{U}{2}\vec{m}_\nu, \\
    \tilde{\vec{a}} &= \vec{a}-\frac{iU'}{2}\tr\left(\rho_{12}\vec{\sigma}\right), \\
    \tilde{t} &= -\frac{U'}{2}\tr\rho_{12}.
\end{align}
\end{subequations}
The Hartree terms shift the orbital energies. The exchange term of~$U$
adds the field~$-U\vec{m}_\nu/2$ to the Zeeman field of each orbital.
The exchange term of~$U'$ adds to the spin-orbit vector~$\vec{a}$ and
generates a spin-independent tunneling~$\tilde{t}$ between the orbitals.
The latter is absent in Eq.~\eqref{eq:dot_hamiltonian}. It can be removed
by a rotation of the orbital basis, which leaves the spin-orbit term
unchanged and only redefines the orbital energies and the tunneling
amplitudes~$t_{\alpha\nu}$.

At the mean-field level, the interacting dot is thus described by a
noninteracting Hamiltonian with self-consistently renormalized
parameters. Without spin polarization, the exchange field~$-U\vec{m}_\nu/2$
vanishes. Moreover,~$\tr\rho_{12}$ is then real
and~$\tr(\rho_{12}\vec{\sigma})$ is imaginary, so that~$\tilde{t}$
and~$\tilde{\vec{a}}$ are real. The interaction then only renormalizes
parameters of the model of Sec.~II. A spin polarization, whether
spontaneous or induced by the Zeeman field, generates two genuinely new
terms. The first is the orbital dependence of the exchange
field~$-U\vec{m}_\nu/2$. The second arises because~$\tilde{\vec{a}}$ can
become complex. Its real part still renormalizes the spin-orbit coupling,
whereas its imaginary part acts as an exchange field that couples the
spins of the two orbitals. They do not constitute an additional source of
time-reversal symmetry breaking. The renormalized parameters depend on the occupations. Through
the term~$U'n_{\bar{\nu}}$, for example, the effective energy of the
second orbital follows the occupation of the first one as~$\epsilon_1$
is swept.

In the Nambu basis of Eq.~\eqref{eq:nambu_dot}, the mean field adds
\begin{equation}\label{eq:delta_hf}
    \delta\mathcal{H}_\mathrm{HF} =
    \begin{pmatrix}
        V & 0 \\
        0 & -\mathcal{T}V^{*}\mathcal{T}
    \end{pmatrix},
    \qquad
    \mathcal{T} = \mathbbm{1}_\nu\otimes\sigma_y,
\end{equation}
to the dot Hamiltonian, so that Eq.~\eqref{eq:g0_inv} is replaced by
\begin{equation}\label{eq:g0_hf}
    (\hat{g}_0^r)^{-1}(\omega) =
    \omega\mathbbm{1}_8 - \mathcal{H}_\mathrm{QD}
    - \delta \mathcal{H}_\mathrm{HF}.
\end{equation}
The form of the hole block follows from the time-reversed hole components
of the spinor. For a state polarized along~$\hat{z}$ without inter-orbital
coherence, Eq.~\eqref{eq:delta_hf} reduces to the familiar renormalized
levels~$\tilde{\epsilon}_{\nu\sigma}=\epsilon_\nu
+U\avg{n_{\nu\bar{\sigma}}}+U'n_{\bar{\nu}}$.

\subsection{Self-consistency and degenerate solutions}
\label{app:hf_selfconsistency}

The density matrix follows from the lesser Green's function,
\begin{equation}\label{eq:occupation}
    \rho_{ab} =
    -i \int \frac{d\omega}{2\pi}\, G_{ab}^{<}(\omega),
\end{equation}
with~$a,b$ restricted to the particle block. Since~$V$ depends on~$\rho$,
Eqs.~\eqref{eq:hf_potential}, \eqref{eq:g0_hf}
and~\eqref{eq:occupation} must be solved self-consistently. For the
symmetric junction at~$\phi=0$, left-right symmetry makes~$\rho$ even
in~$\phi$ and~$\Delta T$. Its linear response therefore vanishes, and the
equilibrium density matrix suffices for the coefficients~$\mathcal{L}$
and~$K_\phi$. We evaluate it as a sum over fermionic Matsubara
frequencies, which is equivalent to Eq.~\eqref{eq:occupation} in
equilibrium. Starting from an initial guess~$\rho^{(0)}$, we iterate
\begin{equation}\label{eq:mixing}
    \rho^{(k+1)} \longleftarrow
    \alpha\,\rho\big[V(\rho^{(k)})\big] + (1-\alpha)\,\rho^{(k)},
\end{equation}
with an adaptive mixing parameter~$\alpha\in(0,1]$, until all elements
change by less than~$10^{-8}$.

\begin{figure*}[t]
    \centering
    \includegraphics[width=.95\textwidth]{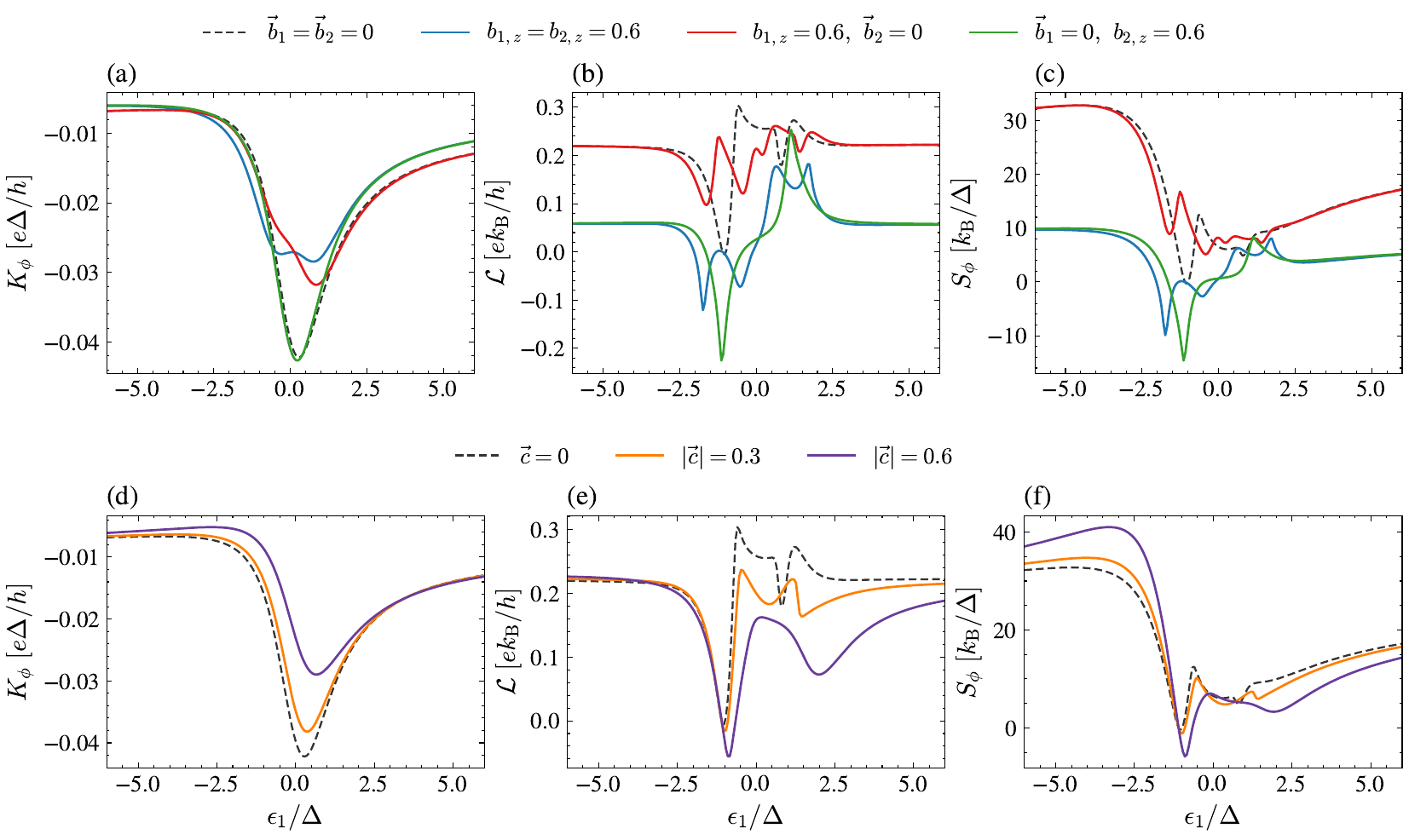}
    \caption{Effect of the two new terms of the mean-field interaction,
    treated as independent parameters of the noninteracting model.
    (a)-(c)~Zeeman fields~$\vec{b}_1$ and~$\vec{b}_2$ acting on the
    individual orbitals. (d)-(f)~Exchange field~$\vec{c}$ coupling the
    spins of the two orbitals. The panels show~$K_\phi$, $\mathcal{L}$
    and~$S_\phi$ from left to right. The second orbital lies
    at~$\epsilon_2=1.1\abs{\Delta}$ without spin-orbit coupling. The
    remaining parameters are those of Fig.~\ref{fig:seebeck}.}
    \label{fig:orbital_zeeman}
\end{figure*}

The mean-field equations can have several solutions. We start the
iteration from the noninteracting density matrix as well as from states
with parallel and antiparallel spin polarizations of the two orbitals along
several axes. Among the converged solutions, we select the one with the
lowest mean-field grand potential
\begin{equation}\label{eq:hf_grand_potential}
    \Omega_\mathrm{HF}[\rho] =
    \Omega_0\big[\mathcal{H}_\mathrm{QD}+\delta\mathcal{H}_\mathrm{HF}\big]
    - \frac{1}{2}\tr\left(V\rho\right),
\end{equation}
where~$\Omega_0[\mathcal{H}]$ denotes the grand potential of the
noninteracting junction with dot Hamiltonian~$\mathcal{H}$. The last term
removes the double counting of the interaction energy.

Without a Zeeman field, the Hamiltonian is invariant under time
reversal~$\Theta=i\sigma_y K$ at $\phi=0,\pi$. A spin-polarized solution~$\rho$ then has a
degenerate partner~$\mathcal{T}\rho^{*}\mathcal{T}$ with reversed
polarizations~$\vec{m}_\nu\to-\vec{m}_\nu$ and the same grand potential.
Each of the two solutions breaks time-reversal symmetry, which is an
artifact of the mean-field approximation. As emphasized in
Ref.~\cite{kleeorinLargeTunableThermophase2016}, both solutions must
therefore be taken into account to preserve the spin symmetry \footnote{At zero Zeeman field, unrestricted Hartree-Fock may yield a pair of degenerate spin-polarized solutions related by time reversal. These solutions represent local-moment formation at the mean-field level, but selecting one of them would assign a permanent spin polarization to the finite dot. We therefore restore the symmetry of spin-even observables by including both time-reversed solutions. For a macroscopic ferromagnet, by contrast, the reversal time between the two sectors becomes effectively infinite, and a single symmetry-broken state is physically realized on experimental time scales.}. We construct the partner explicitly, verify that it is self-consistent and
average the current over the pair. The thermophase is then defined by the
vanishing of the averaged current, and the thermophase Seebeck coefficient
follows from the averaged coefficients,
\begin{equation}\label{eq:seebeck_branch_average}
    S_\phi = -\frac{\mathcal{L}_+ + \mathcal{L}_-}{K_{\phi,+}+K_{\phi,-}},
\end{equation}
where~$\pm$ labels the two solutions. In effect, we assume that fluctuations between the two time-reversed mean-field solutions are fast compared with the measurement time, so that the experiment probes their ensemble-averaged current and the zero-current condition must be imposed only after this averaging. Averaging the density matrices
instead would give a paramagnetic state that is not self-consistent. For
the symmetric couplings considered here, both solutions give the
same~$\mathcal{L}$ and~$K_\phi$. The average
becomes essential for asymmetric couplings. There, a single solution
carries an equilibrium current at~$\phi=0$ that would appear as a spurious
anomalous phase. This current cancels between the two solutions.

\subsection{Orbital-dependent exchange field}\label{app:hf_results}

Within mean-field theory, the Coulomb interaction thus renormalizes the
orbital energies, the spin-orbit vector and the Zeeman field. The two new
terms  are the orbital-dependent exchange field~$-U\vec{m}_\nu/2$ and the exchange field between the orbitals. The orbital-dependent exchange field is
proportional to the spin polarization~$\vec{m}_\nu$ of orbital~$\nu$.
The eigenvalues~$(n_\nu\pm\abs{\vec{m}_\nu})/2$ of~$\rho_{\nu\nu}$ are the
occupations of the spin states parallel and antiparallel
to~$\vec{m}_\nu$. Since both lie between zero and one, the polarization
is bounded by~$\abs{\vec{m}_\nu}\le\min(n_\nu,2-n_\nu)$. An empty or full
orbital therefore cannot be spin polarized. The exchange fields of the
two orbitals differ in particular when one orbital is close to half
filling while the other is nearly empty or full. In spin-polarized states, the exchange term of~$U'$ additionally gives
the spin-orbit vector an imaginary part. It acts as a coupling
$d_1^\dagger(\vec{c}\cdot\vec{\sigma})d_2+\mathrm{h.c.}$ with a real
vector~$\vec{c}$. In terms of the combinations~$d_\pm=(d_1\pm d_2)/\sqrt{2}$
it reads~$d_+^\dagger(\vec{c}\cdot\vec{\sigma})d_+-d_-^\dagger(\vec{c}\cdot\vec{\sigma})d_-$.
It is thus a Zeeman field of opposite sign on the combination coupled
to the leads and on the decoupled one. Both new terms are therefore
orbital-dependent Zeeman fields. We treat them as independent parameters
of the noninteracting model in Fig.~\ref{fig:orbital_zeeman}.

Figs.~\ref{fig:orbital_zeeman}(a)-\ref{fig:orbital_zeeman}(c) show that the dominant effect of an orbital-selective field is controlled by the transport role of the orbital on which it acts. For~$\epsilon_1\lesssim-3\abs{\Delta}$, the first orbital
is doubly occupied and the second orbital near the coherence peaks
carries the quasiparticle current. A field on the first orbital then
leaves all coefficients essentially unchanged. A field on the second orbital splits
it. This suppresses~$\mathcal{L}$ and strongly reduces~$S_\phi$.
Near~$\epsilon_1=0$, the first orbital lies at the Fermi energy and
dominates the pair response. A field on this orbital alone
reduces~$\abs{K_\phi}$ while~$\mathcal{L}$ remains large, which
enhances~$S_\phi$. A common field instead also splits the second orbital
and suppresses~$\mathcal{L}$. An orbital-dependent field can thus soften
the junction without reducing its quasiparticle response. Within mean
field, such a field arises in the orbital close to half filling, which
is the one that controls the pair response. The inter-orbital exchange field in
Figs.~\ref{fig:orbital_zeeman}(d)-\ref{fig:orbital_zeeman}(f) splits the
spin states of the combination coupled to the leads. Like a Zeeman field,
it suppresses the pair response, most strongly near~$\epsilon_1=0$. At large negative $\epsilon_1$, the fixed second orbital dominates the quasiparticle coefficient, and the numerical results show that $\mathcal L$ is only weakly modified by the interorbital exchange field. The same field produces a somewhat larger relative reduction of $|K_\phi|$, causing $|S_\phi|=|\mathcal L/K_\phi|$ to increase with $|\mathbf c|$. Closer to the negative coherence peak, both coefficients change substantially and the thermophase reflects their competition. For
positive~$\epsilon_1$, the field also reduces~$\mathcal{L}$, which
lowers~$S_\phi$ slightly. Both terms thus act on the thermophase like
Zeeman fields restricted to particular orbitals or orbital combinations.
Correlation effects beyond mean field, such as Coulomb blockade or the Kondo effect, are not
captured by this treatment.

\bibliography{references}

\begin{thebibliography}{88}%
\makeatletter
\providecommand \@ifxundefined [1]{%
 \@ifx{#1\undefined}
}%
\providecommand \@ifnum [1]{%
 \ifnum #1\expandafter \@firstoftwo
 \else \expandafter \@secondoftwo
 \fi
}%
\providecommand \@ifx [1]{%
 \ifx #1\expandafter \@firstoftwo
 \else \expandafter \@secondoftwo
 \fi
}%
\providecommand \natexlab [1]{#1}%
\providecommand \enquote  [1]{``#1''}%
\providecommand \bibnamefont  [1]{#1}%
\providecommand \bibfnamefont [1]{#1}%
\providecommand \citenamefont [1]{#1}%
\providecommand \href@noop [0]{\@secondoftwo}%
\providecommand \href [0]{\begingroup \@sanitize@url \@href}%
\providecommand \@href[1]{\@@startlink{#1}\@@href}%
\providecommand \@@href[1]{\endgroup#1\@@endlink}%
\providecommand \@sanitize@url [0]{\catcode `\\12\catcode `\$12\catcode
  `\&12\catcode `\#12\catcode `\^12\catcode `\_12\catcode `\%12\relax}%
\providecommand \@@startlink[1]{}%
\providecommand \@@endlink[0]{}%
\providecommand \url  [0]{\begingroup\@sanitize@url \@url }%
\providecommand \@url [1]{\endgroup\@href {#1}{\urlprefix }}%
\providecommand \urlprefix  [0]{URL }%
\providecommand \Eprint [0]{\href }%
\providecommand \doibase [0]{https://doi.org/}%
\providecommand \selectlanguage [0]{\@gobble}%
\providecommand \bibinfo  [0]{\@secondoftwo}%
\providecommand \bibfield  [0]{\@secondoftwo}%
\providecommand \translation [1]{[#1]}%
\providecommand \BibitemOpen [0]{}%
\providecommand \bibitemStop [0]{}%
\providecommand \bibitemNoStop [0]{.\EOS\space}%
\providecommand \EOS [0]{\spacefactor3000\relax}%
\providecommand \BibitemShut  [1]{\csname bibitem#1\endcsname}%
\let\auto@bib@innerbib\@empty
\bibitem [{\citenamefont {Krinner}\ \emph {et~al.}(2019)\citenamefont
  {Krinner}, \citenamefont {Storz}, \citenamefont {Kurpiers}, \citenamefont
  {Magnard}, \citenamefont {Heinsoo}, \citenamefont {Keller}, \citenamefont
  {L{\"u}tolf}, \citenamefont {Eichler},\ and\ \citenamefont
  {Wallraff}}]{krinnerEngineeringCryogenicSetups2019}%
  \BibitemOpen
  \bibfield  {author} {\bibinfo {author} {\bibfnamefont {S.}~\bibnamefont
  {Krinner}}, \bibinfo {author} {\bibfnamefont {S.}~\bibnamefont {Storz}},
  \bibinfo {author} {\bibfnamefont {P.}~\bibnamefont {Kurpiers}}, \bibinfo
  {author} {\bibfnamefont {P.}~\bibnamefont {Magnard}}, \bibinfo {author}
  {\bibfnamefont {J.}~\bibnamefont {Heinsoo}}, \bibinfo {author} {\bibfnamefont
  {R.}~\bibnamefont {Keller}}, \bibinfo {author} {\bibfnamefont
  {J.}~\bibnamefont {L{\"u}tolf}}, \bibinfo {author} {\bibfnamefont
  {C.}~\bibnamefont {Eichler}},\ and\ \bibinfo {author} {\bibfnamefont
  {A.}~\bibnamefont {Wallraff}},\ }\bibfield  {title} {\bibinfo {title}
  {Engineering cryogenic setups for 100-qubit scale superconducting circuit
  systems},\ }\href {https://doi.org/10.1140/epjqt/s40507-019-0072-0}
  {\bibfield  {journal} {\bibinfo  {journal} {EPJ Quantum Technol.}\ }\textbf
  {\bibinfo {volume} {6}},\ \bibinfo {pages} {2} (\bibinfo {year}
  {2019})}\BibitemShut {NoStop}%
\bibitem [{\citenamefont {Pekola}\ and\ \citenamefont
  {Karimi}(2021)}]{pekolaColloquiumQuantumHeat2021}%
  \BibitemOpen
  \bibfield  {author} {\bibinfo {author} {\bibfnamefont {J.~P.}\ \bibnamefont
  {Pekola}}\ and\ \bibinfo {author} {\bibfnamefont {B.}~\bibnamefont
  {Karimi}},\ }\bibfield  {title} {\bibinfo {title} {Colloquium: {{Quantum}}
  heat transport in condensed matter systems},\ }\href
  {https://doi.org/10.1103/RevModPhys.93.041001} {\bibfield  {journal}
  {\bibinfo  {journal} {Rev. Mod. Phys.}\ }\textbf {\bibinfo {volume} {93}},\
  \bibinfo {pages} {041001} (\bibinfo {year} {2021})}\BibitemShut {NoStop}%
\bibitem [{\citenamefont {Giazotto}\ \emph {et~al.}(2006)\citenamefont
  {Giazotto}, \citenamefont {Heikkil{\"a}}, \citenamefont {Luukanen},
  \citenamefont {Savin},\ and\ \citenamefont
  {Pekola}}]{giazottoOpportunitiesMesoscopicsThermometry2006}%
  \BibitemOpen
  \bibfield  {author} {\bibinfo {author} {\bibfnamefont {F.}~\bibnamefont
  {Giazotto}}, \bibinfo {author} {\bibfnamefont {T.~T.}\ \bibnamefont
  {Heikkil{\"a}}}, \bibinfo {author} {\bibfnamefont {A.}~\bibnamefont
  {Luukanen}}, \bibinfo {author} {\bibfnamefont {A.~M.}\ \bibnamefont
  {Savin}},\ and\ \bibinfo {author} {\bibfnamefont {J.~P.}\ \bibnamefont
  {Pekola}},\ }\bibfield  {title} {\bibinfo {title} {Opportunities for
  mesoscopics in thermometry and refrigeration: {{Physics}} and applications},\
  }\href {https://doi.org/10.1103/RevModPhys.78.217} {\bibfield  {journal}
  {\bibinfo  {journal} {Rev. Mod. Phys.}\ }\textbf {\bibinfo {volume} {78}},\
  \bibinfo {pages} {217} (\bibinfo {year} {2006})}\BibitemShut {NoStop}%
\bibitem [{\citenamefont {Gasparinetti}\ \emph {et~al.}(2015)\citenamefont
  {Gasparinetti}, \citenamefont {Viisanen}, \citenamefont {Saira},
  \citenamefont {Faivre}, \citenamefont {Arzeo}, \citenamefont {Meschke},\ and\
  \citenamefont {Pekola}}]{gasparinettiFastElectronThermometry2015}%
  \BibitemOpen
  \bibfield  {author} {\bibinfo {author} {\bibfnamefont {S.}~\bibnamefont
  {Gasparinetti}}, \bibinfo {author} {\bibfnamefont {K.~L.}\ \bibnamefont
  {Viisanen}}, \bibinfo {author} {\bibfnamefont {O.-P.}\ \bibnamefont {Saira}},
  \bibinfo {author} {\bibfnamefont {T.}~\bibnamefont {Faivre}}, \bibinfo
  {author} {\bibfnamefont {M.}~\bibnamefont {Arzeo}}, \bibinfo {author}
  {\bibfnamefont {M.}~\bibnamefont {Meschke}},\ and\ \bibinfo {author}
  {\bibfnamefont {J.~P.}\ \bibnamefont {Pekola}},\ }\bibfield  {title}
  {\bibinfo {title} {Fast electron thermometry for ultrasensitive calorimetric
  detection},\ }\href {https://doi.org/10.1103/PhysRevApplied.3.014007}
  {\bibfield  {journal} {\bibinfo  {journal} {Phys. Rev. Appl.}\ }\textbf
  {\bibinfo {volume} {3}},\ \bibinfo {pages} {014007} (\bibinfo {year}
  {2015})}\BibitemShut {NoStop}%
\bibitem [{\citenamefont {Saira}\ \emph {et~al.}(2016)\citenamefont {Saira},
  \citenamefont {Zgirski}, \citenamefont {Viisanen}, \citenamefont {Golubev},\
  and\ \citenamefont {Pekola}}]{sairaDispersiveThermometryJosephson2016}%
  \BibitemOpen
  \bibfield  {author} {\bibinfo {author} {\bibfnamefont {O.-P.}\ \bibnamefont
  {Saira}}, \bibinfo {author} {\bibfnamefont {M.}~\bibnamefont {Zgirski}},
  \bibinfo {author} {\bibfnamefont {K.~L.}\ \bibnamefont {Viisanen}}, \bibinfo
  {author} {\bibfnamefont {D.~S.}\ \bibnamefont {Golubev}},\ and\ \bibinfo
  {author} {\bibfnamefont {J.~P.}\ \bibnamefont {Pekola}},\ }\bibfield  {title}
  {\bibinfo {title} {Dispersive {{Thermometry}} with a {{Josephson Junction
  Coupled}} to a {{Resonator}}},\ }\href
  {https://doi.org/10.1103/PhysRevApplied.6.024005} {\bibfield  {journal}
  {\bibinfo  {journal} {Phys. Rev. Appl.}\ }\textbf {\bibinfo {volume} {6}},\
  \bibinfo {pages} {024005} (\bibinfo {year} {2016})}\BibitemShut {NoStop}%
\bibitem [{\citenamefont {Zgirski}\ \emph {et~al.}(2018)\citenamefont
  {Zgirski}, \citenamefont {Foltyn}, \citenamefont {Savin}, \citenamefont
  {Norowski}, \citenamefont {Meschke},\ and\ \citenamefont
  {Pekola}}]{zgirskiNanosecondThermometryJosephson2018}%
  \BibitemOpen
  \bibfield  {author} {\bibinfo {author} {\bibfnamefont {M.}~\bibnamefont
  {Zgirski}}, \bibinfo {author} {\bibfnamefont {M.}~\bibnamefont {Foltyn}},
  \bibinfo {author} {\bibfnamefont {A.}~\bibnamefont {Savin}}, \bibinfo
  {author} {\bibfnamefont {K.}~\bibnamefont {Norowski}}, \bibinfo {author}
  {\bibfnamefont {M.}~\bibnamefont {Meschke}},\ and\ \bibinfo {author}
  {\bibfnamefont {J.~P.}\ \bibnamefont {Pekola}},\ }\bibfield  {title}
  {\bibinfo {title} {Nanosecond thermometry with {{Josephson}} junctions},\
  }\href {https://doi.org/10.1103/PhysRevApplied.10.044068} {\bibfield
  {journal} {\bibinfo  {journal} {Phys. Rev. Appl.}\ }\textbf {\bibinfo
  {volume} {10}},\ \bibinfo {pages} {044068} (\bibinfo {year}
  {2018})}\BibitemShut {NoStop}%
\bibitem [{\citenamefont {Machon}\ \emph {et~al.}(2013)\citenamefont {Machon},
  \citenamefont {Eschrig},\ and\ \citenamefont
  {Belzig}}]{machonNonlocalThermoelectricEffects2013}%
  \BibitemOpen
  \bibfield  {author} {\bibinfo {author} {\bibfnamefont {P.}~\bibnamefont
  {Machon}}, \bibinfo {author} {\bibfnamefont {M.}~\bibnamefont {Eschrig}},\
  and\ \bibinfo {author} {\bibfnamefont {W.}~\bibnamefont {Belzig}},\
  }\bibfield  {title} {\bibinfo {title} {Nonlocal thermoelectric effects and
  nonlocal {{Onsager}} relations in a three-terminal proximity-coupled
  superconductor-ferromagnet device},\ }\href
  {https://doi.org/10.1103/PhysRevLett.110.047002} {\bibfield  {journal}
  {\bibinfo  {journal} {Phys. Rev. Lett.}\ }\textbf {\bibinfo {volume} {110}},\
  \bibinfo {pages} {047002} (\bibinfo {year} {2013})}\BibitemShut {NoStop}%
\bibitem [{\citenamefont {Ozaeta}\ \emph {et~al.}(2014)\citenamefont {Ozaeta},
  \citenamefont {Virtanen}, \citenamefont {Bergeret},\ and\ \citenamefont
  {Heikkil{\"a}}}]{ozaetaPredictedVeryLarge2014}%
  \BibitemOpen
  \bibfield  {author} {\bibinfo {author} {\bibfnamefont {A.}~\bibnamefont
  {Ozaeta}}, \bibinfo {author} {\bibfnamefont {P.}~\bibnamefont {Virtanen}},
  \bibinfo {author} {\bibfnamefont {F.~S.}\ \bibnamefont {Bergeret}},\ and\
  \bibinfo {author} {\bibfnamefont {T.~T.}\ \bibnamefont {Heikkil{\"a}}},\
  }\bibfield  {title} {\bibinfo {title} {Predicted {{Very Large Thermoelectric
  Effect}} in {{Ferromagnet-Superconductor Junctions}} in the {{Presence}} of a
  {{Spin-Splitting Magnetic Field}}},\ }\href
  {https://doi.org/10.1103/PhysRevLett.112.057001} {\bibfield  {journal}
  {\bibinfo  {journal} {Phys. Rev. Lett.}\ }\textbf {\bibinfo {volume} {112}},\
  \bibinfo {pages} {057001} (\bibinfo {year} {2014})}\BibitemShut {NoStop}%
\bibitem [{\citenamefont {Linder}\ and\ \citenamefont
  {Bathen}(2016)}]{linderSpinCaloritronicsSuperconductors2016}%
  \BibitemOpen
  \bibfield  {author} {\bibinfo {author} {\bibfnamefont {J.}~\bibnamefont
  {Linder}}\ and\ \bibinfo {author} {\bibfnamefont {M.~E.}\ \bibnamefont
  {Bathen}},\ }\bibfield  {title} {\bibinfo {title} {Spin caloritronics with
  superconductors: {{Enhanced}} thermoelectric effects, generalized {{Onsager}}
  response-matrix, and thermal spin currents},\ }\href
  {https://doi.org/10.1103/PhysRevB.93.224509} {\bibfield  {journal} {\bibinfo
  {journal} {Phys. Rev. B}\ }\textbf {\bibinfo {volume} {93}},\ \bibinfo
  {pages} {224509} (\bibinfo {year} {2016})}\BibitemShut {NoStop}%
\bibitem [{\citenamefont {Kolenda}\ \emph {et~al.}(2016)\citenamefont
  {Kolenda}, \citenamefont {Wolf},\ and\ \citenamefont
  {Beckmann}}]{kolendaObservationThermoelectricCurrents2016}%
  \BibitemOpen
  \bibfield  {author} {\bibinfo {author} {\bibfnamefont {S.}~\bibnamefont
  {Kolenda}}, \bibinfo {author} {\bibfnamefont {M.~J.}\ \bibnamefont {Wolf}},\
  and\ \bibinfo {author} {\bibfnamefont {D.}~\bibnamefont {Beckmann}},\
  }\bibfield  {title} {\bibinfo {title} {Observation of {{Thermoelectric
  Currents}} in {{High-Field Superconductor-Ferromagnet Tunnel Junctions}}},\
  }\href {https://doi.org/10.1103/PhysRevLett.116.097001} {\bibfield  {journal}
  {\bibinfo  {journal} {Phys. Rev. Lett.}\ }\textbf {\bibinfo {volume} {116}},\
  \bibinfo {pages} {097001} (\bibinfo {year} {2016})}\BibitemShut {NoStop}%
\bibitem [{\citenamefont {Bergeret}\ \emph {et~al.}(2018)\citenamefont
  {Bergeret}, \citenamefont {Silaev}, \citenamefont {Virtanen},\ and\
  \citenamefont {Heikkil{\"a}}}]{bergeretColloquiumNonequilibriumEffects2018}%
  \BibitemOpen
  \bibfield  {author} {\bibinfo {author} {\bibfnamefont {F.~S.}\ \bibnamefont
  {Bergeret}}, \bibinfo {author} {\bibfnamefont {M.}~\bibnamefont {Silaev}},
  \bibinfo {author} {\bibfnamefont {P.}~\bibnamefont {Virtanen}},\ and\
  \bibinfo {author} {\bibfnamefont {T.~T.}\ \bibnamefont {Heikkil{\"a}}},\
  }\bibfield  {title} {\bibinfo {title} {Colloquium: {{Nonequilibrium}} effects
  in superconductors with a spin-splitting field},\ }\href
  {https://doi.org/10.1103/RevModPhys.90.041001} {\bibfield  {journal}
  {\bibinfo  {journal} {Rev. Mod. Phys.}\ }\textbf {\bibinfo {volume} {90}},\
  \bibinfo {pages} {041001} (\bibinfo {year} {2018})}\BibitemShut {NoStop}%
\bibitem [{\citenamefont {Heikkil{\"a}}\ \emph {et~al.}(2019)\citenamefont
  {Heikkil{\"a}}, \citenamefont {Silaev}, \citenamefont {Virtanen},\ and\
  \citenamefont {Bergeret}}]{heikkilaThermalElectricSpin2019}%
  \BibitemOpen
  \bibfield  {author} {\bibinfo {author} {\bibfnamefont {T.~T.}\ \bibnamefont
  {Heikkil{\"a}}}, \bibinfo {author} {\bibfnamefont {M.}~\bibnamefont
  {Silaev}}, \bibinfo {author} {\bibfnamefont {P.}~\bibnamefont {Virtanen}},\
  and\ \bibinfo {author} {\bibfnamefont {F.~S.}\ \bibnamefont {Bergeret}},\
  }\bibfield  {title} {\bibinfo {title} {Thermal, electric and spin transport
  in superconductor/ferromagnetic-insulator structures},\ }\href
  {https://doi.org/10.1016/j.progsurf.2019.100540} {\bibfield  {journal}
  {\bibinfo  {journal} {Prog. Surf. Sci.}\ }\textbf {\bibinfo {volume} {94}},\
  \bibinfo {pages} {100540} (\bibinfo {year} {2019})}\BibitemShut {NoStop}%
\bibitem [{\citenamefont {de~Araujo}\ \emph {et~al.}(2024)\citenamefont
  {de~Araujo}, \citenamefont {Virtanen}, \citenamefont {Spies}, \citenamefont
  {{Gonz{\'a}lez-Orellana}}, \citenamefont {Kerschbaumer}, \citenamefont
  {Ilyn}, \citenamefont {Rogero}, \citenamefont {Heikkil{\"a}}, \citenamefont
  {Giazotto},\ and\ \citenamefont
  {Strambini}}]{araujoSuperconductingSpintronicHeat2024}%
  \BibitemOpen
  \bibfield  {author} {\bibinfo {author} {\bibfnamefont {C.~I.~L.}\
  \bibnamefont {de~Araujo}}, \bibinfo {author} {\bibfnamefont {P.}~\bibnamefont
  {Virtanen}}, \bibinfo {author} {\bibfnamefont {M.}~\bibnamefont {Spies}},
  \bibinfo {author} {\bibfnamefont {C.}~\bibnamefont
  {{Gonz{\'a}lez-Orellana}}}, \bibinfo {author} {\bibfnamefont
  {S.}~\bibnamefont {Kerschbaumer}}, \bibinfo {author} {\bibfnamefont
  {M.}~\bibnamefont {Ilyn}}, \bibinfo {author} {\bibfnamefont {C.}~\bibnamefont
  {Rogero}}, \bibinfo {author} {\bibfnamefont {T.~T.}\ \bibnamefont
  {Heikkil{\"a}}}, \bibinfo {author} {\bibfnamefont {F.}~\bibnamefont
  {Giazotto}},\ and\ \bibinfo {author} {\bibfnamefont {E.}~\bibnamefont
  {Strambini}},\ }\bibfield  {title} {\bibinfo {title} {Superconducting
  spintronic heat engine},\ }\href {https://doi.org/10.1038/s41467-024-49052-z}
  {\bibfield  {journal} {\bibinfo  {journal} {Nat. Commun.}\ }\textbf {\bibinfo
  {volume} {15}},\ \bibinfo {pages} {4823} (\bibinfo {year}
  {2024})}\BibitemShut {NoStop}%
\bibitem [{\citenamefont {Tuero}\ \emph {et~al.}(2025)\citenamefont {Tuero},
  \citenamefont {Tjernshaugen}, \citenamefont {Sanchez}, \citenamefont
  {{Gonzalez-Ruano}}, \citenamefont {Lu}, \citenamefont {Linder},\ and\
  \citenamefont {Aliev}}]{tueroLargeTunableThermoelectric2025}%
  \BibitemOpen
  \bibfield  {author} {\bibinfo {author} {\bibfnamefont {P.}~\bibnamefont
  {Tuero}}, \bibinfo {author} {\bibfnamefont {J.~B.}\ \bibnamefont
  {Tjernshaugen}}, \bibinfo {author} {\bibfnamefont {C.}~\bibnamefont
  {Sanchez}}, \bibinfo {author} {\bibfnamefont {C.}~\bibnamefont
  {{Gonzalez-Ruano}}}, \bibinfo {author} {\bibfnamefont {Y.}~\bibnamefont
  {Lu}}, \bibinfo {author} {\bibfnamefont {J.}~\bibnamefont {Linder}},\ and\
  \bibinfo {author} {\bibfnamefont {F.~G.}\ \bibnamefont {Aliev}},\ }\bibfield
  {title} {\bibinfo {title} {Large {{Tunable Thermoelectric Effects}} in
  {{Superconducting Spin Valves}} with {{Commercially Available Materials}}},\
  }\href {https://doi.org/10.1103/xbgg-gttc} {\bibfield  {journal} {\bibinfo
  {journal} {PRX Energy}\ }\textbf {\bibinfo {volume} {4}},\ \bibinfo {pages}
  {033003} (\bibinfo {year} {2025})}\BibitemShut {NoStop}%
\bibitem [{\citenamefont {Hwang}\ \emph {et~al.}(2015)\citenamefont {Hwang},
  \citenamefont {Lopez},\ and\ \citenamefont
  {Sanchez}}]{hwangCrossThermoelectricCoupling2015}%
  \BibitemOpen
  \bibfield  {author} {\bibinfo {author} {\bibfnamefont {S.-Y.}\ \bibnamefont
  {Hwang}}, \bibinfo {author} {\bibfnamefont {R.}~\bibnamefont {Lopez}},\ and\
  \bibinfo {author} {\bibfnamefont {D.}~\bibnamefont {Sanchez}},\ }\bibfield
  {title} {\bibinfo {title} {Cross thermoelectric coupling in
  normal-superconductor quantum dots},\ }\href
  {https://doi.org/10.1103/PhysRevB.91.104518} {\bibfield  {journal} {\bibinfo
  {journal} {Phys. Rev. B}\ }\textbf {\bibinfo {volume} {91}},\ \bibinfo
  {pages} {104518} (\bibinfo {year} {2015})}\BibitemShut {NoStop}%
\bibitem [{\citenamefont {Kleeorin}\ \emph {et~al.}(2016)\citenamefont
  {Kleeorin}, \citenamefont {Meir}, \citenamefont {Giazotto},\ and\
  \citenamefont {Dubi}}]{kleeorinLargeTunableThermophase2016}%
  \BibitemOpen
  \bibfield  {author} {\bibinfo {author} {\bibfnamefont {Y.}~\bibnamefont
  {Kleeorin}}, \bibinfo {author} {\bibfnamefont {Y.}~\bibnamefont {Meir}},
  \bibinfo {author} {\bibfnamefont {F.}~\bibnamefont {Giazotto}},\ and\
  \bibinfo {author} {\bibfnamefont {Y.}~\bibnamefont {Dubi}},\ }\bibfield
  {title} {\bibinfo {title} {Large {{Tunable Thermophase}} in
  {{Superconductor}} -- {{Quantum Dot}} -- {{Superconductor Josephson
  Junctions}}},\ }\href {https://doi.org/10.1038/srep35116} {\bibfield
  {journal} {\bibinfo  {journal} {Sci. Rep.}\ }\textbf {\bibinfo {volume}
  {6}},\ \bibinfo {pages} {35116} (\bibinfo {year} {2016})}\BibitemShut
  {NoStop}%
\bibitem [{\citenamefont {Kamp}\ and\ \citenamefont
  {Sothmann}(2019)}]{kampPhasedependentHeatCharge2019}%
  \BibitemOpen
  \bibfield  {author} {\bibinfo {author} {\bibfnamefont {M.}~\bibnamefont
  {Kamp}}\ and\ \bibinfo {author} {\bibfnamefont {B.}~\bibnamefont
  {Sothmann}},\ }\bibfield  {title} {\bibinfo {title} {Phase-dependent heat and
  charge transport through superconductor-quantum dot hybrids},\ }\href
  {https://doi.org/10.1103/PhysRevB.99.045428} {\bibfield  {journal} {\bibinfo
  {journal} {Phys. Rev. B}\ }\textbf {\bibinfo {volume} {99}},\ \bibinfo
  {pages} {045428} (\bibinfo {year} {2019})}\BibitemShut {NoStop}%
\bibitem [{\citenamefont {Kalenkov}\ \emph {et~al.}(2012)\citenamefont
  {Kalenkov}, \citenamefont {Zaikin},\ and\ \citenamefont
  {Kuzmin}}]{kalenkovTheoryLargeThermoelectric2012}%
  \BibitemOpen
  \bibfield  {author} {\bibinfo {author} {\bibfnamefont {M.~S.}\ \bibnamefont
  {Kalenkov}}, \bibinfo {author} {\bibfnamefont {A.~D.}\ \bibnamefont
  {Zaikin}},\ and\ \bibinfo {author} {\bibfnamefont {L.~S.}\ \bibnamefont
  {Kuzmin}},\ }\bibfield  {title} {\bibinfo {title} {Theory of a large
  thermoelectric effect in superconductors doped with magnetic impurities},\
  }\href {https://doi.org/10.1103/PhysRevLett.109.147004} {\bibfield  {journal}
  {\bibinfo  {journal} {Phys. Rev. Lett.}\ }\textbf {\bibinfo {volume} {109}},\
  \bibinfo {pages} {147004} (\bibinfo {year} {2012})}\BibitemShut {NoStop}%
\bibitem [{\citenamefont {Bathen}\ and\ \citenamefont
  {Linder}(2017)}]{bathenSpinSeebeckEffect2017}%
  \BibitemOpen
  \bibfield  {author} {\bibinfo {author} {\bibfnamefont {M.~E.}\ \bibnamefont
  {Bathen}}\ and\ \bibinfo {author} {\bibfnamefont {J.}~\bibnamefont
  {Linder}},\ }\bibfield  {title} {\bibinfo {title} {Spin {{Seebeck}} effect
  and thermoelectric phenomena in superconducting hybrids with magnetic
  textures or spin-orbit coupling},\ }\href {https://doi.org/10.1038/srep41409}
  {\bibfield  {journal} {\bibinfo  {journal} {Sci. Rep.}\ }\textbf {\bibinfo
  {volume} {7}},\ \bibinfo {pages} {41409} (\bibinfo {year}
  {2017})}\BibitemShut {NoStop}%
\bibitem [{\citenamefont {Fornieri}\ and\ \citenamefont
  {Giazotto}(2017)}]{fornieriPhasecoherentCaloritronicsSuperconducting2017}%
  \BibitemOpen
  \bibfield  {author} {\bibinfo {author} {\bibfnamefont {A.}~\bibnamefont
  {Fornieri}}\ and\ \bibinfo {author} {\bibfnamefont {F.}~\bibnamefont
  {Giazotto}},\ }\bibfield  {title} {\bibinfo {title} {Towards phase-coherent
  caloritronics in superconducting circuits},\ }\href
  {https://doi.org/10.1038/nnano.2017.204} {\bibfield  {journal} {\bibinfo
  {journal} {Nat. Nanotechnol.}\ }\textbf {\bibinfo {volume} {12}},\ \bibinfo
  {pages} {944} (\bibinfo {year} {2017})}\BibitemShut {NoStop}%
\bibitem [{\citenamefont {Giazotto}\ and\ \citenamefont
  {{Mart{\'i}nez-P{\'e}rez}}(2012)}]{giazottoJosephsonHeatInterferometer2012}%
  \BibitemOpen
  \bibfield  {author} {\bibinfo {author} {\bibfnamefont {F.}~\bibnamefont
  {Giazotto}}\ and\ \bibinfo {author} {\bibfnamefont {M.~J.}\ \bibnamefont
  {{Mart{\'i}nez-P{\'e}rez}}},\ }\bibfield  {title} {\bibinfo {title} {The
  {{Josephson}} heat interferometer},\ }\href
  {https://doi.org/10.1038/nature11702} {\bibfield  {journal} {\bibinfo
  {journal} {Nature}\ }\textbf {\bibinfo {volume} {492}},\ \bibinfo {pages}
  {401} (\bibinfo {year} {2012})}\BibitemShut {NoStop}%
\bibitem [{\citenamefont {Marchegiani}\ \emph {et~al.}(2020)\citenamefont
  {Marchegiani}, \citenamefont {Braggio},\ and\ \citenamefont
  {Giazotto}}]{marchegianiPhasetunableThermoelectricityJosephson2020}%
  \BibitemOpen
  \bibfield  {author} {\bibinfo {author} {\bibfnamefont {G.}~\bibnamefont
  {Marchegiani}}, \bibinfo {author} {\bibfnamefont {A.}~\bibnamefont
  {Braggio}},\ and\ \bibinfo {author} {\bibfnamefont {F.}~\bibnamefont
  {Giazotto}},\ }\bibfield  {title} {\bibinfo {title} {Phase-tunable
  thermoelectricity in a {{Josephson}} junction},\ }\href
  {https://doi.org/10.1103/PhysRevResearch.2.043091} {\bibfield  {journal}
  {\bibinfo  {journal} {Phys. Rev. Res.}\ }\textbf {\bibinfo {volume} {2}},\
  \bibinfo {pages} {043091} (\bibinfo {year} {2020})}\BibitemShut {NoStop}%
\bibitem [{\citenamefont {Germanese}\ \emph {et~al.}(2022)\citenamefont
  {Germanese}, \citenamefont {Paolucci}, \citenamefont {Marchegiani},
  \citenamefont {Braggio},\ and\ \citenamefont
  {Giazotto}}]{germaneseBipolarThermoelectricJosephson2022}%
  \BibitemOpen
  \bibfield  {author} {\bibinfo {author} {\bibfnamefont {G.}~\bibnamefont
  {Germanese}}, \bibinfo {author} {\bibfnamefont {F.}~\bibnamefont {Paolucci}},
  \bibinfo {author} {\bibfnamefont {G.}~\bibnamefont {Marchegiani}}, \bibinfo
  {author} {\bibfnamefont {A.}~\bibnamefont {Braggio}},\ and\ \bibinfo {author}
  {\bibfnamefont {F.}~\bibnamefont {Giazotto}},\ }\bibfield  {title} {\bibinfo
  {title} {Bipolar thermoelectric {{Josephson}} engine},\ }\href
  {https://doi.org/10.1038/s41565-022-01208-y} {\bibfield  {journal} {\bibinfo
  {journal} {Nat. Nanotechnol.}\ }\textbf {\bibinfo {volume} {17}},\ \bibinfo
  {pages} {1084} (\bibinfo {year} {2022})}\BibitemShut {NoStop}%
\bibitem [{\citenamefont
  {Ginzburg}(1991)}]{ginzburgThermoelectricEffectsSuperconducting1991}%
  \BibitemOpen
  \bibfield  {author} {\bibinfo {author} {\bibfnamefont {V.~L.}\ \bibnamefont
  {Ginzburg}},\ }\bibfield  {title} {\bibinfo {title} {Thermoelectric effects
  in the superconducting state},\ }\href
  {https://doi.org/10.1070/PU1991v034n02ABEH002338} {\bibfield  {journal}
  {\bibinfo  {journal} {Sov. Phys. Uspekhi}\ }\textbf {\bibinfo {volume}
  {34}},\ \bibinfo {pages} {101} (\bibinfo {year} {1991})}\BibitemShut
  {NoStop}%
\bibitem [{\citenamefont {Guttman}\ \emph {et~al.}(1997)\citenamefont
  {Guttman}, \citenamefont {Nathanson}, \citenamefont {{Ben-Jacob}},\ and\
  \citenamefont {Bergman}}]{guttmanThermoelectricThermophaseEffects1997}%
  \BibitemOpen
  \bibfield  {author} {\bibinfo {author} {\bibfnamefont {G.~D.}\ \bibnamefont
  {Guttman}}, \bibinfo {author} {\bibfnamefont {B.}~\bibnamefont {Nathanson}},
  \bibinfo {author} {\bibfnamefont {E.}~\bibnamefont {{Ben-Jacob}}},\ and\
  \bibinfo {author} {\bibfnamefont {D.~J.}\ \bibnamefont {Bergman}},\
  }\bibfield  {title} {\bibinfo {title} {Thermoelectric and thermophase effects
  in {{Josephson}} junctions},\ }\href
  {https://doi.org/10.1103/PhysRevB.55.12691} {\bibfield  {journal} {\bibinfo
  {journal} {Phys. Rev. B}\ }\textbf {\bibinfo {volume} {55}},\ \bibinfo
  {pages} {12691} (\bibinfo {year} {1997})}\BibitemShut {NoStop}%
\bibitem [{\citenamefont {Giazotto}\ \emph
  {et~al.}(2015{\natexlab{a}})\citenamefont {Giazotto}, \citenamefont
  {Heikkil{\"a}},\ and\ \citenamefont
  {Bergeret}}]{giazottoVeryLargeThermophase2015}%
  \BibitemOpen
  \bibfield  {author} {\bibinfo {author} {\bibfnamefont {F.}~\bibnamefont
  {Giazotto}}, \bibinfo {author} {\bibfnamefont {T.~T.}\ \bibnamefont
  {Heikkil{\"a}}},\ and\ \bibinfo {author} {\bibfnamefont {F.~S.}\ \bibnamefont
  {Bergeret}},\ }\bibfield  {title} {\bibinfo {title} {Very {{Large
  Thermophase}} in {{Ferromagnetic Josephson Junctions}}},\ }\href
  {https://doi.org/10.1103/PhysRevLett.114.067001} {\bibfield  {journal}
  {\bibinfo  {journal} {Phys. Rev. Lett.}\ }\textbf {\bibinfo {volume} {114}},\
  \bibinfo {pages} {067001} (\bibinfo {year} {2015}{\natexlab{a}})}\BibitemShut
  {NoStop}%
\bibitem [{\citenamefont {Doh}\ \emph {et~al.}(2005)\citenamefont {Doh},
  \citenamefont {{van Dam}}, \citenamefont {Roest}, \citenamefont {Bakkers},
  \citenamefont {Kouwenhoven},\ and\ \citenamefont
  {De~Franceschi}}]{dohTunableSupercurrentSemiconductor2005}%
  \BibitemOpen
  \bibfield  {author} {\bibinfo {author} {\bibfnamefont {Y.-J.}\ \bibnamefont
  {Doh}}, \bibinfo {author} {\bibfnamefont {J.~A.}\ \bibnamefont {{van Dam}}},
  \bibinfo {author} {\bibfnamefont {A.~L.}\ \bibnamefont {Roest}}, \bibinfo
  {author} {\bibfnamefont {E.~P. A.~M.}\ \bibnamefont {Bakkers}}, \bibinfo
  {author} {\bibfnamefont {L.~P.}\ \bibnamefont {Kouwenhoven}},\ and\ \bibinfo
  {author} {\bibfnamefont {S.}~\bibnamefont {De~Franceschi}},\ }\bibfield
  {title} {\bibinfo {title} {Tunable supercurrent through semiconductor
  nanowires},\ }\href {https://doi.org/10.1126/science.1113523} {\bibfield
  {journal} {\bibinfo  {journal} {Science}\ }\textbf {\bibinfo {volume}
  {309}},\ \bibinfo {pages} {272} (\bibinfo {year} {2005})}\BibitemShut
  {NoStop}%
\bibitem [{\citenamefont {{Jarillo-Herrero}}\ \emph {et~al.}(2006)\citenamefont
  {{Jarillo-Herrero}}, \citenamefont {{van Dam}},\ and\ \citenamefont
  {Kouwenhoven}}]{jarilloherreroQuantumSupercurrentTransistors2006}%
  \BibitemOpen
  \bibfield  {author} {\bibinfo {author} {\bibfnamefont {P.}~\bibnamefont
  {{Jarillo-Herrero}}}, \bibinfo {author} {\bibfnamefont {J.~A.}\ \bibnamefont
  {{van Dam}}},\ and\ \bibinfo {author} {\bibfnamefont {L.~P.}\ \bibnamefont
  {Kouwenhoven}},\ }\bibfield  {title} {\bibinfo {title} {Quantum supercurrent
  transistors in carbon nanotubes},\ }\href
  {https://doi.org/10.1038/nature04550} {\bibfield  {journal} {\bibinfo
  {journal} {Nature}\ }\textbf {\bibinfo {volume} {439}},\ \bibinfo {pages}
  {953} (\bibinfo {year} {2006})}\BibitemShut {NoStop}%
\bibitem [{\citenamefont {De~Franceschi}\ \emph {et~al.}(2010)\citenamefont
  {De~Franceschi}, \citenamefont {Kouwenhoven}, \citenamefont
  {Sch{\"o}nenberger},\ and\ \citenamefont
  {Wernsdorfer}}]{defranceschiHybridSuperconductorQuantum2010}%
  \BibitemOpen
  \bibfield  {author} {\bibinfo {author} {\bibfnamefont {S.}~\bibnamefont
  {De~Franceschi}}, \bibinfo {author} {\bibfnamefont {L.}~\bibnamefont
  {Kouwenhoven}}, \bibinfo {author} {\bibfnamefont {C.}~\bibnamefont
  {Sch{\"o}nenberger}},\ and\ \bibinfo {author} {\bibfnamefont
  {W.}~\bibnamefont {Wernsdorfer}},\ }\bibfield  {title} {\bibinfo {title}
  {Hybrid superconductor--quantum dot devices},\ }\href
  {https://doi.org/10.1038/nnano.2010.173} {\bibfield  {journal} {\bibinfo
  {journal} {Nat. Nanotechnol.}\ }\textbf {\bibinfo {volume} {5}},\ \bibinfo
  {pages} {703} (\bibinfo {year} {2010})}\BibitemShut {NoStop}%
\bibitem [{\citenamefont {Kumar}\ \emph {et~al.}(2023)\citenamefont {Kumar},
  \citenamefont {Verma},\ and\ \citenamefont
  {{Ajay}}}]{kumarPhaseThermalDrivenTransport2023}%
  \BibitemOpen
  \bibfield  {author} {\bibinfo {author} {\bibfnamefont {B.}~\bibnamefont
  {Kumar}}, \bibinfo {author} {\bibfnamefont {S.}~\bibnamefont {Verma}},\ and\
  \bibinfo {author} {\bibnamefont {{Ajay}}},\ }\bibfield  {title} {\bibinfo
  {title} {Phase and {{Thermal-Driven Transport Across T-Shaped Double Quantum
  Dot Josephson Junction}}},\ }\href
  {https://doi.org/10.1007/s10948-023-06526-3} {\bibfield  {journal} {\bibinfo
  {journal} {J. Supercond. Nov. Magn.}\ }\textbf {\bibinfo {volume} {36}},\
  \bibinfo {pages} {831} (\bibinfo {year} {2023})}\BibitemShut {NoStop}%
\bibitem [{\citenamefont {Kumar}\ \emph {et~al.}(2025)\citenamefont {Kumar},
  \citenamefont {Verma},\ and\ \citenamefont
  {{Ajay}}}]{kumarJosephsonThermophaseEffect2025}%
  \BibitemOpen
  \bibfield  {author} {\bibinfo {author} {\bibfnamefont {B.}~\bibnamefont
  {Kumar}}, \bibinfo {author} {\bibfnamefont {S.}~\bibnamefont {Verma}},\ and\
  \bibinfo {author} {\bibnamefont {{Ajay}}},\ }\bibfield  {title} {\bibinfo
  {title} {Josephson and thermophase effect in interacting {{T}}-shaped double
  quantum dots system},\ }\href {https://doi.org/10.1016/j.physe.2024.116142}
  {\bibfield  {journal} {\bibinfo  {journal} {Phys. E Low-Dimens. Syst.
  Nanostructures}\ }\textbf {\bibinfo {volume} {166}},\ \bibinfo {pages}
  {116142} (\bibinfo {year} {2025})}\BibitemShut {NoStop}%
\bibitem [{\citenamefont {Fasth}\ \emph {et~al.}(2007)\citenamefont {Fasth},
  \citenamefont {Fuhrer}, \citenamefont {Samuelson}, \citenamefont {Golovach},\
  and\ \citenamefont {Loss}}]{fasthDirectMeasurementSpinOrbit2007}%
  \BibitemOpen
  \bibfield  {author} {\bibinfo {author} {\bibfnamefont {C.}~\bibnamefont
  {Fasth}}, \bibinfo {author} {\bibfnamefont {A.}~\bibnamefont {Fuhrer}},
  \bibinfo {author} {\bibfnamefont {L.}~\bibnamefont {Samuelson}}, \bibinfo
  {author} {\bibfnamefont {V.~N.}\ \bibnamefont {Golovach}},\ and\ \bibinfo
  {author} {\bibfnamefont {D.}~\bibnamefont {Loss}},\ }\bibfield  {title}
  {\bibinfo {title} {Direct {{Measurement}} of the {{Spin-Orbit Interaction}}
  in a {{Two-Electron InAs Nanowire Quantum Dot}}},\ }\href
  {https://doi.org/10.1103/PhysRevLett.98.266801} {\bibfield  {journal}
  {\bibinfo  {journal} {Phys. Rev. Lett.}\ }\textbf {\bibinfo {volume} {98}},\
  \bibinfo {pages} {266801} (\bibinfo {year} {2007})}\BibitemShut {NoStop}%
\bibitem [{\citenamefont {{Nadj-Perge}}\ \emph {et~al.}(2010)\citenamefont
  {{Nadj-Perge}}, \citenamefont {Frolov}, \citenamefont {Bakkers},\ and\
  \citenamefont {Kouwenhoven}}]{nadj-pergeSpinOrbitQubit2010}%
  \BibitemOpen
  \bibfield  {author} {\bibinfo {author} {\bibfnamefont {S.}~\bibnamefont
  {{Nadj-Perge}}}, \bibinfo {author} {\bibfnamefont {S.~M.}\ \bibnamefont
  {Frolov}}, \bibinfo {author} {\bibfnamefont {E.~P. A.~M.}\ \bibnamefont
  {Bakkers}},\ and\ \bibinfo {author} {\bibfnamefont {L.~P.}\ \bibnamefont
  {Kouwenhoven}},\ }\bibfield  {title} {\bibinfo {title} {Spin-orbit qubit in a
  semiconductor nanowire},\ }\href {https://doi.org/10.1038/nature09682}
  {\bibfield  {journal} {\bibinfo  {journal} {Nature}\ }\textbf {\bibinfo
  {volume} {468}},\ \bibinfo {pages} {1084} (\bibinfo {year}
  {2010})}\BibitemShut {NoStop}%
\bibitem [{\citenamefont {Han}\ \emph {et~al.}(2023)\citenamefont {Han},
  \citenamefont {Chan}, \citenamefont {{de Jong}}, \citenamefont {Prosko},
  \citenamefont {Badawy}, \citenamefont {Gazibegovic}, \citenamefont {Bakkers},
  \citenamefont {Kouwenhoven}, \citenamefont {Malinowski},\ and\ \citenamefont
  {Pfaff}}]{hanVariableOrbitalDependent2023}%
  \BibitemOpen
  \bibfield  {author} {\bibinfo {author} {\bibfnamefont {L.}~\bibnamefont
  {Han}}, \bibinfo {author} {\bibfnamefont {M.}~\bibnamefont {Chan}}, \bibinfo
  {author} {\bibfnamefont {D.}~\bibnamefont {{de Jong}}}, \bibinfo {author}
  {\bibfnamefont {C.}~\bibnamefont {Prosko}}, \bibinfo {author} {\bibfnamefont
  {G.}~\bibnamefont {Badawy}}, \bibinfo {author} {\bibfnamefont
  {S.}~\bibnamefont {Gazibegovic}}, \bibinfo {author} {\bibfnamefont {E.~P.
  A.~M.}\ \bibnamefont {Bakkers}}, \bibinfo {author} {\bibfnamefont {L.~P.}\
  \bibnamefont {Kouwenhoven}}, \bibinfo {author} {\bibfnamefont {F.~K.}\
  \bibnamefont {Malinowski}},\ and\ \bibinfo {author} {\bibfnamefont
  {W.}~\bibnamefont {Pfaff}},\ }\bibfield  {title} {\bibinfo {title} {Variable
  and orbital-dependent spin-orbit field orientations in an {{InSb}} double
  quantum dot characterized via dispersive gate sensing},\ }\href
  {https://doi.org/10.1103/PhysRevApplied.19.014063} {\bibfield  {journal}
  {\bibinfo  {journal} {Phys. Rev. Appl.}\ }\textbf {\bibinfo {volume} {19}},\
  \bibinfo {pages} {014063} (\bibinfo {year} {2023})}\BibitemShut {NoStop}%
\bibitem [{\citenamefont {Csonka}\ \emph {et~al.}(2008)\citenamefont {Csonka},
  \citenamefont {Hofstetter}, \citenamefont {Freitag}, \citenamefont
  {Oberholzer}, \citenamefont {Sch{\"o}nenberger}, \citenamefont {Jespersen},
  \citenamefont {Aagesen},\ and\ \citenamefont
  {Nyg{\aa}rd}}]{csonkaGiantFluctuationsGate2008}%
  \BibitemOpen
  \bibfield  {author} {\bibinfo {author} {\bibfnamefont {S.}~\bibnamefont
  {Csonka}}, \bibinfo {author} {\bibfnamefont {L.}~\bibnamefont {Hofstetter}},
  \bibinfo {author} {\bibfnamefont {F.}~\bibnamefont {Freitag}}, \bibinfo
  {author} {\bibfnamefont {S.}~\bibnamefont {Oberholzer}}, \bibinfo {author}
  {\bibfnamefont {C.}~\bibnamefont {Sch{\"o}nenberger}}, \bibinfo {author}
  {\bibfnamefont {T.~S.}\ \bibnamefont {Jespersen}}, \bibinfo {author}
  {\bibfnamefont {M.}~\bibnamefont {Aagesen}},\ and\ \bibinfo {author}
  {\bibfnamefont {J.}~\bibnamefont {Nyg{\aa}rd}},\ }\bibfield  {title}
  {\bibinfo {title} {Giant fluctuations and gate control of the {{g}}-factor in
  {{InAs}} nanowire quantum dots},\ }\href {https://doi.org/10.1021/nl802418w}
  {\bibfield  {journal} {\bibinfo  {journal} {Nano Lett.}\ }\textbf {\bibinfo
  {volume} {8}},\ \bibinfo {pages} {3932} (\bibinfo {year} {2008})}\BibitemShut
  {NoStop}%
\bibitem [{\citenamefont {Nilsson}\ \emph {et~al.}(2009)\citenamefont
  {Nilsson}, \citenamefont {Caroff}, \citenamefont {Thelander}, \citenamefont
  {Larsson}, \citenamefont {Wagner}, \citenamefont {Wernersson}, \citenamefont
  {Samuelson},\ and\ \citenamefont {Xu}}]{nilssonGiantLevelDependent2009}%
  \BibitemOpen
  \bibfield  {author} {\bibinfo {author} {\bibfnamefont {H.~A.}\ \bibnamefont
  {Nilsson}}, \bibinfo {author} {\bibfnamefont {P.}~\bibnamefont {Caroff}},
  \bibinfo {author} {\bibfnamefont {C.}~\bibnamefont {Thelander}}, \bibinfo
  {author} {\bibfnamefont {M.}~\bibnamefont {Larsson}}, \bibinfo {author}
  {\bibfnamefont {J.~B.}\ \bibnamefont {Wagner}}, \bibinfo {author}
  {\bibfnamefont {L.-E.}\ \bibnamefont {Wernersson}}, \bibinfo {author}
  {\bibfnamefont {L.}~\bibnamefont {Samuelson}},\ and\ \bibinfo {author}
  {\bibfnamefont {H.~Q.}\ \bibnamefont {Xu}},\ }\bibfield  {title} {\bibinfo
  {title} {Giant, level-dependent {{g}} factors in {{InSb}} nanowire quantum
  dots},\ }\href {https://doi.org/10.1021/nl901333a} {\bibfield  {journal}
  {\bibinfo  {journal} {Nano Lett.}\ }\textbf {\bibinfo {volume} {9}},\
  \bibinfo {pages} {3151} (\bibinfo {year} {2009})}\BibitemShut {NoStop}%
\bibitem [{\citenamefont {Mu}\ \emph {et~al.}(2020)\citenamefont {Mu},
  \citenamefont {Huang}, \citenamefont {Wang}, \citenamefont {Huang},
  \citenamefont {Wang},\ and\ \citenamefont
  {Xu}}]{muMeasurementsAnisotropicGfactors2020}%
  \BibitemOpen
  \bibfield  {author} {\bibinfo {author} {\bibfnamefont {J.}~\bibnamefont
  {Mu}}, \bibinfo {author} {\bibfnamefont {S.}~\bibnamefont {Huang}}, \bibinfo
  {author} {\bibfnamefont {J.-Y.}\ \bibnamefont {Wang}}, \bibinfo {author}
  {\bibfnamefont {G.-Y.}\ \bibnamefont {Huang}}, \bibinfo {author}
  {\bibfnamefont {X.}~\bibnamefont {Wang}},\ and\ \bibinfo {author}
  {\bibfnamefont {H.~Q.}\ \bibnamefont {Xu}},\ }\bibfield  {title} {\bibinfo
  {title} {Measurements of anisotropic g-factors for electrons in {{InSb}}
  nanowire quantum dots},\ }\href {https://doi.org/10.1088/1361-6528/abbc24}
  {\bibfield  {journal} {\bibinfo  {journal} {Nanotechnology}\ }\textbf
  {\bibinfo {volume} {32}},\ \bibinfo {pages} {020002} (\bibinfo {year}
  {2020})}\BibitemShut {NoStop}%
\bibitem [{\citenamefont {Dell'Anna}\ \emph {et~al.}(2007)\citenamefont
  {Dell'Anna}, \citenamefont {Zazunov}, \citenamefont {Egger},\ and\
  \citenamefont {Martin}}]{dellannaJosephsonCurrentQuantum2007}%
  \BibitemOpen
  \bibfield  {author} {\bibinfo {author} {\bibfnamefont {L.}~\bibnamefont
  {Dell'Anna}}, \bibinfo {author} {\bibfnamefont {A.}~\bibnamefont {Zazunov}},
  \bibinfo {author} {\bibfnamefont {R.}~\bibnamefont {Egger}},\ and\ \bibinfo
  {author} {\bibfnamefont {T.}~\bibnamefont {Martin}},\ }\bibfield  {title}
  {\bibinfo {title} {Josephson current through a quantum dot with spin-orbit
  coupling},\ }\href {https://doi.org/10.1103/PhysRevB.75.085305} {\bibfield
  {journal} {\bibinfo  {journal} {Phys. Rev. B}\ }\textbf {\bibinfo {volume}
  {75}},\ \bibinfo {pages} {085305} (\bibinfo {year} {2007})}\BibitemShut
  {NoStop}%
\bibitem [{\citenamefont {Zazunov}\ \emph {et~al.}(2009)\citenamefont
  {Zazunov}, \citenamefont {Egger}, \citenamefont {Jonckheere},\ and\
  \citenamefont {Martin}}]{zazunovAnomalousJosephsonCurrent2009}%
  \BibitemOpen
  \bibfield  {author} {\bibinfo {author} {\bibfnamefont {A.}~\bibnamefont
  {Zazunov}}, \bibinfo {author} {\bibfnamefont {R.}~\bibnamefont {Egger}},
  \bibinfo {author} {\bibfnamefont {T.}~\bibnamefont {Jonckheere}},\ and\
  \bibinfo {author} {\bibfnamefont {T.}~\bibnamefont {Martin}},\ }\bibfield
  {title} {\bibinfo {title} {Anomalous {{Josephson Current}} through a
  {{Spin-Orbit Coupled Quantum Dot}}},\ }\href
  {https://doi.org/10.1103/PhysRevLett.103.147004} {\bibfield  {journal}
  {\bibinfo  {journal} {Phys. Rev. Lett.}\ }\textbf {\bibinfo {volume} {103}},\
  \bibinfo {pages} {147004} (\bibinfo {year} {2009})}\BibitemShut {NoStop}%
\bibitem [{\citenamefont {Lim}\ \emph {et~al.}(2011)\citenamefont {Lim},
  \citenamefont {L{\'o}pez},\ and\ \citenamefont
  {Aguado}}]{limJosephsonCurrentCarbon2011}%
  \BibitemOpen
  \bibfield  {author} {\bibinfo {author} {\bibfnamefont {J.~S.}\ \bibnamefont
  {Lim}}, \bibinfo {author} {\bibfnamefont {R.}~\bibnamefont {L{\'o}pez}},\
  and\ \bibinfo {author} {\bibfnamefont {R.}~\bibnamefont {Aguado}},\
  }\bibfield  {title} {\bibinfo {title} {Josephson current in carbon nanotubes
  with spin-orbit interaction},\ }\href
  {https://doi.org/10.1103/PhysRevLett.107.196801} {\bibfield  {journal}
  {\bibinfo  {journal} {Phys. Rev. Lett.}\ }\textbf {\bibinfo {volume} {107}},\
  \bibinfo {pages} {196801} (\bibinfo {year} {2011})}\BibitemShut {NoStop}%
\bibitem [{\citenamefont {Brunetti}\ \emph {et~al.}(2013)\citenamefont
  {Brunetti}, \citenamefont {Zazunov}, \citenamefont {Kundu},\ and\
  \citenamefont {Egger}}]{brunettiAnomalousJosephsonCurrent2013}%
  \BibitemOpen
  \bibfield  {author} {\bibinfo {author} {\bibfnamefont {A.}~\bibnamefont
  {Brunetti}}, \bibinfo {author} {\bibfnamefont {A.}~\bibnamefont {Zazunov}},
  \bibinfo {author} {\bibfnamefont {A.}~\bibnamefont {Kundu}},\ and\ \bibinfo
  {author} {\bibfnamefont {R.}~\bibnamefont {Egger}},\ }\bibfield  {title}
  {\bibinfo {title} {Anomalous {{Josephson}} current, incipient time-reversal
  symmetry breaking, and {{Majorana}} bound states in interacting multilevel
  dots},\ }\href {https://doi.org/10.1103/PhysRevB.88.144515} {\bibfield
  {journal} {\bibinfo  {journal} {Phys. Rev. B}\ }\textbf {\bibinfo {volume}
  {88}},\ \bibinfo {pages} {144515} (\bibinfo {year} {2013})}\BibitemShut
  {NoStop}%
\bibitem [{\citenamefont {Yokoyama}\ \emph {et~al.}(2014)\citenamefont
  {Yokoyama}, \citenamefont {Eto},\ and\ \citenamefont
  {Nazarov}}]{yokoyamaAnomalousJosephsonEffect2014}%
  \BibitemOpen
  \bibfield  {author} {\bibinfo {author} {\bibfnamefont {T.}~\bibnamefont
  {Yokoyama}}, \bibinfo {author} {\bibfnamefont {M.}~\bibnamefont {Eto}},\ and\
  \bibinfo {author} {\bibfnamefont {Y.~V.}\ \bibnamefont {Nazarov}},\
  }\bibfield  {title} {\bibinfo {title} {Anomalous {{Josephson}} effect induced
  by spin-orbit interaction and {{Zeeman}} effect in semiconductor nanowires},\
  }\href {https://doi.org/10.1103/PhysRevB.89.195407} {\bibfield  {journal}
  {\bibinfo  {journal} {Phys. Rev. B}\ }\textbf {\bibinfo {volume} {89}},\
  \bibinfo {pages} {195407} (\bibinfo {year} {2014})}\BibitemShut {NoStop}%
\bibitem [{\citenamefont {Lee}\ \emph {et~al.}(2014)\citenamefont {Lee},
  \citenamefont {Jiang}, \citenamefont {Houzet}, \citenamefont {Aguado},
  \citenamefont {Lieber},\ and\ \citenamefont
  {De~Franceschi}}]{leeSpinresolvedAndreevLevels2014}%
  \BibitemOpen
  \bibfield  {author} {\bibinfo {author} {\bibfnamefont {E.~J.~H.}\
  \bibnamefont {Lee}}, \bibinfo {author} {\bibfnamefont {X.}~\bibnamefont
  {Jiang}}, \bibinfo {author} {\bibfnamefont {M.}~\bibnamefont {Houzet}},
  \bibinfo {author} {\bibfnamefont {R.}~\bibnamefont {Aguado}}, \bibinfo
  {author} {\bibfnamefont {C.~M.}\ \bibnamefont {Lieber}},\ and\ \bibinfo
  {author} {\bibfnamefont {S.}~\bibnamefont {De~Franceschi}},\ }\bibfield
  {title} {\bibinfo {title} {Spin-resolved {{Andreev}} levels and parity
  crossings in hybrid superconductor--semiconductor nanostructures},\ }\href
  {https://doi.org/10.1038/nnano.2013.267} {\bibfield  {journal} {\bibinfo
  {journal} {Nat. Nanotechnol.}\ }\textbf {\bibinfo {volume} {9}},\ \bibinfo
  {pages} {79} (\bibinfo {year} {2014})}\BibitemShut {NoStop}%
\bibitem [{\citenamefont {Szombati}\ \emph {et~al.}(2016)\citenamefont
  {Szombati}, \citenamefont {{Nadj-Perge}}, \citenamefont {Car}, \citenamefont
  {Plissard}, \citenamefont {Bakkers},\ and\ \citenamefont
  {Kouwenhoven}}]{szombatiJosephsonF0junctionNanowire2016}%
  \BibitemOpen
  \bibfield  {author} {\bibinfo {author} {\bibfnamefont {D.}~\bibnamefont
  {Szombati}}, \bibinfo {author} {\bibfnamefont {S.}~\bibnamefont
  {{Nadj-Perge}}}, \bibinfo {author} {\bibfnamefont {D.}~\bibnamefont {Car}},
  \bibinfo {author} {\bibfnamefont {S.}~\bibnamefont {Plissard}}, \bibinfo
  {author} {\bibfnamefont {E.}~\bibnamefont {Bakkers}},\ and\ \bibinfo {author}
  {\bibfnamefont {L.}~\bibnamefont {Kouwenhoven}},\ }\bibfield  {title}
  {\bibinfo {title} {Josephson {$\varphi_0$}-junction in nanowire quantum
  dots},\ }\href {https://doi.org/10.1038/nphys3742} {\bibfield  {journal}
  {\bibinfo  {journal} {Nat. Phys.}\ }\textbf {\bibinfo {volume} {12}},\
  \bibinfo {pages} {568} (\bibinfo {year} {2016})}\BibitemShut {NoStop}%
\bibitem [{\citenamefont {Whiticar}\ \emph {et~al.}(2021)\citenamefont
  {Whiticar}, \citenamefont {Fornieri}, \citenamefont {Banerjee}, \citenamefont
  {Drachmann}, \citenamefont {Gronin}, \citenamefont {Gardner}, \citenamefont
  {Lindemann}, \citenamefont {Manfra},\ and\ \citenamefont
  {Marcus}}]{whiticarZeemandrivenParityTransitions2021}%
  \BibitemOpen
  \bibfield  {author} {\bibinfo {author} {\bibfnamefont {A.~M.}\ \bibnamefont
  {Whiticar}}, \bibinfo {author} {\bibfnamefont {A.}~\bibnamefont {Fornieri}},
  \bibinfo {author} {\bibfnamefont {A.}~\bibnamefont {Banerjee}}, \bibinfo
  {author} {\bibfnamefont {A.~C.~C.}\ \bibnamefont {Drachmann}}, \bibinfo
  {author} {\bibfnamefont {S.}~\bibnamefont {Gronin}}, \bibinfo {author}
  {\bibfnamefont {G.~C.}\ \bibnamefont {Gardner}}, \bibinfo {author}
  {\bibfnamefont {T.}~\bibnamefont {Lindemann}}, \bibinfo {author}
  {\bibfnamefont {M.~J.}\ \bibnamefont {Manfra}},\ and\ \bibinfo {author}
  {\bibfnamefont {C.~M.}\ \bibnamefont {Marcus}},\ }\bibfield  {title}
  {\bibinfo {title} {Zeeman-driven parity transitions in an {{Andreev}} quantum
  dot},\ }\href {https://doi.org/10.1103/PhysRevB.103.245308} {\bibfield
  {journal} {\bibinfo  {journal} {Phys. Rev. B}\ }\textbf {\bibinfo {volume}
  {103}},\ \bibinfo {pages} {245308} (\bibinfo {year} {2021})}\BibitemShut
  {NoStop}%
\bibitem [{\citenamefont {Bargerbos}\ \emph {et~al.}(2023)\citenamefont
  {Bargerbos}, \citenamefont {{Pita-Vidal}}, \citenamefont {{\v Z}itko},
  \citenamefont {Splitthoff}, \citenamefont {Gr{\"u}nhaupt}, \citenamefont
  {Wesdorp}, \citenamefont {Liu}, \citenamefont {Kouwenhoven}, \citenamefont
  {Aguado}, \citenamefont {Andersen}, \citenamefont {Kou},\ and\ \citenamefont
  {{van Heck}}}]{bargerbosSpectroscopySpinsplitAndreev2023}%
  \BibitemOpen
  \bibfield  {author} {\bibinfo {author} {\bibfnamefont {A.}~\bibnamefont
  {Bargerbos}}, \bibinfo {author} {\bibfnamefont {M.}~\bibnamefont
  {{Pita-Vidal}}}, \bibinfo {author} {\bibfnamefont {R.}~\bibnamefont {{\v
  Z}itko}}, \bibinfo {author} {\bibfnamefont {L.~J.}\ \bibnamefont
  {Splitthoff}}, \bibinfo {author} {\bibfnamefont {L.}~\bibnamefont
  {Gr{\"u}nhaupt}}, \bibinfo {author} {\bibfnamefont {J.~J.}\ \bibnamefont
  {Wesdorp}}, \bibinfo {author} {\bibfnamefont {Y.}~\bibnamefont {Liu}},
  \bibinfo {author} {\bibfnamefont {L.~P.}\ \bibnamefont {Kouwenhoven}},
  \bibinfo {author} {\bibfnamefont {R.}~\bibnamefont {Aguado}}, \bibinfo
  {author} {\bibfnamefont {C.~K.}\ \bibnamefont {Andersen}}, \bibinfo {author}
  {\bibfnamefont {A.}~\bibnamefont {Kou}},\ and\ \bibinfo {author}
  {\bibfnamefont {B.}~\bibnamefont {{van Heck}}},\ }\bibfield  {title}
  {\bibinfo {title} {Spectroscopy of spin-split {{Andreev}} levels in a quantum
  dot with superconducting leads},\ }\href
  {https://doi.org/10.1103/PhysRevLett.131.097001} {\bibfield  {journal}
  {\bibinfo  {journal} {Phys. Rev. Lett.}\ }\textbf {\bibinfo {volume} {131}},\
  \bibinfo {pages} {097001} (\bibinfo {year} {2023})}\BibitemShut {NoStop}%
\bibitem [{\citenamefont {Della~Rocca}\ \emph {et~al.}(2007)\citenamefont
  {Della~Rocca}, \citenamefont {Chauvin}, \citenamefont {Huard}, \citenamefont
  {Pothier}, \citenamefont {Esteve},\ and\ \citenamefont
  {Urbina}}]{dellaroccaMeasurementCurrentphaseRelation2007}%
  \BibitemOpen
  \bibfield  {author} {\bibinfo {author} {\bibfnamefont {M.~L.}\ \bibnamefont
  {Della~Rocca}}, \bibinfo {author} {\bibfnamefont {M.}~\bibnamefont
  {Chauvin}}, \bibinfo {author} {\bibfnamefont {B.}~\bibnamefont {Huard}},
  \bibinfo {author} {\bibfnamefont {H.}~\bibnamefont {Pothier}}, \bibinfo
  {author} {\bibfnamefont {D.}~\bibnamefont {Esteve}},\ and\ \bibinfo {author}
  {\bibfnamefont {C.}~\bibnamefont {Urbina}},\ }\bibfield  {title} {\bibinfo
  {title} {Measurement of the current-phase relation of superconducting atomic
  contacts},\ }\href {https://doi.org/10.1103/PhysRevLett.99.127005} {\bibfield
   {journal} {\bibinfo  {journal} {Phys. Rev. Lett.}\ }\textbf {\bibinfo
  {volume} {99}},\ \bibinfo {pages} {127005} (\bibinfo {year}
  {2007})}\BibitemShut {NoStop}%
\bibitem [{\citenamefont {Spanton}\ \emph {et~al.}(2017)\citenamefont
  {Spanton}, \citenamefont {Deng}, \citenamefont {Vaitiek{\.e}nas},
  \citenamefont {Krogstrup}, \citenamefont {Nyg{\aa}rd}, \citenamefont
  {Marcus},\ and\ \citenamefont
  {Moler}}]{spantonCurrentphaseRelationsFewmode2017}%
  \BibitemOpen
  \bibfield  {author} {\bibinfo {author} {\bibfnamefont {E.~M.}\ \bibnamefont
  {Spanton}}, \bibinfo {author} {\bibfnamefont {M.}~\bibnamefont {Deng}},
  \bibinfo {author} {\bibfnamefont {S.}~\bibnamefont {Vaitiek{\.e}nas}},
  \bibinfo {author} {\bibfnamefont {P.}~\bibnamefont {Krogstrup}}, \bibinfo
  {author} {\bibfnamefont {J.}~\bibnamefont {Nyg{\aa}rd}}, \bibinfo {author}
  {\bibfnamefont {C.~M.}\ \bibnamefont {Marcus}},\ and\ \bibinfo {author}
  {\bibfnamefont {K.~A.}\ \bibnamefont {Moler}},\ }\bibfield  {title} {\bibinfo
  {title} {Current--phase relations of few-mode {{InAs}} nanowire {{Josephson}}
  junctions},\ }\href {https://doi.org/10.1038/nphys4224} {\bibfield  {journal}
  {\bibinfo  {journal} {Nat. Phys.}\ }\textbf {\bibinfo {volume} {13}},\
  \bibinfo {pages} {1177} (\bibinfo {year} {2017})}\BibitemShut {NoStop}%
\bibitem [{\citenamefont {Nichele}\ \emph {et~al.}(2020)\citenamefont
  {Nichele}, \citenamefont {Portol{\'e}s}, \citenamefont {Fornieri},
  \citenamefont {Whiticar}, \citenamefont {Drachmann}, \citenamefont {Gronin},
  \citenamefont {Wang}, \citenamefont {Gardner}, \citenamefont {Thomas},
  \citenamefont {Hatke}, \citenamefont {Manfra},\ and\ \citenamefont
  {Marcus}}]{nicheleRelatingAndreevBound2020}%
  \BibitemOpen
  \bibfield  {author} {\bibinfo {author} {\bibfnamefont {F.}~\bibnamefont
  {Nichele}}, \bibinfo {author} {\bibfnamefont {E.}~\bibnamefont
  {Portol{\'e}s}}, \bibinfo {author} {\bibfnamefont {A.}~\bibnamefont
  {Fornieri}}, \bibinfo {author} {\bibfnamefont {A.~M.}\ \bibnamefont
  {Whiticar}}, \bibinfo {author} {\bibfnamefont {A.~C.~C.}\ \bibnamefont
  {Drachmann}}, \bibinfo {author} {\bibfnamefont {S.}~\bibnamefont {Gronin}},
  \bibinfo {author} {\bibfnamefont {T.}~\bibnamefont {Wang}}, \bibinfo {author}
  {\bibfnamefont {G.~C.}\ \bibnamefont {Gardner}}, \bibinfo {author}
  {\bibfnamefont {C.}~\bibnamefont {Thomas}}, \bibinfo {author} {\bibfnamefont
  {A.~T.}\ \bibnamefont {Hatke}}, \bibinfo {author} {\bibfnamefont {M.~J.}\
  \bibnamefont {Manfra}},\ and\ \bibinfo {author} {\bibfnamefont {C.~M.}\
  \bibnamefont {Marcus}},\ }\bibfield  {title} {\bibinfo {title} {Relating
  {{Andreev Bound States}} and {{Supercurrents}} in {{Hybrid Josephson
  Junctions}}},\ }\href {https://doi.org/10.1103/PhysRevLett.124.226801}
  {\bibfield  {journal} {\bibinfo  {journal} {Phys. Rev. Lett.}\ }\textbf
  {\bibinfo {volume} {124}},\ \bibinfo {pages} {226801} (\bibinfo {year}
  {2020})}\BibitemShut {NoStop}%
\bibitem [{\citenamefont {Assouline}\ \emph {et~al.}(2019)\citenamefont
  {Assouline}, \citenamefont {{Feuillet-Palma}}, \citenamefont {Bergeal},
  \citenamefont {Zhang}, \citenamefont {Mottaghizadeh}, \citenamefont
  {Zimmers}, \citenamefont {Lhuillier}, \citenamefont {Eddrie}, \citenamefont
  {Atkinson}, \citenamefont {Aprili},\ and\ \citenamefont
  {Aubin}}]{assoulineSpinorbitInducedPhaseshift2019}%
  \BibitemOpen
  \bibfield  {author} {\bibinfo {author} {\bibfnamefont {A.}~\bibnamefont
  {Assouline}}, \bibinfo {author} {\bibfnamefont {C.}~\bibnamefont
  {{Feuillet-Palma}}}, \bibinfo {author} {\bibfnamefont {N.}~\bibnamefont
  {Bergeal}}, \bibinfo {author} {\bibfnamefont {T.}~\bibnamefont {Zhang}},
  \bibinfo {author} {\bibfnamefont {A.}~\bibnamefont {Mottaghizadeh}}, \bibinfo
  {author} {\bibfnamefont {A.}~\bibnamefont {Zimmers}}, \bibinfo {author}
  {\bibfnamefont {E.}~\bibnamefont {Lhuillier}}, \bibinfo {author}
  {\bibfnamefont {M.}~\bibnamefont {Eddrie}}, \bibinfo {author} {\bibfnamefont
  {P.}~\bibnamefont {Atkinson}}, \bibinfo {author} {\bibfnamefont
  {M.}~\bibnamefont {Aprili}},\ and\ \bibinfo {author} {\bibfnamefont
  {H.}~\bibnamefont {Aubin}},\ }\bibfield  {title} {\bibinfo {title}
  {Spin-orbit induced phase-shift in {{Bi}}{$_{2}$}{{Se}}{$_3$} {{Josephson}}
  junctions},\ }\href {https://doi.org/10.1038/s41467-018-08022-y} {\bibfield
  {journal} {\bibinfo  {journal} {Nat. Commun.}\ }\textbf {\bibinfo {volume}
  {10}},\ \bibinfo {pages} {126} (\bibinfo {year} {2019})}\BibitemShut
  {NoStop}%
\bibitem [{\citenamefont {Mayer}\ \emph {et~al.}(2020)\citenamefont {Mayer},
  \citenamefont {Dartiailh}, \citenamefont {Yuan}, \citenamefont
  {Wickramasinghe}, \citenamefont {Rossi},\ and\ \citenamefont
  {Shabani}}]{mayerGateControlledAnomalous2020}%
  \BibitemOpen
  \bibfield  {author} {\bibinfo {author} {\bibfnamefont {W.}~\bibnamefont
  {Mayer}}, \bibinfo {author} {\bibfnamefont {M.~C.}\ \bibnamefont
  {Dartiailh}}, \bibinfo {author} {\bibfnamefont {J.}~\bibnamefont {Yuan}},
  \bibinfo {author} {\bibfnamefont {K.~S.}\ \bibnamefont {Wickramasinghe}},
  \bibinfo {author} {\bibfnamefont {E.}~\bibnamefont {Rossi}},\ and\ \bibinfo
  {author} {\bibfnamefont {J.}~\bibnamefont {Shabani}},\ }\bibfield  {title}
  {\bibinfo {title} {Gate controlled anomalous phase shift in {{Al}}/{{InAs
  Josephson}} junctions},\ }\href {https://doi.org/10.1038/s41467-019-14094-1}
  {\bibfield  {journal} {\bibinfo  {journal} {Nat. Commun.}\ }\textbf {\bibinfo
  {volume} {11}},\ \bibinfo {pages} {212} (\bibinfo {year} {2020})}\BibitemShut
  {NoStop}%
\bibitem [{\citenamefont {Strambini}\ \emph {et~al.}(2020)\citenamefont
  {Strambini}, \citenamefont {Iorio}, \citenamefont {Durante}, \citenamefont
  {Citro}, \citenamefont {{Sanz-Fern{\'a}ndez}}, \citenamefont {Guarcello},
  \citenamefont {Tokatly}, \citenamefont {Braggio}, \citenamefont {Rocci},
  \citenamefont {Ligato}, \citenamefont {Zannier}, \citenamefont {Sorba},
  \citenamefont {Bergeret},\ and\ \citenamefont
  {Giazotto}}]{strambiniJosephsonPhaseBattery2020}%
  \BibitemOpen
  \bibfield  {author} {\bibinfo {author} {\bibfnamefont {E.}~\bibnamefont
  {Strambini}}, \bibinfo {author} {\bibfnamefont {A.}~\bibnamefont {Iorio}},
  \bibinfo {author} {\bibfnamefont {O.}~\bibnamefont {Durante}}, \bibinfo
  {author} {\bibfnamefont {R.}~\bibnamefont {Citro}}, \bibinfo {author}
  {\bibfnamefont {C.}~\bibnamefont {{Sanz-Fern{\'a}ndez}}}, \bibinfo {author}
  {\bibfnamefont {C.}~\bibnamefont {Guarcello}}, \bibinfo {author}
  {\bibfnamefont {I.~V.}\ \bibnamefont {Tokatly}}, \bibinfo {author}
  {\bibfnamefont {A.}~\bibnamefont {Braggio}}, \bibinfo {author} {\bibfnamefont
  {M.}~\bibnamefont {Rocci}}, \bibinfo {author} {\bibfnamefont
  {N.}~\bibnamefont {Ligato}}, \bibinfo {author} {\bibfnamefont
  {V.}~\bibnamefont {Zannier}}, \bibinfo {author} {\bibfnamefont
  {L.}~\bibnamefont {Sorba}}, \bibinfo {author} {\bibfnamefont {F.~S.}\
  \bibnamefont {Bergeret}},\ and\ \bibinfo {author} {\bibfnamefont
  {F.}~\bibnamefont {Giazotto}},\ }\bibfield  {title} {\bibinfo {title} {A
  {{Josephson}} phase battery},\ }\href
  {https://doi.org/10.1038/s41565-020-0712-7} {\bibfield  {journal} {\bibinfo
  {journal} {Nat. Nanotechnol.}\ }\textbf {\bibinfo {volume} {15}},\ \bibinfo
  {pages} {656} (\bibinfo {year} {2020})}\BibitemShut {NoStop}%
\bibitem [{\citenamefont {Zavaritskii}(1974)}]{zavaritskiiThermoelectric1974}%
  \BibitemOpen
  \bibfield  {author} {\bibinfo {author} {\bibfnamefont {N.~V.}\ \bibnamefont
  {Zavaritskii}},\ }\bibfield  {title} {\bibinfo {title} {Observation of
  superconducting current excited in a superconductor by heat flow},\
  }\href@noop {} {\bibfield  {journal} {\bibinfo  {journal} {JETP Lett.}\
  }\textbf {\bibinfo {volume} {19}},\ \bibinfo {pages} {126} (\bibinfo {year}
  {1974})}\BibitemShut {NoStop}%
\bibitem [{\citenamefont {Van~Harlingen}\ \emph {et~al.}(1980)\citenamefont
  {Van~Harlingen}, \citenamefont {Heidel},\ and\ \citenamefont
  {Garland}}]{vanharlingenExperimentalStudyThermoelectricity1980}%
  \BibitemOpen
  \bibfield  {author} {\bibinfo {author} {\bibfnamefont {D.~J.}\ \bibnamefont
  {Van~Harlingen}}, \bibinfo {author} {\bibfnamefont {D.~F.}\ \bibnamefont
  {Heidel}},\ and\ \bibinfo {author} {\bibfnamefont {J.~C.}\ \bibnamefont
  {Garland}},\ }\bibfield  {title} {\bibinfo {title} {Experimental study of
  thermoelectricity in superconducting indium},\ }\href
  {https://doi.org/10.1103/PhysRevB.21.1842} {\bibfield  {journal} {\bibinfo
  {journal} {Phys. Rev. B}\ }\textbf {\bibinfo {volume} {21}},\ \bibinfo
  {pages} {1842} (\bibinfo {year} {1980})}\BibitemShut {NoStop}%
\bibitem [{\citenamefont {Giazotto}\ \emph
  {et~al.}(2015{\natexlab{b}})\citenamefont {Giazotto}, \citenamefont
  {Solinas}, \citenamefont {Braggio},\ and\ \citenamefont
  {Bergeret}}]{giazottoFerromagneticInsulatorBasedSuperconductingJunctions2015}%
  \BibitemOpen
  \bibfield  {author} {\bibinfo {author} {\bibfnamefont {F.}~\bibnamefont
  {Giazotto}}, \bibinfo {author} {\bibfnamefont {P.}~\bibnamefont {Solinas}},
  \bibinfo {author} {\bibfnamefont {A.}~\bibnamefont {Braggio}},\ and\ \bibinfo
  {author} {\bibfnamefont {F.~S.}\ \bibnamefont {Bergeret}},\ }\bibfield
  {title} {\bibinfo {title} {Ferromagnetic-{{Insulator-Based Superconducting
  Junctions}} as {{Sensitive Electron Thermometers}}},\ }\href
  {https://doi.org/10.1103/PhysRevApplied.4.044016} {\bibfield  {journal}
  {\bibinfo  {journal} {Phys. Rev. Appl.}\ }\textbf {\bibinfo {volume} {4}},\
  \bibinfo {pages} {044016} (\bibinfo {year} {2015}{\natexlab{b}})}\BibitemShut
  {NoStop}%
\bibitem [{\citenamefont {Karimi}\ and\ \citenamefont
  {Pekola}(2018)}]{karimiNoninvasiveThermometerBased2018}%
  \BibitemOpen
  \bibfield  {author} {\bibinfo {author} {\bibfnamefont {B.}~\bibnamefont
  {Karimi}}\ and\ \bibinfo {author} {\bibfnamefont {J.~P.}\ \bibnamefont
  {Pekola}},\ }\bibfield  {title} {\bibinfo {title} {Noninvasive {{Thermometer
  Based}} on the {{Zero-Bias Anomaly}} of a {{Superconducting Junction}} for
  {{Ultrasensitive Calorimetry}}},\ }\href
  {https://doi.org/10.1103/PhysRevApplied.10.054048} {\bibfield  {journal}
  {\bibinfo  {journal} {Phys. Rev. Appl.}\ }\textbf {\bibinfo {volume} {10}},\
  \bibinfo {pages} {054048} (\bibinfo {year} {2018})}\BibitemShut {NoStop}%
\bibitem [{\citenamefont {Winkler}(2003)}]{winklerSpinOrbitCoupling2003}%
  \BibitemOpen
  \bibfield  {author} {\bibinfo {author} {\bibfnamefont {R.}~\bibnamefont
  {Winkler}},\ }\href {https://doi.org/10.1007/b13586} {\emph {\bibinfo {title}
  {Spin-Orbit Coupling Effects in Two-Dimensional Electron and Hole
  Systems}}},\ \bibinfo {series} {Springer Tracts in Modern Physics}, Vol.\
  \bibinfo {volume} {191}\ (\bibinfo  {publisher} {Springer},\ \bibinfo
  {address} {Berlin, Heidelberg},\ \bibinfo {year} {2003})\BibitemShut
  {NoStop}%
\bibitem [{\citenamefont {Danon}\ and\ \citenamefont
  {Nazarov}(2009)}]{danonPauliSpinBlockade2009}%
  \BibitemOpen
  \bibfield  {author} {\bibinfo {author} {\bibfnamefont {J.}~\bibnamefont
  {Danon}}\ and\ \bibinfo {author} {\bibfnamefont {{\relax Yu}.~V.}\
  \bibnamefont {Nazarov}},\ }\bibfield  {title} {\bibinfo {title} {Pauli spin
  blockade in the presence of strong spin-orbit coupling},\ }\href
  {https://doi.org/10.1103/PhysRevB.80.041301} {\bibfield  {journal} {\bibinfo
  {journal} {Phys. Rev. B}\ }\textbf {\bibinfo {volume} {80}},\ \bibinfo
  {pages} {041301(R)} (\bibinfo {year} {2009})}\BibitemShut {NoStop}%
\bibitem [{\citenamefont {Kouwenhoven}\ \emph {et~al.}(2001)\citenamefont
  {Kouwenhoven}, \citenamefont {Austing},\ and\ \citenamefont
  {Tarucha}}]{kouwenhovenFewelectronQuantumDots2001}%
  \BibitemOpen
  \bibfield  {author} {\bibinfo {author} {\bibfnamefont {L.~P.}\ \bibnamefont
  {Kouwenhoven}}, \bibinfo {author} {\bibfnamefont {D.~G.}\ \bibnamefont
  {Austing}},\ and\ \bibinfo {author} {\bibfnamefont {S.}~\bibnamefont
  {Tarucha}},\ }\bibfield  {title} {\bibinfo {title} {Few-electron quantum
  dots},\ }\href {https://doi.org/10.1088/0034-4885/64/6/201} {\bibfield
  {journal} {\bibinfo  {journal} {Rep. Prog. Phys.}\ }\textbf {\bibinfo
  {volume} {64}},\ \bibinfo {pages} {701} (\bibinfo {year} {2001})}\BibitemShut
  {NoStop}%
\bibitem [{\citenamefont {Vecino}\ \emph {et~al.}(2003)\citenamefont {Vecino},
  \citenamefont {{Mart{\'i}n-Rodero}},\ and\ \citenamefont
  {Levy~Yeyati}}]{vecinoJosephsonCurrentCorrelated2003}%
  \BibitemOpen
  \bibfield  {author} {\bibinfo {author} {\bibfnamefont {E.}~\bibnamefont
  {Vecino}}, \bibinfo {author} {\bibfnamefont {A.}~\bibnamefont
  {{Mart{\'i}n-Rodero}}},\ and\ \bibinfo {author} {\bibfnamefont
  {A.}~\bibnamefont {Levy~Yeyati}},\ }\bibfield  {title} {\bibinfo {title}
  {Josephson current through a correlated quantum level: {{Andreev}} states and
  {{$\pi$}} junction behavior},\ }\href
  {https://doi.org/10.1103/PhysRevB.68.035105} {\bibfield  {journal} {\bibinfo
  {journal} {Phys. Rev. B}\ }\textbf {\bibinfo {volume} {68}},\ \bibinfo
  {pages} {035105} (\bibinfo {year} {2003})}\BibitemShut {NoStop}%
\bibitem [{\citenamefont {Meng}\ \emph {et~al.}(2009)\citenamefont {Meng},
  \citenamefont {Florens},\ and\ \citenamefont
  {Simon}}]{mengSelfconsistentDescriptionAndreev2009}%
  \BibitemOpen
  \bibfield  {author} {\bibinfo {author} {\bibfnamefont {T.}~\bibnamefont
  {Meng}}, \bibinfo {author} {\bibfnamefont {S.}~\bibnamefont {Florens}},\ and\
  \bibinfo {author} {\bibfnamefont {P.}~\bibnamefont {Simon}},\ }\bibfield
  {title} {\bibinfo {title} {Self-consistent description of {{Andreev}} bound
  states in {{Josephson}} quantum dot devices},\ }\href
  {https://doi.org/10.1103/PhysRevB.79.224521} {\bibfield  {journal} {\bibinfo
  {journal} {Phys. Rev. B}\ }\textbf {\bibinfo {volume} {79}},\ \bibinfo
  {pages} {224521} (\bibinfo {year} {2009})}\BibitemShut {NoStop}%
\bibitem [{\citenamefont {{Mart{\'i}n-Rodero}}\ and\ \citenamefont
  {Levy~Yeyati}(2012)}]{martinroderoAndreevStatesSuperconducting2012}%
  \BibitemOpen
  \bibfield  {author} {\bibinfo {author} {\bibfnamefont {A.}~\bibnamefont
  {{Mart{\'i}n-Rodero}}}\ and\ \bibinfo {author} {\bibfnamefont
  {A.}~\bibnamefont {Levy~Yeyati}},\ }\bibfield  {title} {\bibinfo {title} {The
  {{Andreev}} states of a superconducting quantum dot: {{Mean}} field versus
  exact numerical results},\ }\href
  {https://doi.org/10.1088/0953-8984/24/38/385303} {\bibfield  {journal}
  {\bibinfo  {journal} {J. Phys. Condens. Matter}\ }\textbf {\bibinfo {volume}
  {24}},\ \bibinfo {pages} {385303} (\bibinfo {year} {2012})}\BibitemShut
  {NoStop}%
\bibitem [{\citenamefont {Sun}\ \emph {et~al.}(2000)\citenamefont {Sun},
  \citenamefont {Wang}, \citenamefont {Wang},\ and\ \citenamefont
  {Lin}}]{sunElectronTransportMesoscopic2000}%
  \BibitemOpen
  \bibfield  {author} {\bibinfo {author} {\bibfnamefont {Q.-f.}\ \bibnamefont
  {Sun}}, \bibinfo {author} {\bibfnamefont {B.-g.}\ \bibnamefont {Wang}},
  \bibinfo {author} {\bibfnamefont {J.}~\bibnamefont {Wang}},\ and\ \bibinfo
  {author} {\bibfnamefont {T.-h.}\ \bibnamefont {Lin}},\ }\bibfield  {title}
  {\bibinfo {title} {Electron transport through a mesoscopic hybrid
  multiterminal resonant-tunneling system},\ }\href
  {https://doi.org/10.1103/PhysRevB.61.4754} {\bibfield  {journal} {\bibinfo
  {journal} {Phys. Rev. B}\ }\textbf {\bibinfo {volume} {61}},\ \bibinfo
  {pages} {4754} (\bibinfo {year} {2000})}\BibitemShut {NoStop}%
\bibitem [{\citenamefont {Mansikkam{\"a}ki}\ \emph {et~al.}(2025)\citenamefont
  {Mansikkam{\"a}ki}, \citenamefont {Giazotto},\ and\ \citenamefont
  {Balatsky}}]{mansikkamakiThermoelectricACJosephson2025}%
  \BibitemOpen
  \bibfield  {author} {\bibinfo {author} {\bibfnamefont {O.}~\bibnamefont
  {Mansikkam{\"a}ki}}, \bibinfo {author} {\bibfnamefont {F.}~\bibnamefont
  {Giazotto}},\ and\ \bibinfo {author} {\bibfnamefont {A.}~\bibnamefont
  {Balatsky}},\ }\href {https://doi.org/10.48550/arXiv.2504.01884} {\bibinfo
  {title} {Thermoelectric {{AC Josephson}} effect}} (\bibinfo {year} {2025}),\
  \Eprint {https://arxiv.org/abs/2504.01884} {arXiv:2504.01884} \BibitemShut
  {NoStop}%
\bibitem [{\citenamefont {Glazman}\ and\ \citenamefont
  {Matveev}(1989)}]{glazmanResonantJosephsonCurrent1989}%
  \BibitemOpen
  \bibfield  {author} {\bibinfo {author} {\bibfnamefont {L.~I.}\ \bibnamefont
  {Glazman}}\ and\ \bibinfo {author} {\bibfnamefont {K.~A.}\ \bibnamefont
  {Matveev}},\ }\bibfield  {title} {\bibinfo {title} {Resonant {{Josephson}}
  current through {{Kondo}} impurities in a tunnel barrier},\ }\href@noop {}
  {\bibfield  {journal} {\bibinfo  {journal} {JETP Lett.}\ }\textbf {\bibinfo
  {volume} {49}},\ \bibinfo {pages} {659} (\bibinfo {year} {1989})}\BibitemShut
  {NoStop}%
\bibitem [{\citenamefont {Spivak}\ and\ \citenamefont
  {Kivelson}(1991)}]{spivakNegativeLocalSuperfluid1991}%
  \BibitemOpen
  \bibfield  {author} {\bibinfo {author} {\bibfnamefont {B.~I.}\ \bibnamefont
  {Spivak}}\ and\ \bibinfo {author} {\bibfnamefont {S.~A.}\ \bibnamefont
  {Kivelson}},\ }\bibfield  {title} {\bibinfo {title} {Negative local
  superfluid densities: {{The}} difference between dirty superconductors and
  dirty {{Bose}} liquids},\ }\href {https://doi.org/10.1103/PhysRevB.43.3740}
  {\bibfield  {journal} {\bibinfo  {journal} {Phys. Rev. B}\ }\textbf {\bibinfo
  {volume} {43}},\ \bibinfo {pages} {3740} (\bibinfo {year}
  {1991})}\BibitemShut {NoStop}%
\bibitem [{\citenamefont {Zhu}\ \emph {et~al.}(2001)\citenamefont {Zhu},
  \citenamefont {Sun},\ and\ \citenamefont {Lin}}]{zhuAndreevBoundStates2001}%
  \BibitemOpen
  \bibfield  {author} {\bibinfo {author} {\bibfnamefont {Y.}~\bibnamefont
  {Zhu}}, \bibinfo {author} {\bibfnamefont {Q.-f.}\ \bibnamefont {Sun}},\ and\
  \bibinfo {author} {\bibfnamefont {T.-h.}\ \bibnamefont {Lin}},\ }\bibfield
  {title} {\bibinfo {title} {Andreev bound states and the {$\pi$}-junction
  transition in a superconductor/quantum-dot/superconductor system},\ }\href
  {https://doi.org/10.1088/0953-8984/13/39/307} {\bibfield  {journal} {\bibinfo
   {journal} {J. Phys. Condens. Matter}\ }\textbf {\bibinfo {volume} {13}},\
  \bibinfo {pages} {8783} (\bibinfo {year} {2001})}\BibitemShut {NoStop}%
\bibitem [{\citenamefont {{van Dam}}\ \emph {et~al.}(2006)\citenamefont {{van
  Dam}}, \citenamefont {Nazarov}, \citenamefont {Bakkers}, \citenamefont
  {De~Franceschi},\ and\ \citenamefont
  {Kouwenhoven}}]{vandamSupercurrentReversalQuantum2006}%
  \BibitemOpen
  \bibfield  {author} {\bibinfo {author} {\bibfnamefont {J.~A.}\ \bibnamefont
  {{van Dam}}}, \bibinfo {author} {\bibfnamefont {Y.~V.}\ \bibnamefont
  {Nazarov}}, \bibinfo {author} {\bibfnamefont {E.~P. A.~M.}\ \bibnamefont
  {Bakkers}}, \bibinfo {author} {\bibfnamefont {S.}~\bibnamefont
  {De~Franceschi}},\ and\ \bibinfo {author} {\bibfnamefont {L.~P.}\
  \bibnamefont {Kouwenhoven}},\ }\bibfield  {title} {\bibinfo {title}
  {Supercurrent reversal in quantum dots},\ }\href
  {https://doi.org/10.1038/nature05018} {\bibfield  {journal} {\bibinfo
  {journal} {Nature}\ }\textbf {\bibinfo {volume} {442}},\ \bibinfo {pages}
  {667} (\bibinfo {year} {2006})}\BibitemShut {NoStop}%
\bibitem [{\citenamefont {Cleuziou}\ \emph {et~al.}(2006)\citenamefont
  {Cleuziou}, \citenamefont {Wernsdorfer}, \citenamefont {Bouchiat},
  \citenamefont {Ondar{\c c}uhu},\ and\ \citenamefont
  {Monthioux}}]{cleuziouCarbonNanotubeSuperconducting2006}%
  \BibitemOpen
  \bibfield  {author} {\bibinfo {author} {\bibfnamefont {J.-P.}\ \bibnamefont
  {Cleuziou}}, \bibinfo {author} {\bibfnamefont {W.}~\bibnamefont
  {Wernsdorfer}}, \bibinfo {author} {\bibfnamefont {V.}~\bibnamefont
  {Bouchiat}}, \bibinfo {author} {\bibfnamefont {T.}~\bibnamefont {Ondar{\c
  c}uhu}},\ and\ \bibinfo {author} {\bibfnamefont {M.}~\bibnamefont
  {Monthioux}},\ }\bibfield  {title} {\bibinfo {title} {Carbon nanotube
  superconducting quantum interference device},\ }\href
  {https://doi.org/10.1038/nnano.2006.54} {\bibfield  {journal} {\bibinfo
  {journal} {Nat. Nanotechnol.}\ }\textbf {\bibinfo {volume} {1}},\ \bibinfo
  {pages} {53} (\bibinfo {year} {2006})}\BibitemShut {NoStop}%
\bibitem [{\citenamefont {J{\o}rgensen}\ \emph {et~al.}(2007)\citenamefont
  {J{\o}rgensen}, \citenamefont {Novotn{\'y}}, \citenamefont
  {{Grove-Rasmussen}}, \citenamefont {Flensberg},\ and\ \citenamefont
  {Lindelof}}]{jorgensenCriticalCurrentTransition2007}%
  \BibitemOpen
  \bibfield  {author} {\bibinfo {author} {\bibfnamefont {H.~I.}\ \bibnamefont
  {J{\o}rgensen}}, \bibinfo {author} {\bibfnamefont {T.}~\bibnamefont
  {Novotn{\'y}}}, \bibinfo {author} {\bibfnamefont {K.}~\bibnamefont
  {{Grove-Rasmussen}}}, \bibinfo {author} {\bibfnamefont {K.}~\bibnamefont
  {Flensberg}},\ and\ \bibinfo {author} {\bibfnamefont {P.~E.}\ \bibnamefont
  {Lindelof}},\ }\bibfield  {title} {\bibinfo {title} {Critical current
  0--{{$\pi$}} transition in designed {{Josephson}} quantum dot junctions},\
  }\href {https://doi.org/10.1021/nl071152w} {\bibfield  {journal} {\bibinfo
  {journal} {Nano Lett.}\ }\textbf {\bibinfo {volume} {7}},\ \bibinfo {pages}
  {2441} (\bibinfo {year} {2007})}\BibitemShut {NoStop}%
\bibitem [{\citenamefont {{Mart{\'i}n-Rodero}}\ and\ \citenamefont
  {Levy~Yeyati}(2011)}]{martin-roderoJosephsonAndreevTransport2011}%
  \BibitemOpen
  \bibfield  {author} {\bibinfo {author} {\bibfnamefont {A.}~\bibnamefont
  {{Mart{\'i}n-Rodero}}}\ and\ \bibinfo {author} {\bibfnamefont
  {A.}~\bibnamefont {Levy~Yeyati}},\ }\bibfield  {title} {\bibinfo {title}
  {Josephson and {{Andreev}} transport through quantum dots},\ }\href
  {https://doi.org/10.1080/00018732.2011.624266} {\bibfield  {journal}
  {\bibinfo  {journal} {Adv. Phys.}\ }\textbf {\bibinfo {volume} {60}},\
  \bibinfo {pages} {899} (\bibinfo {year} {2011})}\BibitemShut {NoStop}%
\bibitem [{\citenamefont {Buzdin}(2008)}]{buzdinDirectCouplingMagnetism2008}%
  \BibitemOpen
  \bibfield  {author} {\bibinfo {author} {\bibfnamefont {A.}~\bibnamefont
  {Buzdin}},\ }\bibfield  {title} {\bibinfo {title} {Direct {{Coupling Between
  Magnetism}} and {{Superconducting Current}} in the {{Josephson}}
  {$\varphi_0$} {{Junction}}},\ }\href
  {https://doi.org/10.1103/PhysRevLett.101.107005} {\bibfield  {journal}
  {\bibinfo  {journal} {Phys. Rev. Lett.}\ }\textbf {\bibinfo {volume} {101}},\
  \bibinfo {pages} {107005} (\bibinfo {year} {2008})}\BibitemShut {NoStop}%
\bibitem [{\citenamefont
  {Beenakker}(1991)}]{beenakkerUniversalLimitCriticalcurrent1991a}%
  \BibitemOpen
  \bibfield  {author} {\bibinfo {author} {\bibfnamefont {C.~W.~J.}\
  \bibnamefont {Beenakker}},\ }\bibfield  {title} {\bibinfo {title} {Universal
  limit of critical-current fluctuations in mesoscopic {{Josephson}}
  junctions},\ }\href {https://doi.org/10.1103/PhysRevLett.67.3836} {\bibfield
  {journal} {\bibinfo  {journal} {Phys. Rev. Lett.}\ }\textbf {\bibinfo
  {volume} {67}},\ \bibinfo {pages} {3836} (\bibinfo {year}
  {1991})}\BibitemShut {NoStop}%
\bibitem [{\citenamefont {Golubov}\ \emph {et~al.}(2004)\citenamefont
  {Golubov}, \citenamefont {Kupriyanov},\ and\ \citenamefont
  {Il'ichev}}]{golubovCurrentphaseRelationJosephson2004}%
  \BibitemOpen
  \bibfield  {author} {\bibinfo {author} {\bibfnamefont {A.~A.}\ \bibnamefont
  {Golubov}}, \bibinfo {author} {\bibfnamefont {M.~{\relax Yu}.}\ \bibnamefont
  {Kupriyanov}},\ and\ \bibinfo {author} {\bibfnamefont {E.}~\bibnamefont
  {Il'ichev}},\ }\bibfield  {title} {\bibinfo {title} {The current-phase
  relation in {{Josephson}} junctions},\ }\href
  {https://doi.org/10.1103/RevModPhys.76.411} {\bibfield  {journal} {\bibinfo
  {journal} {Rev. Mod. Phys.}\ }\textbf {\bibinfo {volume} {76}},\ \bibinfo
  {pages} {411} (\bibinfo {year} {2004})}\BibitemShut {NoStop}%
\bibitem [{\citenamefont
  {Buzdin}(2005)}]{buzdinProximityEffectsSuperconductorferromagnet2005}%
  \BibitemOpen
  \bibfield  {author} {\bibinfo {author} {\bibfnamefont {A.~I.}\ \bibnamefont
  {Buzdin}},\ }\bibfield  {title} {\bibinfo {title} {Proximity effects in
  superconductor-ferromagnet heterostructures},\ }\href
  {https://doi.org/10.1103/RevModPhys.77.935} {\bibfield  {journal} {\bibinfo
  {journal} {Rev. Mod. Phys.}\ }\textbf {\bibinfo {volume} {77}},\ \bibinfo
  {pages} {935} (\bibinfo {year} {2005})}\BibitemShut {NoStop}%
\bibitem [{\citenamefont {Padurariu}\ and\ \citenamefont
  {Nazarov}(2010)}]{padurariuTheoreticalProposalSuperconducting2010}%
  \BibitemOpen
  \bibfield  {author} {\bibinfo {author} {\bibfnamefont {C.}~\bibnamefont
  {Padurariu}}\ and\ \bibinfo {author} {\bibfnamefont {{\relax Yu}.~V.}\
  \bibnamefont {Nazarov}},\ }\bibfield  {title} {\bibinfo {title} {Theoretical
  proposal for superconducting spin qubits},\ }\href
  {https://doi.org/10.1103/PhysRevB.81.144519} {\bibfield  {journal} {\bibinfo
  {journal} {Phys. Rev. B}\ }\textbf {\bibinfo {volume} {81}},\ \bibinfo
  {pages} {144519} (\bibinfo {year} {2010})}\BibitemShut {NoStop}%
\bibitem [{\citenamefont {Jabdaraghi}\ \emph {et~al.}(2017)\citenamefont
  {Jabdaraghi}, \citenamefont {Golubev}, \citenamefont {Pekola},\ and\
  \citenamefont {Peltonen}}]{jabdaraghiNoiseSuperconductingMagnetic2017}%
  \BibitemOpen
  \bibfield  {author} {\bibinfo {author} {\bibfnamefont {R.~N.}\ \bibnamefont
  {Jabdaraghi}}, \bibinfo {author} {\bibfnamefont {D.~S.}\ \bibnamefont
  {Golubev}}, \bibinfo {author} {\bibfnamefont {J.~P.}\ \bibnamefont
  {Pekola}},\ and\ \bibinfo {author} {\bibfnamefont {J.~T.}\ \bibnamefont
  {Peltonen}},\ }\bibfield  {title} {\bibinfo {title} {Noise of a
  superconducting magnetic flux sensor based on a proximity {{Josephson}}
  junction},\ }\href {https://doi.org/10.1038/s41598-017-08710-7} {\bibfield
  {journal} {\bibinfo  {journal} {Sci. Rep.}\ }\textbf {\bibinfo {volume}
  {7}},\ \bibinfo {pages} {8011} (\bibinfo {year} {2017})}\BibitemShut
  {NoStop}%
\bibitem [{\citenamefont {Jekat}\ \emph {et~al.}(2020)\citenamefont {Jekat},
  \citenamefont {Pestka}, \citenamefont {Car}, \citenamefont {Gazibegovi{\'c}},
  \citenamefont {Fl{\"o}hr}, \citenamefont {Heedt}, \citenamefont {Schubert},
  \citenamefont {Liebmann}, \citenamefont {Bakkers}, \citenamefont
  {Sch{\"a}pers},\ and\ \citenamefont
  {Morgenstern}}]{jekatExfoliatedHexagonalBN2020}%
  \BibitemOpen
  \bibfield  {author} {\bibinfo {author} {\bibfnamefont {F.}~\bibnamefont
  {Jekat}}, \bibinfo {author} {\bibfnamefont {B.}~\bibnamefont {Pestka}},
  \bibinfo {author} {\bibfnamefont {D.}~\bibnamefont {Car}}, \bibinfo {author}
  {\bibfnamefont {S.}~\bibnamefont {Gazibegovi{\'c}}}, \bibinfo {author}
  {\bibfnamefont {K.}~\bibnamefont {Fl{\"o}hr}}, \bibinfo {author}
  {\bibfnamefont {S.}~\bibnamefont {Heedt}}, \bibinfo {author} {\bibfnamefont
  {J.}~\bibnamefont {Schubert}}, \bibinfo {author} {\bibfnamefont
  {M.}~\bibnamefont {Liebmann}}, \bibinfo {author} {\bibfnamefont {E.~P.
  A.~M.}\ \bibnamefont {Bakkers}}, \bibinfo {author} {\bibfnamefont
  {T.}~\bibnamefont {Sch{\"a}pers}},\ and\ \bibinfo {author} {\bibfnamefont
  {M.}~\bibnamefont {Morgenstern}},\ }\bibfield  {title} {\bibinfo {title}
  {Exfoliated hexagonal {{BN}} as gate dielectric for {{InSb}} nanowire quantum
  dots with improved gate hysteresis and charge noise},\ }\href
  {https://doi.org/10.1063/5.0002112} {\bibfield  {journal} {\bibinfo
  {journal} {Appl. Phys. Lett.}\ }\textbf {\bibinfo {volume} {116}},\ \bibinfo
  {pages} {253101} (\bibinfo {year} {2020})}\BibitemShut {NoStop}%
\bibitem [{\citenamefont {Hendrickx}\ \emph {et~al.}(2018)\citenamefont
  {Hendrickx}, \citenamefont {Franke}, \citenamefont {Sammak}, \citenamefont
  {Kouwenhoven}, \citenamefont {Sabbagh}, \citenamefont {Yeoh}, \citenamefont
  {Li}, \citenamefont {Tagliaferri}, \citenamefont {Virgilio}, \citenamefont
  {Capellini}, \citenamefont {Scappucci},\ and\ \citenamefont
  {Veldhorst}}]{hendrickxGatecontrolledQuantumDots2018}%
  \BibitemOpen
  \bibfield  {author} {\bibinfo {author} {\bibfnamefont {N.~W.}\ \bibnamefont
  {Hendrickx}}, \bibinfo {author} {\bibfnamefont {D.~P.}\ \bibnamefont
  {Franke}}, \bibinfo {author} {\bibfnamefont {A.}~\bibnamefont {Sammak}},
  \bibinfo {author} {\bibfnamefont {M.}~\bibnamefont {Kouwenhoven}}, \bibinfo
  {author} {\bibfnamefont {D.}~\bibnamefont {Sabbagh}}, \bibinfo {author}
  {\bibfnamefont {L.}~\bibnamefont {Yeoh}}, \bibinfo {author} {\bibfnamefont
  {R.}~\bibnamefont {Li}}, \bibinfo {author} {\bibfnamefont {M.~L.~V.}\
  \bibnamefont {Tagliaferri}}, \bibinfo {author} {\bibfnamefont
  {M.}~\bibnamefont {Virgilio}}, \bibinfo {author} {\bibfnamefont
  {G.}~\bibnamefont {Capellini}}, \bibinfo {author} {\bibfnamefont
  {G.}~\bibnamefont {Scappucci}},\ and\ \bibinfo {author} {\bibfnamefont
  {M.}~\bibnamefont {Veldhorst}},\ }\bibfield  {title} {\bibinfo {title}
  {Gate-controlled quantum dots and superconductivity in planar germanium},\
  }\href {https://doi.org/10.1038/s41467-018-05299-x} {\bibfield  {journal}
  {\bibinfo  {journal} {Nat. Commun.}\ }\textbf {\bibinfo {volume} {9}},\
  \bibinfo {pages} {2835} (\bibinfo {year} {2018})}\BibitemShut {NoStop}%
\bibitem [{\citenamefont {Yoneda}\ \emph {et~al.}(2018)\citenamefont {Yoneda},
  \citenamefont {Takeda}, \citenamefont {Otsuka}, \citenamefont {Nakajima},
  \citenamefont {Delbecq}, \citenamefont {Allison}, \citenamefont {Honda},
  \citenamefont {Kodera}, \citenamefont {Oda}, \citenamefont {Hoshi},
  \citenamefont {Usami}, \citenamefont {Itoh},\ and\ \citenamefont
  {Tarucha}}]{yonedaQuantumdotSpinQubit2018}%
  \BibitemOpen
  \bibfield  {author} {\bibinfo {author} {\bibfnamefont {J.}~\bibnamefont
  {Yoneda}}, \bibinfo {author} {\bibfnamefont {K.}~\bibnamefont {Takeda}},
  \bibinfo {author} {\bibfnamefont {T.}~\bibnamefont {Otsuka}}, \bibinfo
  {author} {\bibfnamefont {T.}~\bibnamefont {Nakajima}}, \bibinfo {author}
  {\bibfnamefont {M.~R.}\ \bibnamefont {Delbecq}}, \bibinfo {author}
  {\bibfnamefont {G.}~\bibnamefont {Allison}}, \bibinfo {author} {\bibfnamefont
  {T.}~\bibnamefont {Honda}}, \bibinfo {author} {\bibfnamefont
  {T.}~\bibnamefont {Kodera}}, \bibinfo {author} {\bibfnamefont
  {S.}~\bibnamefont {Oda}}, \bibinfo {author} {\bibfnamefont {Y.}~\bibnamefont
  {Hoshi}}, \bibinfo {author} {\bibfnamefont {N.}~\bibnamefont {Usami}},
  \bibinfo {author} {\bibfnamefont {K.~M.}\ \bibnamefont {Itoh}},\ and\
  \bibinfo {author} {\bibfnamefont {S.}~\bibnamefont {Tarucha}},\ }\bibfield
  {title} {\bibinfo {title} {A quantum-dot spin qubit with coherence limited by
  charge noise and fidelity higher than 99.9\%},\ }\href
  {https://doi.org/10.1038/s41565-017-0014-x} {\bibfield  {journal} {\bibinfo
  {journal} {Nat. Nanotechnol.}\ }\textbf {\bibinfo {volume} {13}},\ \bibinfo
  {pages} {102} (\bibinfo {year} {2018})}\BibitemShut {NoStop}%
\bibitem [{\citenamefont {Bj{\"o}rk}\ \emph {et~al.}(2005)\citenamefont
  {Bj{\"o}rk}, \citenamefont {Fuhrer}, \citenamefont {Hansen}, \citenamefont
  {Larsson}, \citenamefont {Fr{\"o}berg},\ and\ \citenamefont
  {Samuelson}}]{bjorkTunableEffectiveFactor2005}%
  \BibitemOpen
  \bibfield  {author} {\bibinfo {author} {\bibfnamefont {M.~T.}\ \bibnamefont
  {Bj{\"o}rk}}, \bibinfo {author} {\bibfnamefont {A.}~\bibnamefont {Fuhrer}},
  \bibinfo {author} {\bibfnamefont {A.~E.}\ \bibnamefont {Hansen}}, \bibinfo
  {author} {\bibfnamefont {M.~W.}\ \bibnamefont {Larsson}}, \bibinfo {author}
  {\bibfnamefont {L.~E.}\ \bibnamefont {Fr{\"o}berg}},\ and\ \bibinfo {author}
  {\bibfnamefont {L.}~\bibnamefont {Samuelson}},\ }\bibfield  {title} {\bibinfo
  {title} {Tunable effective {{g}} factor in {{InAs}} nanowire quantum dots},\
  }\href {https://doi.org/10.1103/PhysRevB.72.201307} {\bibfield  {journal}
  {\bibinfo  {journal} {Phys. Rev. B}\ }\textbf {\bibinfo {volume} {72}},\
  \bibinfo {pages} {201307(R)} (\bibinfo {year} {2005})}\BibitemShut {NoStop}%
\bibitem [{\citenamefont {Vasyukov}\ \emph {et~al.}(2013)\citenamefont
  {Vasyukov}, \citenamefont {Anahory}, \citenamefont {Embon}, \citenamefont
  {Halbertal}, \citenamefont {Cuppens}, \citenamefont {Neeman}, \citenamefont
  {Finkler}, \citenamefont {Segev}, \citenamefont {Myasoedov}, \citenamefont
  {Rappaport}, \citenamefont {Huber},\ and\ \citenamefont
  {Zeldov}}]{vasyukovScanningSuperconductingQuantum2013}%
  \BibitemOpen
  \bibfield  {author} {\bibinfo {author} {\bibfnamefont {D.}~\bibnamefont
  {Vasyukov}}, \bibinfo {author} {\bibfnamefont {Y.}~\bibnamefont {Anahory}},
  \bibinfo {author} {\bibfnamefont {L.}~\bibnamefont {Embon}}, \bibinfo
  {author} {\bibfnamefont {D.}~\bibnamefont {Halbertal}}, \bibinfo {author}
  {\bibfnamefont {J.}~\bibnamefont {Cuppens}}, \bibinfo {author} {\bibfnamefont
  {L.}~\bibnamefont {Neeman}}, \bibinfo {author} {\bibfnamefont
  {A.}~\bibnamefont {Finkler}}, \bibinfo {author} {\bibfnamefont
  {Y.}~\bibnamefont {Segev}}, \bibinfo {author} {\bibfnamefont
  {Y.}~\bibnamefont {Myasoedov}}, \bibinfo {author} {\bibfnamefont {M.~L.}\
  \bibnamefont {Rappaport}}, \bibinfo {author} {\bibfnamefont {M.~E.}\
  \bibnamefont {Huber}},\ and\ \bibinfo {author} {\bibfnamefont
  {E.}~\bibnamefont {Zeldov}},\ }\bibfield  {title} {\bibinfo {title} {A
  scanning superconducting quantum interference device with single electron
  spin sensitivity},\ }\href {https://doi.org/10.1038/nnano.2013.169}
  {\bibfield  {journal} {\bibinfo  {journal} {Nat. Nanotechnol.}\ }\textbf
  {\bibinfo {volume} {8}},\ \bibinfo {pages} {639} (\bibinfo {year}
  {2013})}\BibitemShut {NoStop}%
\bibitem [{\citenamefont {Lodari}\ \emph {et~al.}(2021)\citenamefont {Lodari},
  \citenamefont {Hendrickx}, \citenamefont {Lawrie}, \citenamefont {Hsiao},
  \citenamefont {Vandersypen}, \citenamefont {Sammak}, \citenamefont
  {Veldhorst},\ and\ \citenamefont
  {Scappucci}}]{lodariLowPercolationDensity2021}%
  \BibitemOpen
  \bibfield  {author} {\bibinfo {author} {\bibfnamefont {M.}~\bibnamefont
  {Lodari}}, \bibinfo {author} {\bibfnamefont {N.~W.}\ \bibnamefont
  {Hendrickx}}, \bibinfo {author} {\bibfnamefont {W.~I.~L.}\ \bibnamefont
  {Lawrie}}, \bibinfo {author} {\bibfnamefont {T.-K.}\ \bibnamefont {Hsiao}},
  \bibinfo {author} {\bibfnamefont {L.~M.~K.}\ \bibnamefont {Vandersypen}},
  \bibinfo {author} {\bibfnamefont {A.}~\bibnamefont {Sammak}}, \bibinfo
  {author} {\bibfnamefont {M.}~\bibnamefont {Veldhorst}},\ and\ \bibinfo
  {author} {\bibfnamefont {G.}~\bibnamefont {Scappucci}},\ }\bibfield  {title}
  {\bibinfo {title} {Low percolation density and charge noise with holes in
  germanium},\ }\href {https://doi.org/10.1088/2633-4356/abcd82} {\bibfield
  {journal} {\bibinfo  {journal} {Mater. Quantum Technol.}\ }\textbf {\bibinfo
  {volume} {1}},\ \bibinfo {pages} {011002} (\bibinfo {year}
  {2021})}\BibitemShut {NoStop}%
\bibitem [{\citenamefont {Paquelet~Wuetz}\ \emph {et~al.}(2023)\citenamefont
  {Paquelet~Wuetz}, \citenamefont {Degli~Esposti}, \citenamefont {Zwerver},
  \citenamefont {Amitonov}, \citenamefont {Botifoll}, \citenamefont {Arbiol},
  \citenamefont {Sammak}, \citenamefont {Vandersypen}, \citenamefont {Russ},\
  and\ \citenamefont {Scappucci}}]{paqueletwuetzReducingChargeNoise2023}%
  \BibitemOpen
  \bibfield  {author} {\bibinfo {author} {\bibfnamefont {B.}~\bibnamefont
  {Paquelet~Wuetz}}, \bibinfo {author} {\bibfnamefont {D.}~\bibnamefont
  {Degli~Esposti}}, \bibinfo {author} {\bibfnamefont {A.-M.~J.}\ \bibnamefont
  {Zwerver}}, \bibinfo {author} {\bibfnamefont {S.~V.}\ \bibnamefont
  {Amitonov}}, \bibinfo {author} {\bibfnamefont {M.}~\bibnamefont {Botifoll}},
  \bibinfo {author} {\bibfnamefont {J.}~\bibnamefont {Arbiol}}, \bibinfo
  {author} {\bibfnamefont {A.}~\bibnamefont {Sammak}}, \bibinfo {author}
  {\bibfnamefont {L.~M.~K.}\ \bibnamefont {Vandersypen}}, \bibinfo {author}
  {\bibfnamefont {M.}~\bibnamefont {Russ}},\ and\ \bibinfo {author}
  {\bibfnamefont {G.}~\bibnamefont {Scappucci}},\ }\bibfield  {title} {\bibinfo
  {title} {Reducing charge noise in quantum dots by using thin silicon quantum
  wells},\ }\href {https://doi.org/10.1038/s41467-023-36951-w} {\bibfield
  {journal} {\bibinfo  {journal} {Nat. Commun.}\ }\textbf {\bibinfo {volume}
  {14}},\ \bibinfo {pages} {1385} (\bibinfo {year} {2023})}\BibitemShut
  {NoStop}%
\bibitem [{\citenamefont {Stefanucci}\ and\ \citenamefont {{van
  Leeuwen}}(2013)}]{stefanucciNonequilibriumManyBodyTheory2013}%
  \BibitemOpen
  \bibfield  {author} {\bibinfo {author} {\bibfnamefont {G.}~\bibnamefont
  {Stefanucci}}\ and\ \bibinfo {author} {\bibfnamefont {R.}~\bibnamefont {{van
  Leeuwen}}},\ }\href {https://doi.org/10.1017/CBO9781139023979} {\emph
  {\bibinfo {title} {Nonequilibrium {{Many-Body Theory}} of {{Quantum
  Systems}}: {{A Modern Introduction}}}}}\ (\bibinfo  {publisher} {Cambridge
  University Press},\ \bibinfo {address} {Cambridge},\ \bibinfo {year}
  {2013})\BibitemShut {NoStop}%
\bibitem [{\citenamefont {Haug}\ and\ \citenamefont
  {Jauho}(2008)}]{QuantumKineticsTransport2008}%
  \BibitemOpen
  \bibfield  {author} {\bibinfo {author} {\bibfnamefont {H.}~\bibnamefont
  {Haug}}\ and\ \bibinfo {author} {\bibfnamefont {A.-P.}\ \bibnamefont
  {Jauho}},\ }\href {https://doi.org/10.1007/978-3-540-73564-9} {\emph
  {\bibinfo {title} {Quantum {{Kinetics}} in {{Transport}} and {{Optics}} of
  {{Semiconductors}}}}},\ \bibinfo {series} {Springer {{Series}} in
  {{Solid-State Sciences}}}, Vol.\ \bibinfo {volume} {123}\ (\bibinfo
  {publisher} {Springer},\ \bibinfo {address} {Berlin, Heidelberg},\ \bibinfo
  {year} {2008})\BibitemShut {NoStop}%
\bibitem [{\citenamefont
  {Langreth}(1976)}]{langrethLinearNonlinearResponse1976}%
  \BibitemOpen
  \bibfield  {author} {\bibinfo {author} {\bibfnamefont {D.~C.}\ \bibnamefont
  {Langreth}},\ }\bibfield  {title} {\bibinfo {title} {Linear and {{Nonlinear
  Response Theory}} with {{Applications}}},\ }in\ \href
  {https://doi.org/10.1007/978-1-4757-0875-2_1} {\emph {\bibinfo {booktitle}
  {Linear and {{Nonlinear Electron Transport}} in {{Solids}}}}},\ \bibinfo
  {editor} {edited by\ \bibinfo {editor} {\bibfnamefont {J.~T.}\ \bibnamefont
  {Devreese}}\ and\ \bibinfo {editor} {\bibfnamefont {V.~E.}\ \bibnamefont
  {{van Doren}}}}\ (\bibinfo  {publisher} {Springer US},\ \bibinfo {address}
  {Boston, MA},\ \bibinfo {year} {1976})\ pp.\ \bibinfo {pages}
  {3--32}\BibitemShut {NoStop}%
\bibitem [{Note1()}]{Note1}%
  \BibitemOpen
  \bibinfo {note} {At zero Zeeman field, unrestricted Hartree-Fock may yield a
  pair of degenerate spin-polarized solutions related by time reversal. These
  solutions represent local-moment formation at the mean-field level, but
  selecting one of them would assign a permanent spin polarization to the
  finite dot. We therefore restore the symmetry of spin-even observables by
  including both time-reversed solutions. For a macroscopic ferromagnet, by
  contrast, the reversal time between the two sectors becomes effectively
  infinite, and a single symmetry-broken state is physically realized on
  experimental time scales.}\BibitemShut {Stop}%
\end{thebibliography}%
        
\end{document}